\documentclass{emulateapj}
\pdfoutput=1

\usepackage{epstopdf}
\usepackage{longtable}
\usepackage{tabularx}

\newcommand{\HST}{{\sl HST}}
\newcommand{\ACS}{{\sl ACS}}

\newcommand{\flt}{{\tt $\ast$\_flt.fits}}
\newcommand{\drz}{{\tt $\ast$\_drz.fits}}

\slugcomment{Accept on The Astrophysical Journal Supplement Series, March 2013}

\shorttitle{\HST\ Treasury Program on Orion}
\shortauthors{Robberto et al.}

\begin{document}

\title{The {\sl Hubble Space Telescope} Treasury Program on the Orion Nebula Cluster
\altaffilmark{$\dagger$}\altaffilmark{$\ddagger$}
}

\author{M.~Robberto\altaffilmark{1},
        D.~R.~Soderblom\altaffilmark{1},
        E.~Bergeron\altaffilmark{1},
        V.~Kozhurina-Platais\altaffilmark{1},
        R.~B.~Makidon\altaffilmark{1},
        P.~R.~McCullough\altaffilmark{1},
        M.~McMaster\altaffilmark{1},
        N.~Panagia\altaffilmark{1,14},
        I.~N.~Reid\altaffilmark{1},
        Z. ~Levay\altaffilmark{1},
        L.~Frattare\altaffilmark{1},
        N.~Da~Rio\altaffilmark{2},
        M.~Andersen\altaffilmark{2},
        C.~R.~O'Dell\altaffilmark{3},
        K.~G.~Stassun\altaffilmark{3,15,16},
        M.~Simon\altaffilmark{4},
        E.~D.~Feigelson\altaffilmark{5},
        J.~R.~Stauffer\altaffilmark{6},
        M.~Meyer\altaffilmark{7},
	M.~Reggiani\altaffilmark{7},
        J.~Krist\altaffilmark{8},
        C.~F.~Manara\altaffilmark{9}
        M.~Romaniello\altaffilmark{9},
        L.~A.~Hillenbrand\altaffilmark{10},
        L.~Ricci\altaffilmark{10},
        F.~Palla\altaffilmark{11},
        J.~R.~Najita\altaffilmark{12},
        T.~T.~Ananna\altaffilmark{13}
        G.~Scandariato\altaffilmark{14}
        K.~Smith\altaffilmark{17}
        }
\altaffiltext{$\dagger$}{This paper is dedicated to the memory of our friend and colleague Russell B. Makidon, passed away on June 22, 2009.}
\altaffiltext{$\ddagger$}{Based on observations with the NASA / ESA Hubble Space Telescope, obtained at the Space Telescope Science Institute, which is operated by Association of Universities for Research in Astronomy, Inc., under NASA contract NAS5-26555.}
\altaffiltext{1}{Space Telescope Science Institute, 3700 San Martin Drive, Baltimore, MD 21218, USA}
\altaffiltext{2}{European Space Agency, Keplerlaan 1, 2200 AG Noordwijk, The Netherlands}
\altaffiltext{3}{Vanderbilt University, Department of Physics \& Astronomy 6301 Stevenson Center, Nashville, TN 37235, USA}
\altaffiltext{4}{Stony Brook University, Department of Physics \& Astronomy, Stony Brook, NY, 11794, USA}
\altaffiltext{5}{Pennsylvania State University, Department of Astronomy and Astrophysics, 518 Davey Lab, University Park, PA 16802, USA}
\altaffiltext{6}{Spitzer Science Center, California Institute of Technology 314-6, Pasadena, CA 91125, USA}
\altaffiltext{7}{ETH Z\"{u}rich, Institut f\"{u}r Astronomie, Wolfgang-Pauli-Strasse 27, CH-8093 Z\"{u}rich, Switzerland}
\altaffiltext{8}{Jet Propulsion Laboratory, California Institute of Technology, 4800 Oak Grove Drive, Pasadena, CA 91109, USA}
\altaffiltext{9}{European Southern Observatory, Karl-Schwarzschild-Strasse 2, D-85748 Garching, Germany}
\altaffiltext{10}{California Institute of Technology, Cahill Center for Astronomy and Astrophysics, 1200 East California Boulervard, 91125 Pasadena, CA, USA}
\altaffiltext{11}{INAF-Osservatorio Astrofisico di Arcetri, Largo Enrico Fermi, 5 I-50125 Firenze, Italy}
\altaffiltext{12}{National Optical Astronomy Observatories, 950 N.~Cherry Ave, Tucson, AZ 85719, USA}
\altaffiltext{13}{Bryn Mawr College, 101 North Merion Avenue, Bryn Mawr, PA 19010, USA}
\altaffiltext{14}{INAF-CT Osservatorio Astrofisico di Catania, Via S. Sofia 79, I-95123 Catania, Italy}
\altaffiltext{15}{Fisk University, Department of Physics, 1000 17th Ave. N., Nashville, TN 37208, USA}
\altaffiltext{16}{Massachusetts Institute of Technology, Department of Physics, 77 Massachusetts Ave., Cambridge, MA 02139, USA}
\altaffiltext{17}{Max Planck Institut f\"ur Astronomie, K\"onigstuhl 17, D-69117 Heidelberg, Germany }
\email{contact author: robberto@stsci.edu}

\begin{abstract}
The {\sl Hubble Space Telescope (HST) Treasury Program on the Orion Nebula Cluster} has used 104 orbits of \HST\ time to image the Great Orion Nebula region with the {\sl Advanced Camera for Surveys (ACS)}, the {\sl Wide-Field/Planetary Camera 2 (WFPC2)} and the {\sl Near Infrared Camera and Multi Object Spectrograph (NICMOS)} instruments in 11 filters ranging from the U-band to the H-band equivalent of \HST. The program has been intended to perform the definitive study of the stellar component of the ONC at visible wavelengths, addressing key questions like the cluster IMF, age spread, mass accretion, binarity and cirumstellar disk evolution. 
The scanning pattern allowed to cover a contiguous field of approximately 600 square arcminutes with both ACS and WFPC2, with a typical exposure time of approximately 11 minutes per ACS filter, corresponding to a point source depth AB(F435W) = 25.8 and AB(F775W)=25.2 with 0.2 magnitudes of photometric error. We describe the observations, data reduction and data products, including images, source catalogs and tools for quick look preview. In particular, we provide ACS photometry for 3399 stars, most of them detected at multiple epochs, WFPC2 photometry for 1643 stars, 1021 of them detected in the U-band, and NICMOS JH photometry for 2116 stars. We summarize the early science results that have been presented  in a number of papers. 
The final set of images and the photometric catalogs are publicly available through the archive as High Level Science Products at the  STScI Multimission Archive hosted by the Space Telescope Science Institute.
\end{abstract}

\keywords{ISM: individual (M42), stars: formation, stars: pre-main sequence, stars: low-mass, brown dwarfs}
 
\section{Introduction}\label{sec:Introduction}
As the nearest active site of massive star formation, the Orion Nebula (Messier 42, NGC1976) and its associated young stellar cluster -- the Orion Nebula Cluster (ONC) --  provide a unique opportunity for studying the star formation process at the present epoch in our Galaxy \citep{pudritz2002}.

Young (a few Myr old), relatively massive ($\gtrsim10^2-10^3$M$_\odot$) stellar clusters like the ONC contain, besides a large number of low mass objects ($M<1$M$_\odot$, \citealt{briceno2007}), also OB stars. These massive, early type stars disrupt the placental molecular cloud and affect the evolution of the multitude of surrounding low mass stars through their strong ionizing radiation, line-driven winds, and induced photoevaporative flows. Mass ejection from low mass cluster members,  close dynamical encounters and substantial and variable X-ray emission also contribute to create a harsh environment which may critically affect planet formation \citep{bally2000,scally2005,eisner2008}. Images of photo-evaporating circumstellar disks taken with the Hubble Space Telescope \citep[e.g.,][]{odell-wong1996} have shown that the canonical scenario for star formation, valid for isolated low-mass  stars (M $\lesssim1$~M$_\odot$) quietly forming in sparse, low mass clusters (T associations), may not adequately account for typical star formation in rich clusters \citep{ladalada2003}. As the Sun appears to have formed in a similar environment \citep{hester2005,williams2007}, understanding the ONC may shed light not only on key passages of the star and planet formation in general, but also on the origin of our own planetary system \citep{looney2006}.

Due to its location and structure \citep{ODell2008}, the ONC can be studied in great detail. It is close ($d\simeq436\pm20$~pc, \citealt{ODellHenney2008}), at relatively high galactic latitude ($b=-19$\degr, corresponding to 135~pc from the Galactic plane), and in an anti-center quadrant ($l=209$\degr) with minimal foreground confusion. The large extinction of OMC-1, (up to $A_V=50^{m}-100^m$) on the main ridge eliminates background confusion down to almost the H-burning limit \citep{hillenbrand2000,scandariato2011}. The moderate extinction ($A_V=1.5^m$) caused by a foreground veil of neutral gas encompassing the nearest members of the ONC \citep{odell-yusefzadeh2000} allows detailed studies at visible wavelengths of individual stars.

The young stellar population of the ONC has been studied for decades. One of the most outstanding works to characterize the individual properties of these pre-main sequance (PMS) members is that of \citet{hillenbrand97}. In particular, she assembled a large photometric and spectroscopic database at visible wavelengths down to $I\sim18$~mag. By estimating the stellar extinction for more than 1000 individual sources down to $\sim1.0$~M$_\odot$,  she derived absolute luminosities, ages and masses. More recently \citep{dario2009,dario2010,2012ApJ...748...14D} the census of PMS stars in Orion has been extended and the stellar parameters derived for each source (e.g., effective temperature $T_{\rm. eff}$; luminosity $L_{\rm bol}$) further refined. The isochronal age of the ONC is about 1-2~Myr, with evidence for a spread in ages \citep{reggiani2011}, although its actual extent in time is still debated \citep{jeffries2011}.  In what concerns the mass distribution, it peaks at about $0.3$~M$_\odot$, but the shape of the stellar initial mass function (IMF) is highly dependent on the assumed evolutionary models \citep{dario2010}.

Over the last decade several near infrared surveys, both in imaging \citep[e.g.,][]{lucasroche2000,hillenbrand2000,luhman2000,kaifu2000,muench2002,lada2000,lucas2006} and in spectroscopy \citep[e.g.,][]{lucas2006,slesnick04,riddick2007} have probed the IMF well into the brown dwarf regime down to the planetary mass range. The combination of photometry and spectroscopy is obviously a most powerful tool, allowing one to estimate the stellar parameters for individual sources; however, the spectroscopic surveys in the ONC have been limited either to the central part of the region, or to a sparse sampling of the entire cluster.

Despite the large amount of data collected over the years in the ONC, the accurate estimate of the properties of its population remains challenging. Fluxes of PMS stars are generally contaminated by UV and optical flux excesses due to accretion processes \citep{appenzeller-mundt1989,gullbring1998} as well as by the near infrared \citep{meyer1997} emission from circumstellar disks. Also, the differential extinction of the different sources partially embedded in the parental cloud, limits a precise estimate of their properties, due to the strong degeneracy between $T_{\rm eff}$ and $A_V$ \citep{hillenbrand97}; this is mitigated by having at disposal multiple observed colors. Moreover, young stars show evident photometric variability \citep{herbst1994,herbst2002}, which also introduces further uncertainties. For what concerns the substellar young members in Orion, their faint luminosity generally limits their optical observational investigation to those least affected by dust extinction. Finally, the non-uniformity of the nebular background, both at large scales and in the local stellar vicinity (due to the possible presence of circumstellar material, either dusty disks or photoevaporating ones) requires imaging with high angular resolution, not achievable using ground-based facilities.

It is clear that under these circumstances, a panchromatic survey of the ONC with the high sensitivity and angular resolution of \HST\ guarantees an unprecedented, outstanding data set which enables to overcome these problems and  attack the open questions about this star forming region. Specifically, the main aspects to understand in depth include: 1) What is the shape of the IMF, down to the lowest substellar masses in Orion? Does it differ from other star forming regions, and from the Galaxy field? 2) What is the age of the system, and what have the time scales of its formation been? 3) How does mass accretion take place? How does it depend on the properties of the central objects? 4) What are the properties of the circumstellar disks, and to what extent are they affected by the harsh environment of the Orion Nebula? Can they form planets? 5) What are the properties of multiple stellar system in Orion, and their circumstellar disks? 

The particular abilities of \HST\ to investigate the ONC were soon recognized in the early years of the mission, when \citet{prosser94} performed a photometric survey of the Trapezium region \citep{herbig86}, the inner 0.3~pc (3.5~arcmin) of the ONC. Unfortunately, the photometry derived from these data, taken in 1991 with the spherically aberrated WFPC1, is subject to uncertainties as large as 0.4 magnitudes. Several observing programs have been executed since then with all instruments onboard \HST, mostly dedicated to high resolution imaging in narrow-band filters of the central region to study the photoionized and dark silhouette disks \citep[e.g.,][]{odell-wen1994, mccaughreanodell1996,bally2000}. A wide area survey (\HST\ program GO-9825, P.I. J.~Bally) has also covered a large part of the Orion Nebula in the H$\alpha$ filter of ACS. 

The \HST\ Treasury Program on the ONC, described in this paper, has been intended to perform the definitive study of the stellar component of the ONC at visible wavelengths. The program, awarded with a total of 104 orbits (Cycle 13, GO program 10246, P.I. M. Robberto), has covered the brightest regions of the Great Orion Nebula at optical wavelengths, using  broad band filters to obtain the most accurate photometry of the largest possible number of pre-main-sequence objects.  All imagers onboard the \HST\ have been used, namely the {\sl Advanced Camera for Surveys} in its Wide Field Channel mode (ACS/WFC) the {\sl Wide Field and Planetary Camera 2 (WFPC2)} and the {\sl Near Infrared Camera and Multi-Object Spectrograph (NICMOS)} in Camera 3 mode. The \HST\ observations have been complemented by a ground-based campaign with obtained nearly {\sl simultaneous} photometry of the cluster from the U-band to the K-band \citep{dario2009,Robberto+2010}.
Given the complexity of the observations, the nonconventional use of the \HST\ and  the large amount of data (more than 2,000 single exposures) that were obtained, we provide in this paper a description of the program and  a guide to the  main data products.
In Section~\ref{section:observations} we describe the observations; in Section~\ref{section:datareduction} the reduction and processing of the imaging data; Section~\ref{section:photometry} presents the details of photometry
performed for each instrument. In Section~\ref{section:dataproducts} we summarize the data products we are publicly releasing, consisting of photometric catalogs and atlases and mosaiced images. Finally, in Section~\ref{section:results} we outline the main results that have been presented so far based on the \HST\ Orion Treasury program. Also, as an appendix, we present in detail the realization of the spectacular color rendering of the region based on our ACS observations.

\section{Observations} \label{section:observations}
\subsection{Instrument characteristics and set-up} \label{section:observations:instruments-setup}

\begin{figure}
\epsscale{1.1}
\plotone{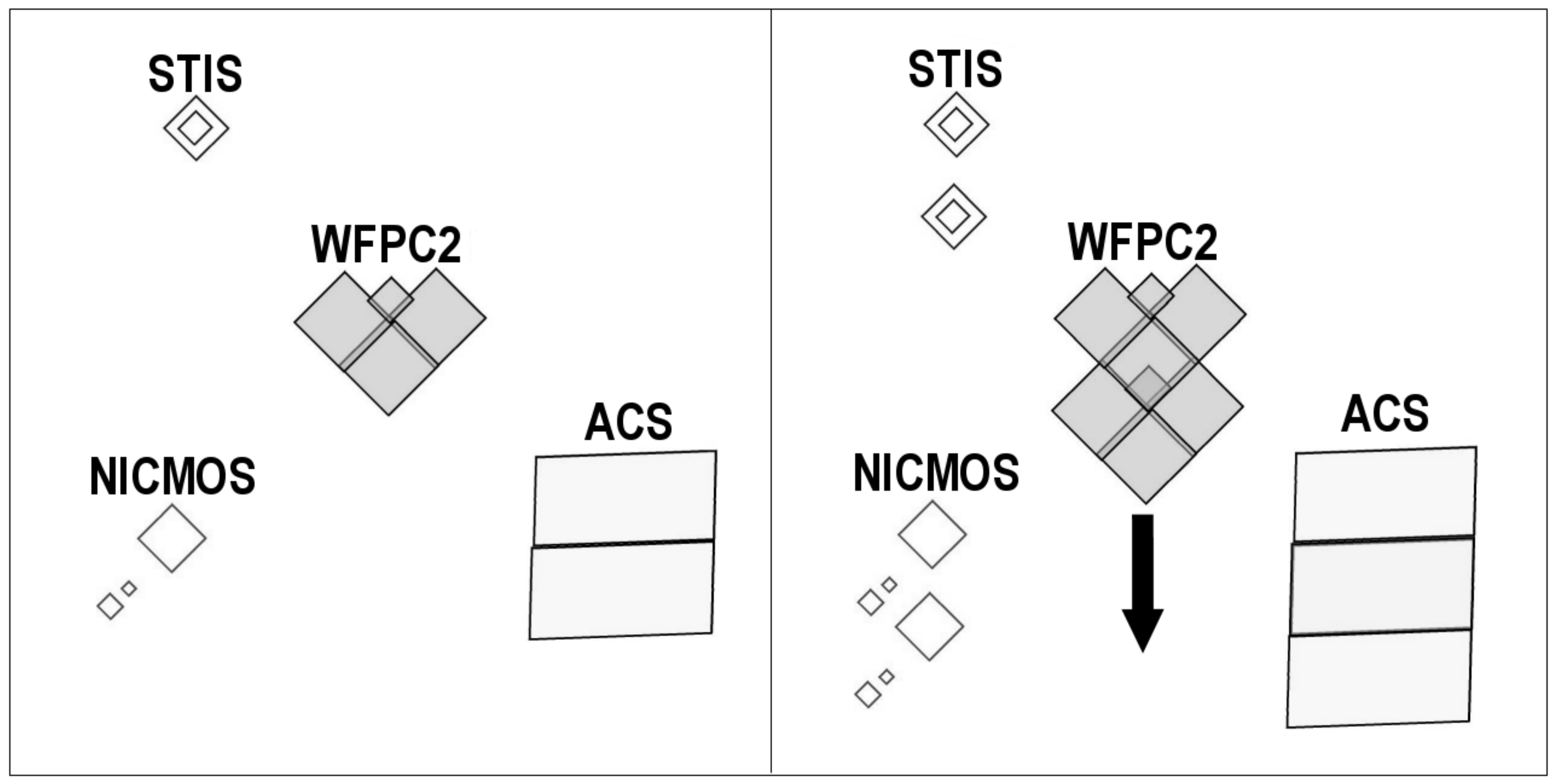}
\caption{Field of view of the \HST\ instruments in the \HST\ focal plane (prior to SM4) ({\em left panel}). The right panel shows the telescope shift between subsequent exposures chosen to optimize the coverage with both ACS and WFPC2.\label{Fig:ScanMode}}
\end{figure}

\begin{figure*}
\plotone{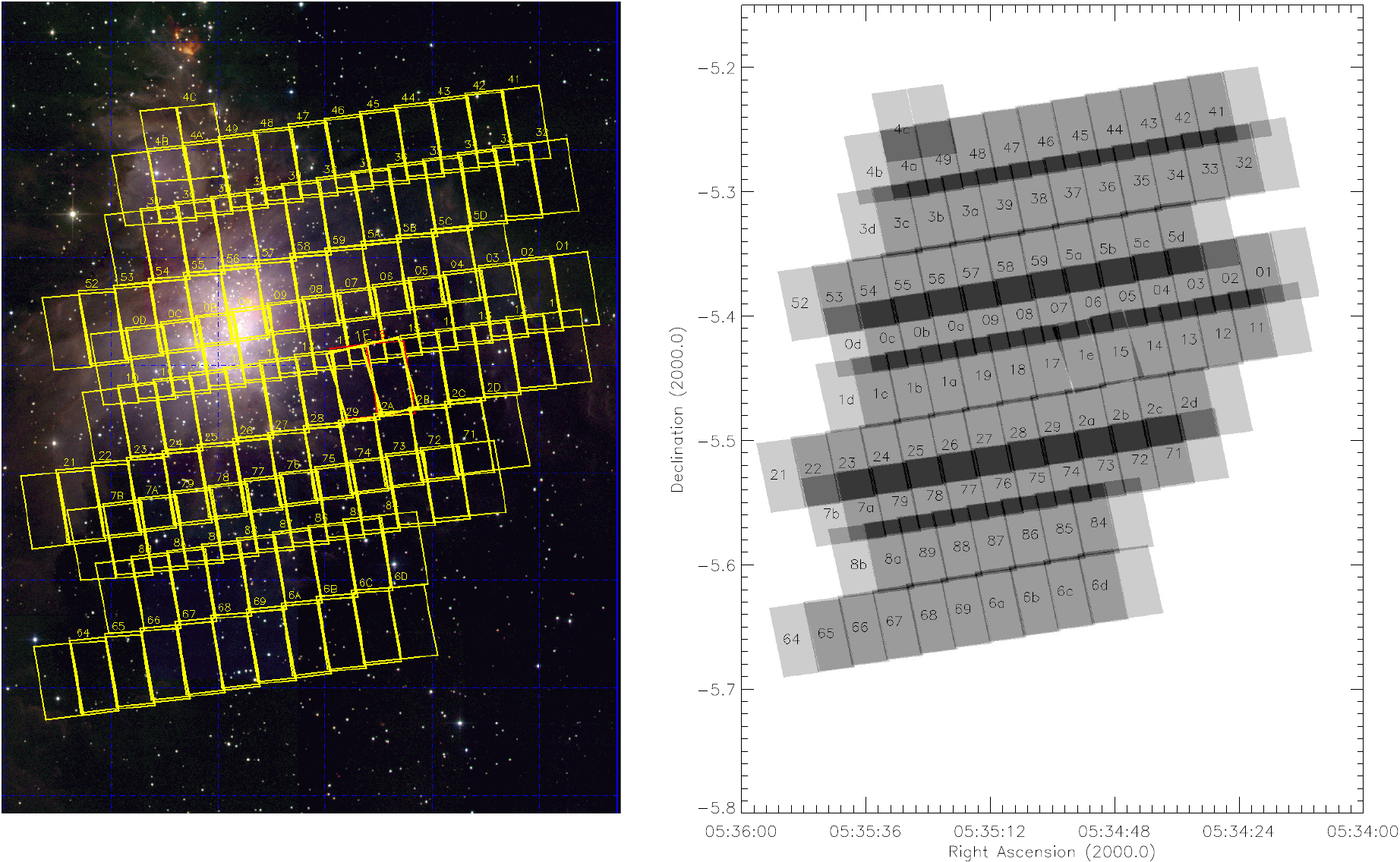}
\caption{Total coverage of the ACS survey, superimposed to a color-composite $JHK$ image of the Orion Nebula from 2MASS ({\em left panel}). The right panel highlights the overlapping regions between neighboring visits (darker gray shades), and reports the reference identification number of each of the individual visits. \label{Fig:ACSfield}}
\end{figure*}

\begin{figure*}
\plotone{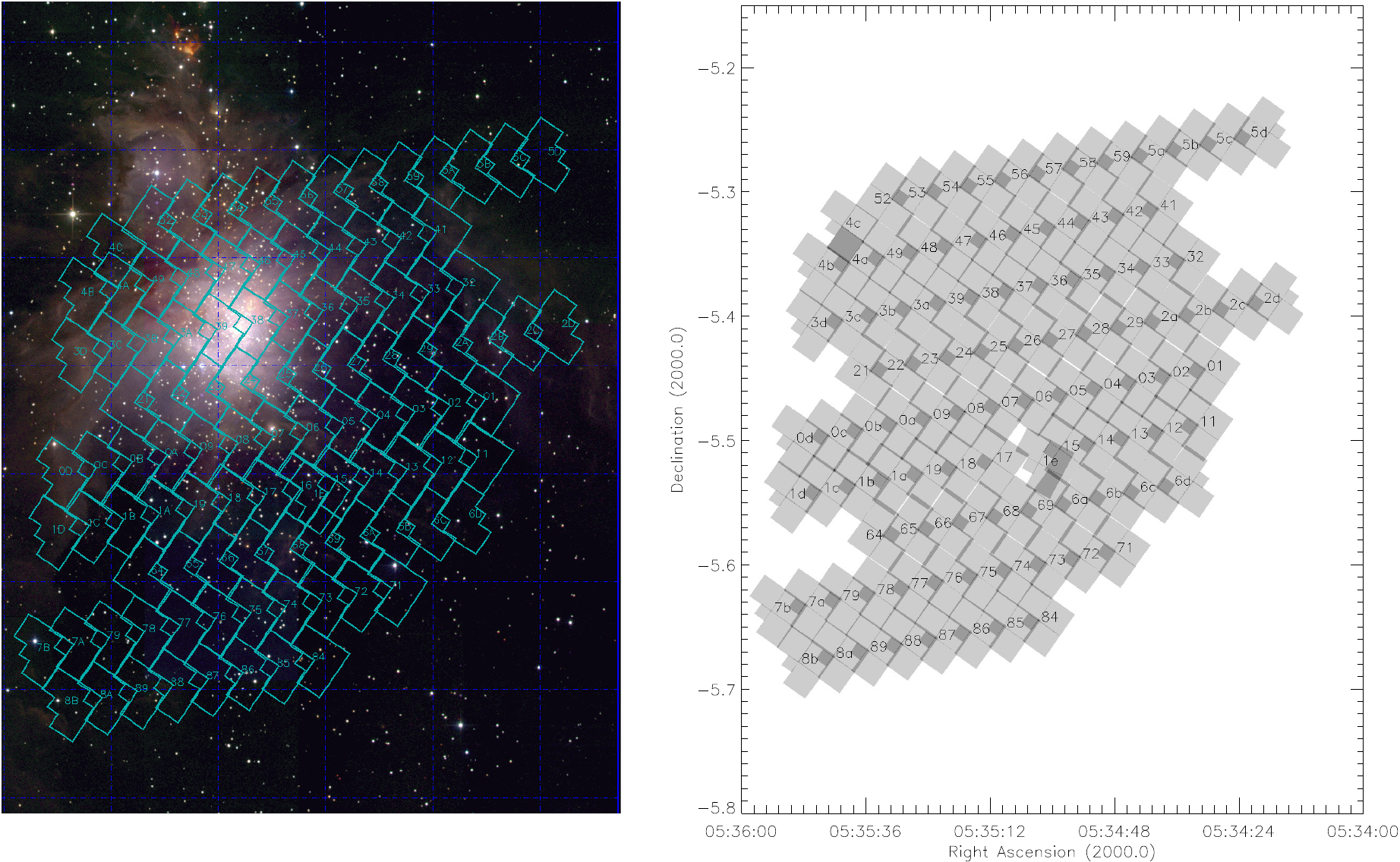}
\caption{Same as Figure \ref{Fig:ACSfield}, for the WFPC2 survey. \label{Fig:WFPC2field}}
\end{figure*}

\begin{figure}
\plotone{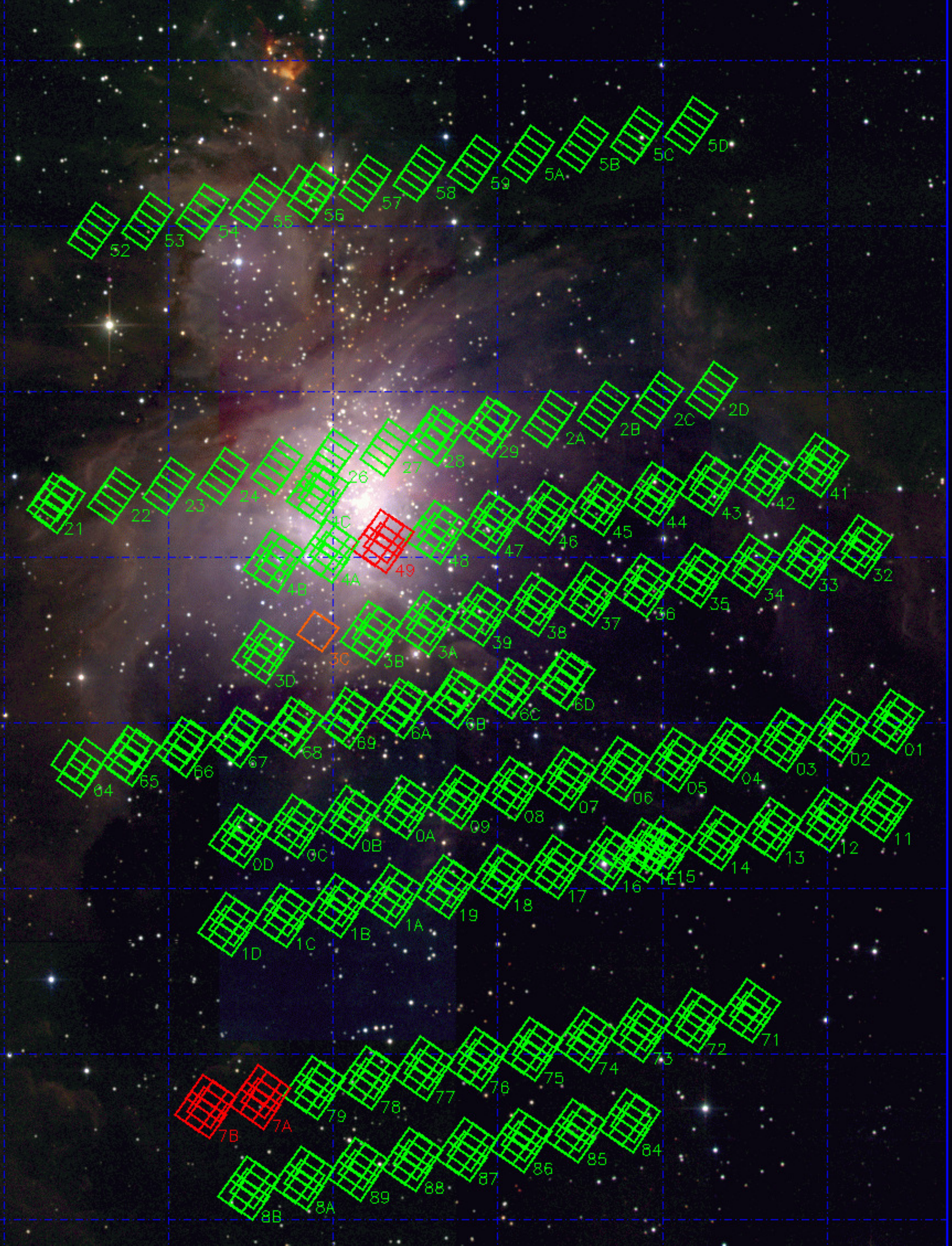}
\caption{Same as Figure \ref{Fig:ACSfield}, left panel, for the NICMOS survey. Note that because of the small field of view of this instrument, the individual visits observed disjoint patches of the region. Fields marked in red denote visits affected by partial or total data loss (see Section \ref{section:observations:LostData}). \label{Fig:actualNICMOS}}
\end{figure}

\begin{figure*}
\plottwo{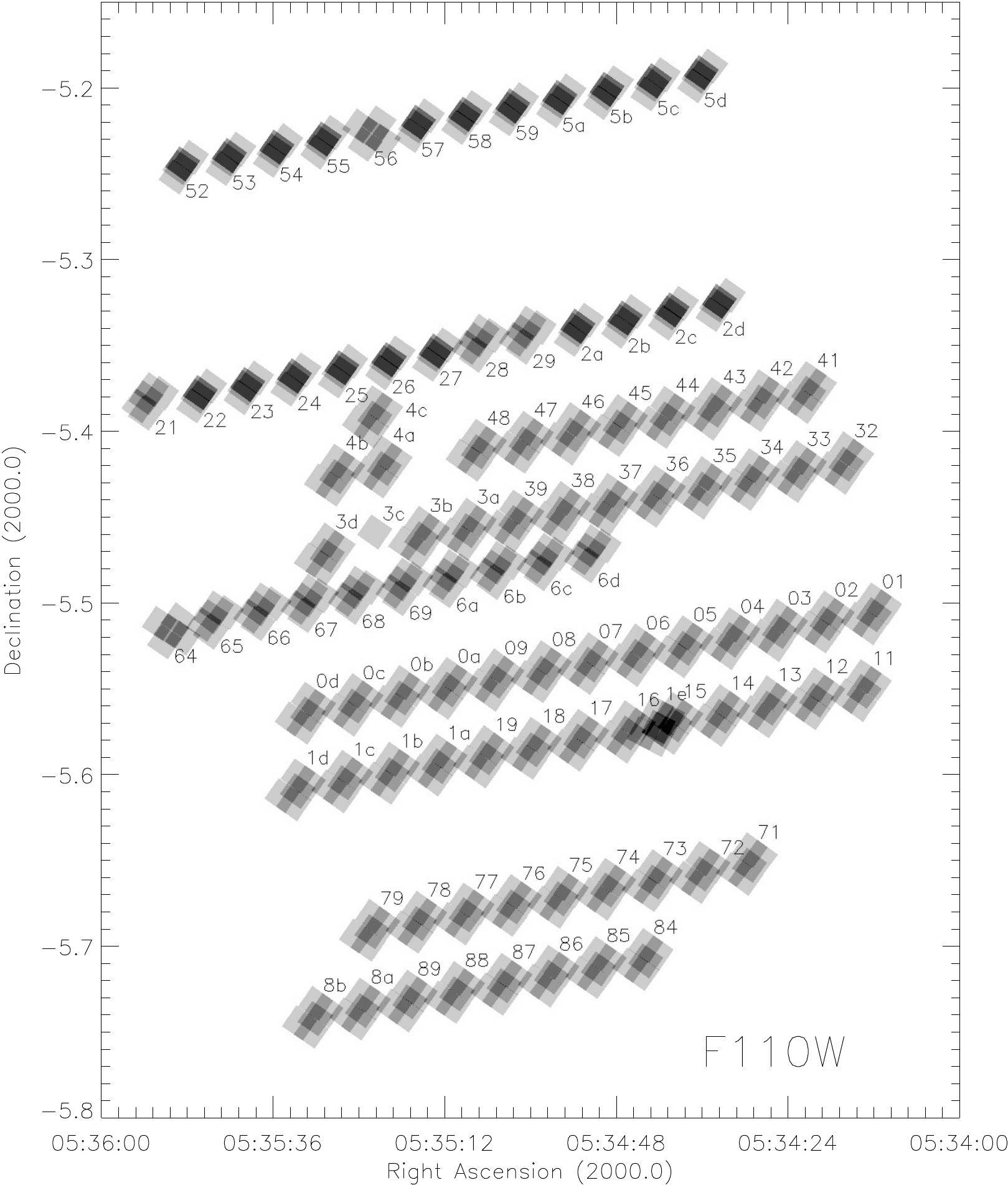}{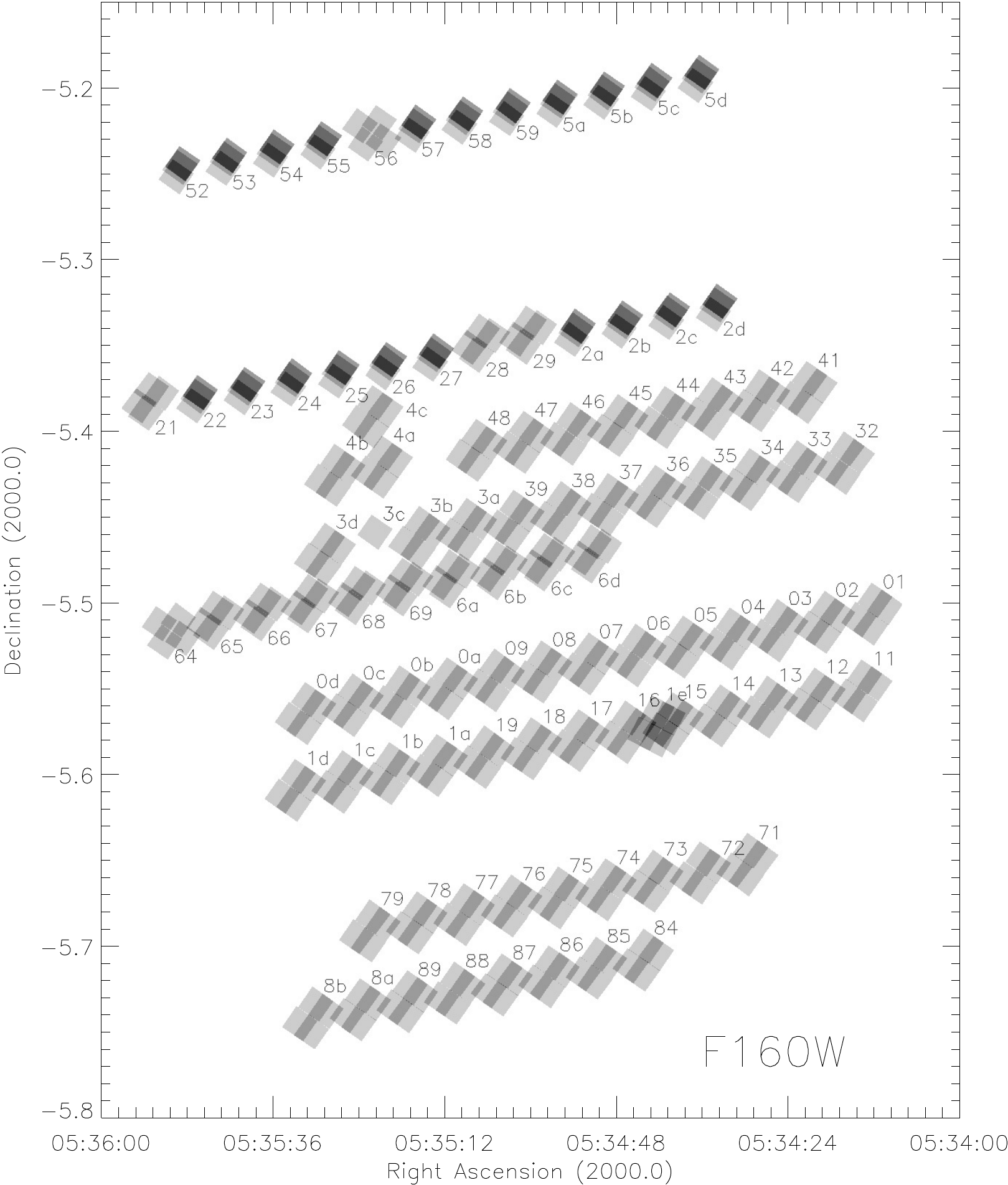}
\caption{Same as Figure \ref{Fig:ACSfield}, right panel, for the NICMOS F110W (\emph{left panel}) and F160W (\emph{right panel}).  \label{Fig:NICMOSfields}}
\end{figure*}

\subsubsection{ACS} \label{section:observations:instruments-setup:ACS}
The ACS Wide Field Channel (WFC) camera is based on a mosaic of two $2048\times4096$-pixel CCD detectors. The instrument optics deliver a spatial scale of $\sim 50$~mas/pixel, corresponding to a nominal field of view of $202\arcsec\times200\arcsec$. The two detectors are butted together on the long side with a interchip gap of approximately 50 pixels ($2\farcs5$). The  readout noise is $\simeq 5$~e pixel$^{-1}$ and the average dark current rate is $0.0038$~e s$^{-1}$pixel$^{-1}$ \citep{Maybathe+2010}.

We used the non-standard electronic gain of 2~e/adu in order to keep the detector saturation level, around 84,700 e$^-$,  within the range of the 16-bit Analog-to-Digital converters.  It has been shown \citep{Gilliland04} that  the  CCDs of ACS remain
almost perfectly linear ($<0.1\%$ discrepancy between relative count levels) up to saturation, and saturated charges spreading in the bleeding trails are fully preserved, allowing to recover  the photometry of saturated sources with very high precision.

The off-axis optics of ACS introduce a skew image distortion between 7\% and 10\%  of the field size. Since the pixels are projected as skew trapezoids on the sky, the fraction of sky area they actually map changes across the detector. This introduces a photometric error which can be corrected, as explained in Section~\ref{section:photometry:ACS}.

\subsubsection{WFPC2} \label{section:observations:instruments-setup:WFCP2}
The WFPC2 camera, now replaced by the Wide Field Camera-3, used four $800\times800$-pixel CCDs. Three of them image a $150\arcsec\times150\arcsec$ chevron-shaped region (WF, Wide Field) with a spatial sampling of 100~mas/pixel, the fourth one (PC, Planetary Camera) images a smaller  $34\arcsec\times34\arcsec$ square field  with 46~mas/pixel nested at the center of the chevron, in the fourth quadrant.
Following the WFPC2 Instrument Handbook \citep{McMasterBiretta2008}, the nominal readout noise  is  7.02, 7.84, 6.99 and 8.32 electrons for the channels WF1 to WF4. The mean dark current rate ($0.0045$~e$^-$s$^{-1}$pixel$^{-1}$) is comparable to the one of ACS.

Our observations were made when the WFPC2 instrument had already accumulated about 11 years of total radiation dose
in a space environment, and was therefore affected by significant Charge Transfer Efficiency (CTE) losses. The brightness of the Orion Nebula background, especially in broad band filters and at the center of the region, mitigates this problem. We applied the prescribed photometric
correction \citep{Dolphin00}\footnote{with updates at http://purcell.as.arizona.edu/wfpc2\_calib/} to our results
(see Section~\ref{section:results}), but  still CTE losses represent the major source of noise and photometric uncertainty of the WFPC2 data. Concerning the gain setting, WFPC2 is equipped with 12 bit Analog-to-Digital Converters. Their 4096 ADUs of  dynamic range does not sample the full dynamic range of the CCD detector even with the highest selectable gain of 14 e/ADU, which we have used. The photometric information of saturated (peak flux higher than  about 53,000 electrons) stars in WFPC2 images is therefore irremediably lost.

\subsubsection{NICMOS} \label{section:observations:instruments-setup:NICMOS}
{The NICMOS instruments has three cameras, all equipped with a  $256\times256$ HgCdTe detector but with different pixel scale. We selected Camera 3 (NIC3) because of its largest field of view, $51\farcs2\times51\farcs2$ with 200~mas/pixel, aiming to maximize the survey efficiency}. NIC3 operates slightly out of focus with a loss in encircled energy beyond one pixel radius  less than $10-15\%$. This small degradation of the image quality partially compensates for the original undersampling of the NIC3 PSF.
The equivalent readout noise of NICMOS depends on the adopted sampling scheme and on the number of reads. We used the sampling sequence STEP64 with NSAMP=12  for the F110W filter and NSAMP=11\ for the F160W filter, providing 255~s and  192~s integration time, respectively.
With these parameters the  readout noise is close to the nominal value of the instrument, around 27 electrons. We used the standard gain of 6.5 e/ADU.

\subsection{Observing Strategy} \label{section:observations:strategy}
Our main scientific goals required a mosaic of the largest possible field in Orion with both ACS and WFPC2 in a number of broad-band filters. The strategy we implemented exploited the relative position and orientation of the ACS/WFC and WFPC2 fields of view. We noticed that the combination of size, offset and position angle of  the fields imaged by two instruments implies that the pointing move needed to image the same field seen by one of the two ACS CCDs with the other CCD causes the chevron pattern of WFPC2 to translate along its symmetry axis almost exactly by 1 ACS/WF chip (see Figure~\ref{Fig:ScanMode}). Therefore, an ACS strip obtained by repeatedly offsetting the telescope by half of the ACS field-of-view produces a parallel, seamless herringbone pattern with WFPC2.
A second scan, adjacent to the first one, can then be performed maintaining full-field coverage with WFPC2 at the price of a relatively modest (approximately $25\%$) overlap of the larger ACS/WFPC2 fields. By performing a third strip with the telescope rotated by $180^\circ$, the relative positions of ACS/WFC and WFPC2 are exchanged. This third passage nicely complements the first two, providing contiguous area coverage of nearly the same field with both instruments.

We have repeated the 2+1 strip pattern 3 times, adjusting the length of each strip to cover the largest area of bright nebular background, where \HST\ provides the greatest gain with respect to ground-based observations.  The nine strips, numbered from 0 to 8, are oriented nearly in the East-West direction with position angles 100$^\circ$ or 280$^\circ$.

The full fields covered by the ACS/WFC and WFPC2 surveys are shown, respectively, in Figures  \ref{Fig:ACSfield} and \ref{Fig:WFPC2field}. Both are approximately centered at $RA=05:35:00$ and $DEC=-5:26:00$ (J2000.0), approximately $200\arcsec$ to the SW of the Trapezium asterism and extend over about $1/6^{th}$ of a square degree. In particular, the full ACS mosaic covers 627 square arcminutes (corresponding to about 904~mega pixel), whereas the WFPC2 mosaic covers 570.5 square arcminutes (corresponding to 205.4 Mpix, assuming the  100~mas/pixel scale of the WFPC2 Wide Field channels).

This scanning pattern typically provides, for each filter, two images with ACS with  a few exceptions: 1) the half-fields at the extremes of each strip, covered by only one CCD, are observed once; 2) the overlay regions between adjacent strips are typically observed 4 times; 3) a small fraction of the ACS field is observed 1 or 3 times, in correspondence of the
detector gaps, or in the case of a visit repeated  with a different orientation (see Section~\ref{section:observations:LostData}). Further, smaller deviations are also occasionally present. For WFPC2, we typically have one passage per filter, except for the F336W band where the observations were split in two equal-length exposures (see Section~\ref{section:observations:filters}). The areas observed with the WFPC2/PC have been generally  observed also with the median detector (nr. 2)\ of the WFPC2/WF channel, with the exception of the very last field of each strip.

With NICMOS, our survey covered only a fraction of the ACS field due to the limited field of view of the instrument (Figure~\ref{Fig:actualNICMOS}).
However, we created a strategy to perform some field dithering and increase the area covered without moving the telescope during the visits.  The  NICMOS instrument is equipped  with a Field Offset Mirror (FOM) which allows selecting the portion of the telescope focal plane
falling on the detector. Moving the FOM, it is possible to  probe a field larger than the nominal $52\farcs6 \times 52\farcs 6$ of NIC3. The default FOM position had been optimized in the early days of instrument commissioning and before our program FOM moves had never been supported for science programs, for a number of good reasons: the image quality degrades at large distances from the optimal position, some vignetting is introduced at one FOM
extreme of the offset range, and the moving mechanisms does not allow
repositioning the FOM to better that 1 or 2 NIC3 pixels. Still,
we were able to move the FOM, building for each visit a 4 piece mosaic in the {F160W} band and a 5 piece mosaic in the {F110W} band. Our request was granted on a shared risk basis, as  the necessary changes to the NICMOS\ control software would have been tested ``on-the-fly'' using our observing time. Our first NICMOS observations, in fact, suffered from a number of FOM\ software limit errors. When this occurred, the FOM returned to its nominal central position and the following NICMOS observations within the same visit were suspended. This problem reoccurred for each visits scheduled using the same pattern. After each failure, the NICMOS operation team immediately investigated the cause of the error implementing the needed changes. Soon we were able to move the FOM anywhere within the limits
presented in Table~\ref{Tab:FOM}, where we also include for reference the values for NIC1 and NIC2.

The improvements made with the scheduling of the NICMOS observations are illustrated in Figure~\ref{Fig:NICMOSfields} by the uneven  shape of the fields covered with NIC3 across the field, the size of each tile being generally proportional to the ``maturity'' of the FOM\ control software. A total of 102 regions were covered in both the filter~{F110W} and filter~{F160W} bands,
corresponding to about 177 square arcminutes (15.9 Mpix)\ and 171.5 square arcminutes (15.4 Mpix) in the F110W\ and F160W\ filters, respectively.


\begin{table}
\begin{center}
\caption{FOM position range \label{Tab:FOM}}
\begin{tabular}{lcc}
\tableline
\tableline
Detector & arcsec & FOMYPOS (arcsec) \\
\tableline
NIC1     & -20.0 to +22.0   & -21.0 to +23.0   \\
NIC2     & -20.0 to +22.0   & -21.0 to +23.0   \\
NIC3     & -21.0 to +23.0   & -37.0 to +8.0    \\
\tableline
\end{tabular}
\end{center}
\end{table}

\subsection{Scheduling} \label{section:observations:scheduling}
In order to exchange the relative position of ACS\ and WFPC2, we rotated by $180^\circ$ the roll angle of the \HST\ by scheduling the observations in two epochs separated by approximately 6 months.
The first epoch, executed in the Fall of 2004 (October 11, 2004 - November 7, 2004), covered strips 2, 5 and 6 with a position angle of $280^\circ$, for a total of 36 orbits. The second epoch, executed in the Spring of 2005 (March 3, 2005 - April 26, 2005)
covered strips 0, 1, 3, 4, 7 and 8 with a position angle of $100^\circ$, for a total of 68 orbits.
Both epochs and position angles were optimized to schedule the observations at the peaks of the visibility periods of the Orion Nebula. 

\begin{figure}
\epsscale{1}
\plotone{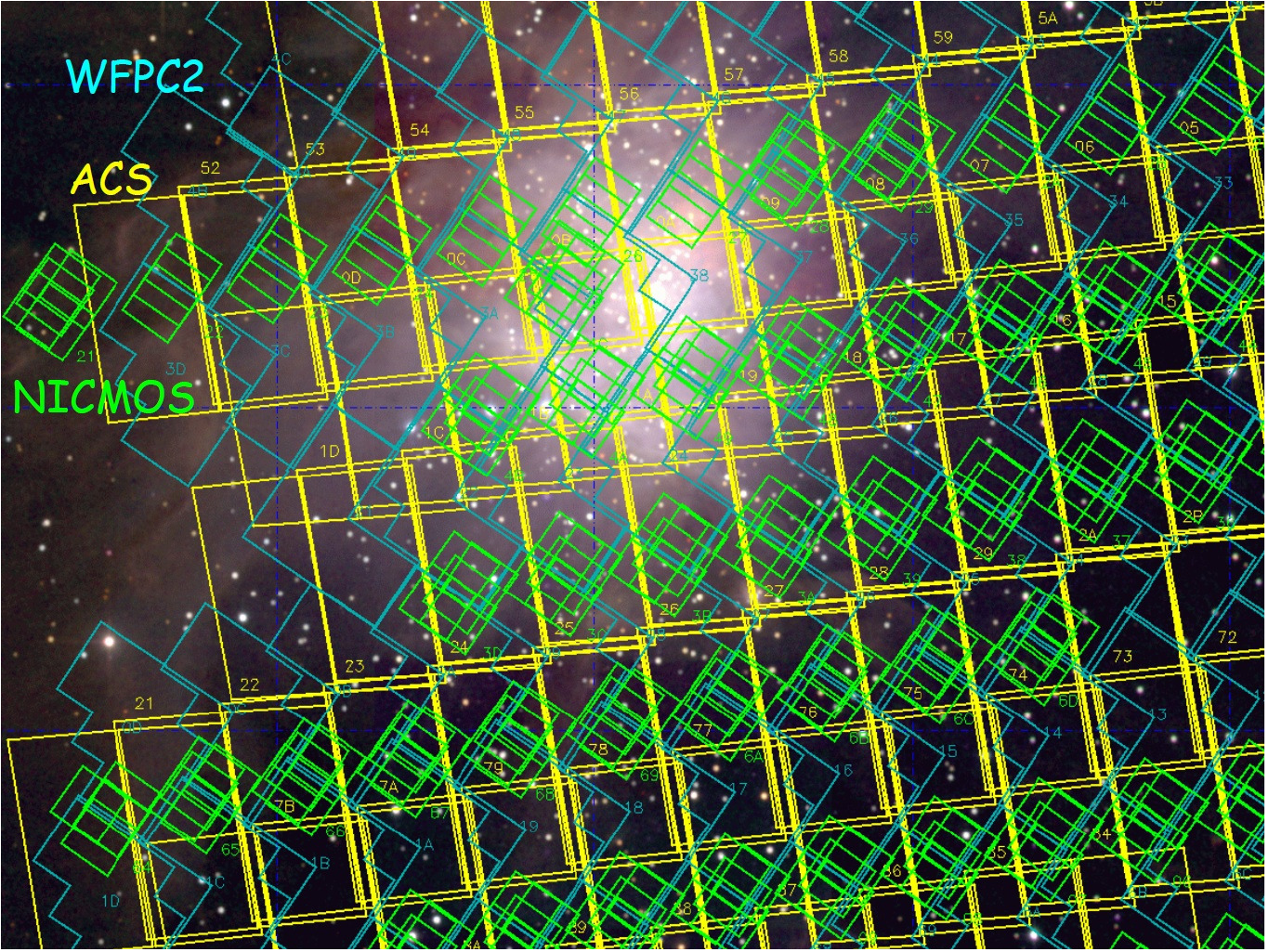}
\caption{Coverage of the survey for all the 3 cameras, limited for the central part of the region. See also Figures \ref{Fig:ACSfield}, \ref{Fig:WFPC2field} and \ref{Fig:NICMOSfields}. \label{Fig:AnOverview}}
\end{figure}

Table \ref{Tab:ObsLog} summarizes the scheduling of our observations with the central coordinates of the field, for each instrument. In Figure \ref{Fig:AnOverview} we present the overall coverage of the survey, limited on the central part of the ONC, with the field of view of all the instruments and for all the visits overlaid.

\begin{longtable*}{ccccccccc}
\caption{Observations log.} \label{Tab:ObsLog}  \\

\hline \hline
\multicolumn{1}{c}{Target Name} & \multicolumn{1}{c}{Orient} & \multicolumn{1}{c}{Visit} & \multicolumn{2}{c}{ACS} & \multicolumn{2}{c}{WFPC2} & \multicolumn{2}{c}{NIC3} \\
\multicolumn{3}{c}{  } & \multicolumn{1}{c}{RA (2000.0)} & \multicolumn{1}{c}{DEC (2000.0)} & \multicolumn{1}{c}{RA (2000.0)} & \multicolumn{1}{c}{DEC (2000.0)} & \multicolumn{1}{c}{RA (2000.0)} & \multicolumn{1}{c}{DEC (2000.0)} \\ \hline
\endfirsthead

\multicolumn{9}{c}{{\bfseries \tablename\ \thetable{} -- continued from previous page}} \\ \hline
\multicolumn{1}{c}{Target Name} & \multicolumn{1}{c}{Orient} & \multicolumn{1}{c}{Visit} & \multicolumn{2}{c}{ACS} & \multicolumn{2}{c}{WFPC2} & \multicolumn{2}{c}{NIC3} \\
\multicolumn{3}{c}{  } & \multicolumn{1}{c}{RA (2000.0)} & \multicolumn{1}{c}{DEC (2000.0)} & \multicolumn{1}{c}{RA (2000.0)} & \multicolumn{1}{c}{DEC (2000.0)} & \multicolumn{1}{c}{RA (2000.0)} & \multicolumn{1}{c}{DEC (2000.0)} \\ \hline
\endhead

\hline
\multicolumn{9}{c}{Continued on the next page} \\ 
\endfoot

\hline \hline
\endlastfoot

3/31/2005  &  100  &  01   &  05:34:16.80  &  -05:21:35.62  &  05:34:30.39  &  -05:26:36.2  &  5:34:12.23  &  -5:29:55.9  \\
4/1/2005   &  100  &  02   &  05:34:23.40  &  -05:21:53.01  &  05:34:36.99  &  -05:26:53.6  &  5:34:18.83  &  -5:30:13.3  \\
3/30/2005  &  100  & 03   &  05:34:29.99  &  -05:22:10.40  &  05:34:43.58  &  -05:27:10.9  &  5:34:25.42  &  -5:30:30.7  \\
4/3/2005   &  100  &  04   &  05:34:36.76  &  -05:22:27.78  &  05:34:50.18  &  -05:27:28.3  &  5:34:32.02  &  -5:30:48.1  \\
4/2/2005   &  100  &  05   &  05:34:43.18  &  -05:22:45.15  &  05:34:56.77  &  -05:27:45.7  &  5:34:38.61  &  -5:31:05.5  \\
3/30/2005  &  100  &  06   &  05:34:49.78  &  -05:23:02.52  &  05:35:03.37  &  -05:28:03.1  &  5:34:45.21  &  -5:31:22.8  \\
4/5/2005   &  100  &  07   &  05:34:56.37  &  -05:23:19.89  &  05:35:09.96  &  -05:28:20.4  &  5:34:51.80  &  -5:31:40.2  \\
4/5/2005   &  100  &  08   &  05:35:02.96  &  -05:23:37.25  &  05:35:16.55  &  -05:28:37.8  &  5:34:58.39  &  -5:31:57.6  \\
4/6/2005   &  100  &  09   &  05:35:09.56  &  -05:23:54.61  &  05:35:23.15  &  -05:28:55.2  &  5:35:04.99  &  -5:32:14.9  \\
4/6/2005   &  100  &  0A  &  05:35:16.15  &  -05:24:11.96  &  05:35:29.74  &  -05:29:12.5  &  5:35:11.58  &  -5:32:32.3  \\
4/7/2005   &  100  &  0B  &  05:35:22.75  &  -05:24:29.31  &  05:35:36.34  &  -05:29:29.9  &  5:35:18.18  &  -5:32:49.6  \\
4/8/2005   &  100  &  0C  &  05:35:29.34  &  -05:24:46.66  &  05:35:42.93  &  -05:29:47.2  &  5:35:24.77  &  -5:33:07.0  \\
4/7/2005   &  100  &  0D  &  05:35:35.94  &  -05:25:04.00  &  05:35:49.53  &  -05:30:04.5  &  5:35:31.37  &  -5:33:24.3  \\
4/7/2005   &  100  &  11  &  05:34:18.22  &  -05:24:16.36  &  05:34:31.81  &  -05:29:16.9  &  5:34:13.65  &  -5:32:36.7  \\
4/7/2005   &  100  &  12  &  05:34:24.82  &  -05:24:33.75  &  05:34:38.41  &  -05:29:34.3  &  5:34:20.25  &  -5:32:54.1  \\
4/4/2005   &  100  &  13  &  05:34:31.41  &  -05:24:51.13  &  05:34:45.00  &  -05:29:51.7  &  5:34:26.84  &  -5:33:11.4  \\
4/8/2005   &  100  &  14  &  05:34:38.00  &  -05:25:08.51  &  05:34:51.59  &  -05:30:09.1  &  5:34:33.43  &  -5:33:28.8  \\
4/8/2005   &  100  &  15  &  05:34:44.60  &  -05:25:25.89  &  05:34:58.19  &  -05:30:26.4  &  5:34:40.03  &  -5:33:46.2  \\
4/9/2005   &  100  &  17  &  05:34:57.79  &  -05:26:00.63  &  05:35:11.38  &  -05:31:01.2  &  5:34:53.22  &  -5:34:20.9  \\
4/9/2005   &  100  &  18  &  05:35:04.39  &  -05:26:17.99  &  05:35:17.98  &  -05:31:18.5  &  5:34:59.82  &  -5:34:38.3  \\
4/9/2005   &  100  &  19  &  05:35:10.98  &  -05:26:35.35  &  05:35:24.57  &  -05:31:35.9  &  5:35:06.41  &  -5:34:55.7  \\
4/9/2005   &  100  &  1A  &  05:35:17.58  &  -05:26:52.70  &  05:35:31.17  &  -05:31:53.2  &  5:35:13.01  &  -5:35:13.0  \\
4/10/2005  &  100  &  1B  &  05:35:24.17  &  -05:27:10.05  &  05:35:37.76  &  -05:32:10.6  &  5:35:19.60  &  -5:35:30.4  \\
4/10/2005  &  100  &  1C  &  05:35:30.77  &  -05:27:27.40  &  05:35:44.36  &  -05:32:27.9  &  5:35:26.20  &  -5:35:47.7  \\
4/10/2005  &  100  &  1D  &  05:35:37.36  &  -05:27:44.74  &  05:35:50.95  &  -05:32:45.3  &  5:35:32.79  &  -5:36:05.1  \\
4/26/2005  &  287  &  1E  &  05:35:37.36  &  -05:27:44.74  &  05:35:02.22  &  -05:31:06.3  &  5:34:42.44  &  -5:33:27.7  \\
11/7/2004  &  280  &  21  &  05:35:48.92  &  -05:31:39.11  &  05:35:35.33  &  -05:26:38.5  &  5:35:53.36  &  -5:23:20.1  \\
10/29/2004 &  280  &  22  &  05:35:42.32  &  -05:31:21.77  &  05:35:28.73  &  -05:26:21.2  &  5:35:46.21  &  -5:23:09.2  \\
10/31/2004 &  280  &  23  &  05:35:35.72  &  -05:31:04.42  &  05:35:22.13  &  -05:26:03.9  &  5:35:39.61  &  -5:22:51.8  \\
10/28/2004 &  280  &  24  &  05:35:29.13  &  -05:30:47.07  &  05:35:15.54  &  -05:25:46.5  &  5:35:33.02  &  -5:22:34.5  \\
10/25/2004 &  280  &  25  &  05:35:22.53  &  -05:30:29.72  &  05:35:08.94  &  -05:25:29.2  &  5:35:26.42  &  -5:22:17.1  \\
10/27/2004 &  280  &  26  &  05:35:15.94  &  -05:30:12.36  &  05:35:02.35  &  -05:25:11.8  &  5:35:19.83  &  -5:21:59.7  \\
10/25/2004 &  280  &  27  &  05:35:09.34  &  -05:29:55.00  &  05:34:55.75  &  -05:24:54.4  &  5:35:13.23  &  -5:21:42.4  \\
11/1/2004  &  280  &  28  &  05:35:02.74  &  -05:29:37.63  &  05:34:49.15  &  -05:24:37.1  &  5:35:07.18  &  -5:21:18.6  \\
11/1/2004  &  280  &  29  &  05:34:56.15  &  -05:29:20.26  &  05:34:42.56  &  -05:24:19.7  &  5:35:00.59  &  -5:21:01.2  \\
10/31/2004 &  280  &  2A  &  05:34:49.55  &  -05:29:02.89  &  05:34:35.96  &  -05:24:02.3  &  5:34:53.44  &  -5:20:50.3  \\
10/31/2004 &  280  &  2B  &  05:34:42.96  &  -05:28:45.50  &  05:34:29.37  &  -05:23:44.9  &  5:34:46.85  &  -5:20:32.9  \\
10/27/2004 &  280  &  2C  &  05:34:36.36  &  -05:28:28.12  &  05:34:22.77  &  -05:23:27.6  &  5:34:40.25  &  -5:20:15.5  \\
10/29/2004 &  280  &  2D  &  05:34:29.77  &  -05:28:10.73  &  05:34:16.18  &  -05:23:10.2  &  5:34:33.66  &  -5:19:58.1  \\
4/10/2005  &  100  &  32  &  05:34:20.69  &  -05:16:19.80  &  05:34:34.27  &  -05:21:20.3  &  5:34:16.12  &  -5:24:40.1  \\
4/9/2005   &  100  &  33  &  05:34:27.29  &  -05:16:37.19  &  05:34:40.88  &  -05:21:37.7  &  5:34:22.72  &  -5:24:57.5  \\
4/9/2005   &  100  &  34  &  05:34:33.88  &  -05:16:54.57  &  05:34:47.47  &  -05:21:55.1  &  5:34:29.31  &  -5:25:14.9  \\
4/4/2005   &  100  &  35  &  05:34:40.47  &  -05:17:11.94  &  05:34:54.06  &  -05:22:12.5  &  5:34:35.90  &  -5:25:32.3  \\
4/4/2005   &  100  &  36  &  05:34:47.07  &  -05:17:29.31  &  05:35:00.66  &  -05:22:29.9  &  5:34:42.50  &  -5:25:49.6  \\
4/10/2005  &  100  &  37  &  05:34:53.66  &  -05:17:46.68  &  05:35:07.25  &  -05:22:47.2  &  5:34:49.09  &  -5:26:07.0  \\
4/10/2005  &  100  &  38  &  05:35:00.25  &  -05:18:04.04  &  05:35:13.84  &  -05:23:04.6  &  5:34:55.68  &  -5:26:24.4  \\
4/10/2005  &  100  &  39  &  05:35:06.85  &  -05:18:21.40  &  05:35:20.44  &  -05:23:21.9  &  5:35:02.28  &  -5:26:41.7  \\
4/10/2005  &  100  &  3A  &  05:35:13.44  &  -05:18:38.76  &  05:35:27.03  &  -05:23:39.3  &  5:35:08.87  &  -5:26:59.1  \\
4/9/2005   &  100  &  3B  &  05:35:20.04  &  -05:18:56.10  &  05:35:33.63  &  -05:23:56.6  &  5:35:15.47  &  -5:27:16.4  \\
4/10/2005  &  100  &  3C  &  05:35:26.63  &  -05:19:13.45  &  05:35:40.22  &  -05:24:14.0  &  5:35:21.74  &  -5:27:10.3  \\
4/9/2005   &  100  &  3D  &  05:35:33.22  &  -05:19:30.79  &  05:35:46.81  &  -05:24:31.3  &  5:35:28.65  &  -5:27:51.1  \\
4/10/2005  &  100  &  41  &  05:34:25.92  &  -05:13:51.70  &  05:34:39.50  &  -05:18:52.2  &  5:34:21.35  &  -5:22:12.0  \\
4/7/2005   &  100  &  42  &  05:34:32.51  &  -05:14:09.09  &  05:34:46.09  &  -05:19:09.6  &  5:34:27.94  &  -5:22:29.4  \\
4/5/2005   &  100  &  43  &  05:34:39.10  &  -05:14:26.47  &  05:34:52.68  &  -05:19:27.0  &  5:34:34.53  &  -5:22:46.8  \\
4/6/2005   &  100  &  44  &  05:34:45.69  &  -05:14:43.84  &  05:34:59.27  &  -05:19:44.4  &  5:34:41.12  &  -5:23:04.2  \\
4/6/2005   &  100  &  45  &  05:34:52.29  &  -05:15:01.21  &  05:35:05.87  &  -05:20:01.8  &  5:34:47.72  &  -5:23:21.5  \\
4/5/2005   &  100  &  46  &  05:34:58.88  &  -05:15:18.58  &  05:35:12.46  &  -05:20:19.1  &  5:34:54.31  &  -5:23:38.9  \\
4/16/2005  &  100  &  47  &  05:35:05.47  &  -05:15:35.94  &  05:35:19.05  &  -05:20:36.5  &  5:35:00.90  &  -5:23:56.3  \\
4/16/2005  &  100  &  48  &  05:35:12.07  &  -05:15:53.30  &  05:35:25.65  &  -05:20:53.8  &  5:35:07.50  &  -5:24:13.6  \\
4/11/2005  &  100  &  49  &  05:35:18.66  &  -05:16:10.65  &  05:35:32.24  &  -05:21:11.2  &              &              \\
4/13/2005  &  100  &  4A  &  05:35:25.25  &  -05:16:28.00  &  05:35:38.83  &  -05:21:28.5  &  5:35:20.68  &  -5:24:48.3  \\
4/13/2005  &  100  &  4B  &  05:35:31.85  &  -05:16:45.35  &  05:35:45.44  &  -05:21:45.9  &  5:35:27.28  &  -5:25:05.7  \\
4/13/2005  &  100  &  4C  &  05:35:26.57  &  -05:14:42.45  &  05:35:40.15  &  -05:19:43.0  &  5:35:22.00  &  -5:23:02.8  \\
10/28/2004 &  280  &  52  &  05:35:44.81  &  -05:23:22.03  &  05:35:31.23  &  -05:18:21.5  &  5:35:48.70  &  -5:15:09.4  \\
10/30/2004 &  280  &  53  &  05:35:38.22  &  -05:23:04.68  &  05:35:24.64  &  -05:18:04.1  &  5:35:42.11  &  -5:14:52.1  \\
10/30/2004 &  280  &  54  &  05:35:31.62  &  -05:22:47.33  &  05:35:18.04  &  -05:17:46.8  &  5:35:35.51  &  -5:14:34.7  \\
10/25/2004 &  280  &  55  &  05:35:25.03  &  -05:22:29.98  &  05:35:11.45  &  -05:17:29.4  &  5:35:28.92  &  -5:14:17.4  \\
10/12/2004 &  280  &  56  &  05:35:18.43  &  -05:22:12.62  &  05:35:04.85  &  -05:17:12.1  &  5:35:22.42  &  -5:13:58.9  \\
10/30/2004 &  280  &  57  &  05:35:11.84  &  -05:21:55.26  &  05:34:58.26  &  -05:16:54.7  &  5:35:15.73  &  -5:13:42.6  \\
10/26/2004 &  280  &  58  &  05:35:05.25  &  -05:21:37.89  &  05:34:51.67  &  -05:16:37.3  &  5:35:09.14  &  -5:13:25.3  \\
10/27/2004 &  280  &  59  &  05:34:58.65  &  -05:21:20.52  &  05:34:45.07  &  -05:16:20.0  &  5:35:02.54  &  -5:13:07.9  \\
10/26/2004 &  280  &  5A  &  05:34:52.06  &  -05:21:03.15  &  05:34:38.48  &  -05:16:02.6  &  5:34:55.95  &  -5:12:50.5  \\
10/26/2004 &  280  &  5B  &  05:34:45.46  &  -05:20:45.77  &  05:34:31.88  &  -05:15:45.2  &  5:34:49.35  &  -5:12:33.2  \\
10/29/2004 &  280  &  5C  &  05:34:38.87  &  -05:20:28.38  &  05:34:25.29  &  -05:15:27.8  &  5:34:42.76  &  -5:12:15.8  \\
10/30/2004 &  280  &  5D  &  05:34:32.28  &  -05:20:10.99  &  05:34:18.70  &  -05:15:10.4  &  5:34:36.17  &  -5:11:58.4  \\
10/11/2004 &  280  &  64  &  05:35:46.44  &  -05:39:34.96  &  05:35:32.85  &  -05:34:34.4  &  5:35:50.43  &  -5:31:21.2  \\
11/7/2004  &  280  &  65  &  05:35:39.84  &  -05:39:17.61  &  05:35:26.25  &  -05:34:17.0  &  5:35:44.28  &  -5:30:58.6  \\
11/7/2004  &  280  &  66  &  05:35:33.25  &  -05:39:00.25  &  05:35:19.66  &  -05:33:59.7  &  5:35:37.69  &  -5:30:41.2  \\
11/1/2004  &  280  &  67  &  05:35:26.65  &  -05:38:42.89  &  05:35:13.06  &  -05:33:42.3  &  5:35:31.09  &  -5:30:23.8  \\
11/7/2004  &  280  &  68  &  05:35:20.05  &  -05:38:25.52  &  05:35:06.46  &  -05:33:25.0  &  5:35:24.49  &  -5:30:06.5  \\
11/7/2004  &  280  &  69  &  05:35:13.44  &  -05:38:08.15  &  05:34:59.85  &  -05:33:07.6  &  5:35:17.88  &  -5:29:49.1  \\
11/7/2004  &  280  &  6A  &  05:35:06.86  &  -05:37:50.77  &  05:34:53.27  &  -05:32:50.2  &  5:35:11.30  &  -5:29:31.7  \\
11/6/2004  &  280  &  6B  &  05:35:00.26  &  -05:37:33.39  &  05:34:46.67  &  -05:32:32.8  &  5:35:04.70  &  -5:29:14.3  \\
11/1/2004  &  280  &  6C  &  05:34:53.66  &  -05:37:16.01  &  05:34:40.07  &  -05:32:15.4  &  5:34:58.10  &  -5:28:57.0  \\
11/1/2004  &  280  &  6D  &  05:34:47.07  &  -05:36:58.62  &  05:34:33.48  &  -05:31:58.1  &  5:34:51.51  &  -5:28:39.6  \\
4/14/2005  &  100  &  71  &  05:34:34.20  &  -05:30:20.73  &  05:34:47.79  &  -05:35:21.3  &  5:34:29.63  &  -5:38:41.0  \\
4/14/2005  &  100  &  72  &  05:34:40.79  &  -05:30:38.12  &  05:34:54.38  &  -05:35:38.7  &  5:34:36.22  &  -5:38:58.4  \\
4/15/2005  &  100  &  73  &  05:34:47.39  &  -05:30:55.50  &  05:35:00.98  &  -05:35:56.0  &  5:34:42.82  &  -5:39:15.8  \\
4/13/2005  &  100  &  74  &  05:34:53.98  &  -05:31:12.89  &  05:35:07.57  &  -05:36:13.4  &  5:34:49.41  &  -5:39:33.2  \\
4/11/2005  &  100  &  75  &  05:35:00.58  &  -05:31:30.26  &  05:35:14.17  &  -05:36:30.8  &  5:34:56.01  &  -5:39:50.6  \\
4/12/2005  &  100  &  76  &  05:35:07.17  &  -05:31:47.63  &  05:35:20.76  &  -05:36:48.2  &  5:35:02.60  &  -5:40:07.9  \\
4/12/2005  &  100  &  77  &  05:35:13.77  &  -05:32:05.00  &  05:35:27.36  &  -05:37:05.5  &  5:35:09.20  &  -5:40:25.3  \\
4/12/2005  &  100  &  78  &  05:35:20.37  &  -05:32:22.36  &  05:35:33.96  &  -05:37:22.9  &  5:35:15.80  &  -5:40:42.7  \\
4/12/2005  &  100  &  79  &  05:35:26.96  &  -05:32:39.72  &  05:35:40.55  &  -05:37:40.3  &  5:35:22.39  &  -5:41:00.0  \\
4/11/2005  &  100  &  7A  &  05:35:33.56  &  -05:32:57.07  &  05:35:47.15  &  -05:37:57.6  &              &              \\
4/11/2005  &  100  &  7B  &  05:35:40.15  &  -05:33:14.42  &  05:35:53.74  &  -05:38:15.0  &              &              \\
4/11/2005  &  100  &  84  &  05:34:48.72  &  -05:33:40.42  &  05:35:02.31  &  -05:38:41.0  &  5:34:44.15  &  -5:42:00.7  \\
4/11/2005  &  100  &  85  &  05:34:55.32  &  -05:33:57.80  &  05:35:08.91  &  -05:38:58.3  &  5:34:50.75  &  -5:42:18.1  \\
4/11/2005  &  100  &  86  &  05:35:01.91  &  -05:34:15.17  &  05:35:15.50  &  -05:39:15.7  &  5:34:57.34  &  -5:42:35.5  \\
4/14/2005  &  100  &  87  &  05:35:08.51  &  -05:34:32.54  &  05:35:22.10  &  -05:39:33.1  &  5:35:03.94  &  -5:42:52.9  \\
4/14/2005  &  100  &  88  &  05:35:15.11  &  -05:34:49.90  &  05:35:28.70  &  -05:39:50.4  &  5:35:10.54  &  -5:43:10.2  \\
4/15/2005  &  100  &  89  &  05:35:21.70  &  -05:35:07.26  &  05:35:35.29  &  -05:40:07.8  &  5:35:17.13  &  -5:43:27.6  \\
4/15/2005  &  100  &  8A  &  05:35:28.30  &  -05:35:24.61  &  05:35:41.89  &  -05:40:25.2  &  5:35:23.73  &  -5:43:44.9  \\
4/15/2005  &  100  &  8B  &  05:35:34.90  &  -05:35:41.96  &  05:35:48.49  &  -05:40:42.5  &  5:35:30.33  &  -5:44:02.3  \\
\end{longtable*}

\subsection{Visit configuration} \label{section:observations:visit-config}
In order to use simultaneously all imaging instruments onboard \HST, we pushed to the limit the capability of the telescope to handle parallel, asynchronous observations. When different \HST\  instruments are operated in parallel, they ultimately compete to access the main Solid State Recorder (SSR)\ of the spacecraft to dump their data. We looked for a timing pattern that provides the highest possible observing efficiency while maintaining in the background a nearly continuous data transfer activity.  

\begin{figure}
\epsscale{1}
\plotone{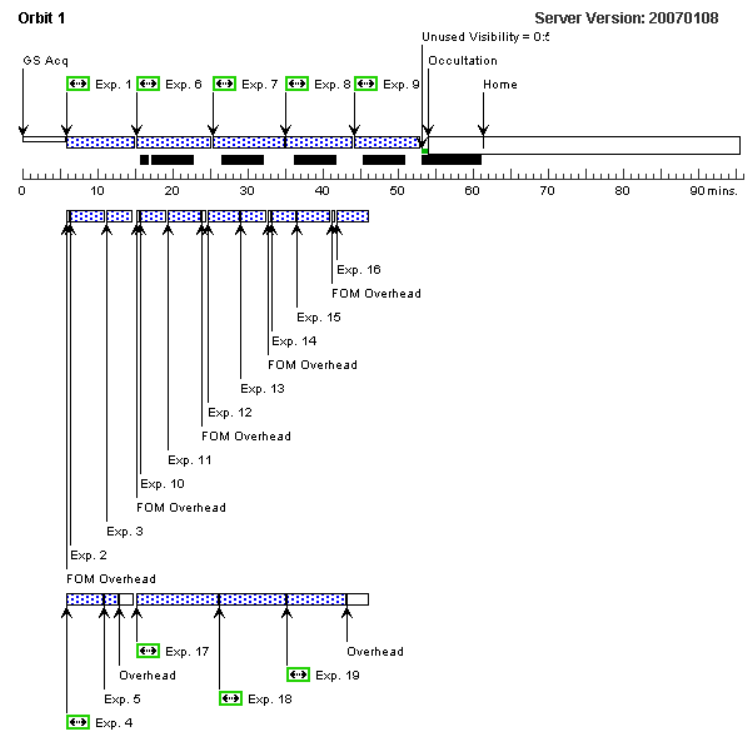}
\caption{Sequence of observations of a typical \HST\ orbit of our program. \label{Fig.visit}}
\end{figure}

We configured each pointing as a one orbit visit. Figure~\ref{Fig.visit} shows how the various exposures were distributed during each visit. After the initial Guide Star acquisition, the three instruments simultaneously started the observations. The exposure time with ACS has been fine tuned to allow continuous operations with buffer dump activity during the successive exposure or, in the case of the last image, during the occultation, leaving just 1~second of unused visibility time.
At the same time NICMOS obtained 9 images and WFPC2
5 images; the first two images on WFPC2 are the short ones, in order to access the SSR before the first data dump from ACS.

\subsection{Filters and exposure times} \label{section:observations:filters}

\begin{figure}
\epsscale{1.0}
\plotone{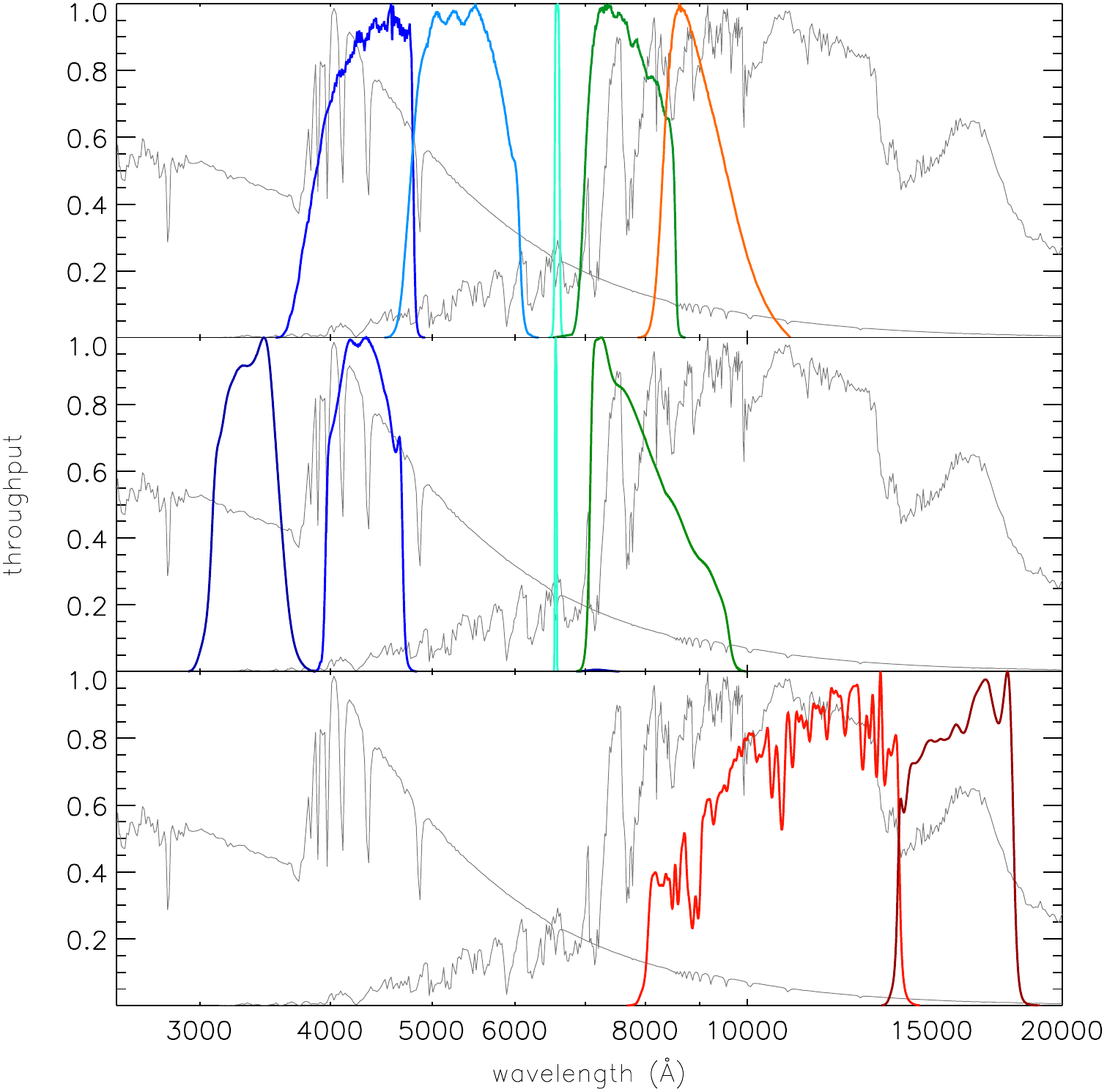}
\caption{Passbands of the 11 filters used for our program. Top diagram: ACS filters, from left to right: F435W, F555W, F658N, F775W and F850LP; middle diagram: WFPC2 filters, from top to right: F336W, F439W, F656N, and F814W; bottom diagram: NICMOS filters, from left to right: F110W and F160W. The gray lines represent the spectrum of an A0V star, according to the Nextgen model \citep{Hauschildt+1999} for a 10,000K atmosphere, and a M6V star according to the AMES-MT model \citep{Allard+2000} at 3,000K.
\label{Fig.filters}}
\end{figure}

As shown in Figure~\ref{Fig.visit}, a total of 19 images have been taken on each visit. The filters, exposure times and observing strategies are given in Table~\ref{Tab:Exposures} and the filter transmission curves are presented in Figure~\ref{Fig.filters}.


\begin{table}
\tiny
\caption{Visit Structure \label{Tab:Exposures}}
\begin{tabular}{lccccc}
\tableline
\tableline
{Instrument} & {Filter} & {Ground Equivalent} & {Exposure} & {Integration time (s)} & {Zero Point (VegaMag)}\\
\tableline
ACS/WFC & F435W & Johnson B & 6 & 420 & 25.779 \\
ACS/WFC & F555W & Johnson V & 9 & 385 & 25.724\\
ACS/WFC & F658N & broad H$\alpha$ & 1 & 340 & 22.365\\
ACS/WFC & F775W & Sloan i & 8 &385 & 25.256\\
ACS/WFC & F850LP & Sloan z & 7 & 385 & 24.326\\
WFPC2 & F336W & Johnson U & 17, 18 &400 $\times$ 2 (CR-SPLIT) & 20.279 \\
WFPC2 & F439W & Johnson B & 4 & 80 & 20.107\\
WFPC2 & F656N & narrow H$\alpha$ &  19 & 400 & 16.819\\
WFPC2 & F814W & Cousin I$_c$ & 5 & 10 & 20.845\\
NIC3 & F110W & J-band & 2, 11, 12, 15, 16& $256 \times 5$-pointing FOM dither & 22.50\\
NIC3 & F160W & H-band & 3, 10, 13, 14 & $192 \times 4$-pointing FOM dither& 21.66\\
\tableline
\end{tabular}
\tablecomments{Actual number of pointing for NIC3 may be lower (see Section~\ref{section:observations:LostData}).}
\tablecomments{To convert from total counts (DN)  to magnitude use the relation  $magnitude = -2.5 \log(DN / Exptime) + ZeroPoint$}
\end{table}

The motivation for the choice of the filters is the following.
The broad-band ACS filters, F435W, F555W, F775W and F850LP have been adopted for the
main broad-band photometric survey. We estimated that all ONC members down to below the substellar masses, with
the exception of the brightest stars whose bleeding trail could
fall out of the detector edges, would have provided enough signal in
at least two ACS\ bands to allow constraining their stellar colors and therefore,
given their spectral types, their reddening and absolute luminosity. The ACS
F658N $H\alpha$ filter is intended for mapping circumstellar matter with
the highest possible resolution, mostly to discriminate extended
sources and assess the presence of circumstellar emission or dark silhouette disks. The WFPC2 F336W filter was intended to probe the accretion
luminosity and therefore the mass accretion rates on the ONC stars,
along the lines of the earlier work by
\citet{Robberto+2004}. Since this filter is affected by red leak,
we have added a short exposure in the F814W filter to simultaneously probe the spectral region of the red leak and get a measure of the contamination unaffected by stellar variability. The F439W filter of WFPC2 also allows for a solid measure of the
Balmer jump. Finally, the WFPC2 F656N filter is a narrow band
$H\alpha$ filter which does not pass the [NII] line and has therefore
stronger diagnostic power than the broader ACS F658N filter, contaminated by the [NII] line at $\lambda=6583$~\AA. For NICMOS, the F110W and
F160W filters on NIC3 were adopted to  provide the deepest images of the fields ever obtained in the near-IR.

Due to FOM dithering pattern, the integration times of the combined NICMOS exposures varies across the mosaics, ranging between 192--762 seconds in filter~{F160W} and 256--1280 seconds in filter~{F110W}, the central parts of each sub-mosaic being imaged for the  full amount of time. About 55\% of the mosaics are covered for
192 seconds in the F160W band, and 40\% for 380 seconds. For the F110W filter, around 45\% is covered for 256 seconds, and
40\% for 512 seconds.

\subsection{Cosmic ray removal strategy} \label{section:observations:CRremoval}
Our observing strategy, optimized to map the largest possible field with all instruments onboard \HST\ is less than ideal in what
concerns the removal of cosmetic defects, cosmic rays in particular. We have therefore adopted the following strategy: for ACS, where
we have typically two (and occasionally more) images, cosmic rays were removed by adopting the value found in the other image(s), when available (see Section~\ref{section:datareduction:ACSdrizzling}).

The same strategy has been adopted for the F336W images of WFPC2, which come in pairs without dithering moves.
For the other filters, taken as single images,  we used the IDL routine \texttt{la\_cosmic.pro}\footnote{Available at \rm http://www.astro.yale.edu/dokkum/lacosmic/la\_cosmic.pro} written by Joshua Bloom (Caltech) based on the Laplacian algorithm originally developed by \citet{vandokkum01}. 

Finally, NICMOS image have been taken in Multiaccum mode sampling the integration signal as it accumulates on the detector. This allowed us to eliminate cosmic ray events using the standard NICMOS pipeline.



\subsection{Anomalies and lost data}\label{section:observations:LostData}
The brightness and stellar density at the core of the Orion Nebula reduces the number and reliability of useful guide stars. Spurious sources (extended, binaries) cause loss of guide on the Fine Guide Sensors of the telescope.  In most cases, the problems we encountered  during the observations were minor. In a number of visits, only one guide star (instead of the 2 usually required) was available and the orientation of the telescope had to be maintained by the less accurate gyroscopes. This, together with the catalog errors in the absolute position of the guide stars, causes the images to be slightly misaligned with respect to their nominal position. On the other hand, due to the relatively short exposure times we generally adopted, it did not result in a significant degradation of the Point Spread Function. Only in a few occasions we encountered major anomalies:
\begin{enumerate}
\item{Visit 01} only one of the two F336W images was taken with WFPC2
\item{Visit 16} lost due to failed guide star acquisition. This visit was successfully repeated on April 26, 2005 as visit 1E, with slightly different orientation (See Figures \ref{Fig:ACSfield} and \ref{Fig:WFPC2field}).
\item{Visit 17} the F435W exposure with ACS was interrupted at 75\% of total exposure time due to the loss of guide star. The previous F658N exposure, also with ACS, was completely lost.
\item{Visit 57} like visit 17, the F435W exposuse with ACS has 80\% of total exposure time due to the loss of guide star.
\end{enumerate}

NICMOS observations also experienced anomalies, related either to the guide star errors or to the errors in the commanding of the FOM:
\begin{enumerate}
\item Visit 49, 7A, 7B: lost due to the NICMOS suspend error triggered by FOM\ error.
\item Visits 3C: partial  data loss due to NIC suspend error for FOM error. Only the first F110W and F160W pattern positions were acquired.
\end{enumerate}

\subsection{File naming convention}\label{section:appendix:naming}
To facilitate data retrieval from the \HST\ archive, we provide a legend for the image file names. The original fits file naming convention follows the general 9-letters {\sl IPPPSSOOT} keyword of the \HST\ datasets, where:
\begin{itemize}
 \item{\sl I}\ refers to the instrument, i.e. {\sl I=j} for \ACS, {\sl I=u}\ for WFPC2 and {\sl I=n} for NICMOS;
 \item{\sl PPP}\ refers to the program identification, which in the case of the \HST\ Treasury Program is {\sl PPP=93k};
 \item{\sl SS}\ refers to the visit number, varying between {\sl SS=01}\ and {\sl SS=8D}.
 \item{\sl OO}\ refers to the observation number during each visit; for ACS\ and NICMOS this is an hexadecimal code which changes with each individual observation (exposure), whereas for WFPC2 it cycles through  the same values according to  the following list: 01=F439W; 02=F814W; 03=F336W; 04=F336W; 05=F656N.
 \item{\sl T}\ refers to a data-transmission code from the \HST\ spacecraft. In our case it is always {\sl T=M} (merged)\ for ACS and NICMOS, whereas  it is  {\sl T=Q} (retransmitted)\ for WFPC2.
\end{itemize}

\section{Data Reduction and Processing} \label{section:datareduction}

\subsection{ACS \flt\  images} \label{section:datareduction:ACSFLT}

Each of the $104 \times 5$ ACS exposures was run through the ACS On-The-Fly Reprocessing (OTFR) pipeline, which delivers $4096\times 4096$ pixel images corrected for bias, dark current  and flat-field. OTFR corrected images take the FLT suffix (\flt). The \flt\  images are still affected by the geometric distortion introduced by the ACS optics, mostly an 8\% compression of the ratio of the diagonals. As a result, WFC pixels project on the sky as rhombuses rather than squares. Varying nonlinearly with field position, geometric distortion introduces a photometric error by up to 9\%,  depending on the sky area imaged by each pixel.

Geometric distortion, and the associated photometric error, can be corrected using e.g. the drizzle package, discussed in Section~\ref{section:datareduction:ACSdrizzling}. However, when only a few dithered images are available (as in our case), it is more appropriate to extract the source photometry from the geometrically distorted \flt\ images, correcting only for the  photometric  error by multiplying the \flt\ images by the Pixel Area Map correction \citep[PAM][]{Gonzaga+2011}.  In the case of  ACS/WFC, the PAM
represents the fraction of nominal sky area ($0\farcs05\times0\farcs05$)\ seen by each pixel.
By avoiding re-sampling the images into a different projection, this approach preserves the original sharpness of both real stellar images and artifacts like bleeding trails and cosmic rays. Working on individual images rather than mosaiced images can also yield better photometry is some cases. For example, stars that are significantly variable can be mistakenly identified as being affected by cosmic rays in the mosaicing process, resulting in pixels near the PSF core being corrupted in the mosaic. Sticking with the individual images for the photometry avoids this problem. Our ACS photometry has therefore been obtained on PAM\ corrected \flt\  images (see Section~\ref{section:photometry}).

\subsection{ACS F850LP fringing} \label{section:datareduction:ACSfringing}
The two CCDs of ACS, thinned and backside illuminated, may be affected by fringing longward of 7500{\AA} due to the interference between the incident light and the light internally reflected at the interface between the thin layers of the chip. This problem affects our F850LP ACS\ images. On the ground, fringing is usually due to atmospheric airglow lines. In our case, it is due to the nebular emission of the Orion Nebula, dominated in the  F850LP pass-band by the [S III] lines at 9069{\AA} and 9532{\AA}, in a ratio of approximately 1:3 \citep{Osterbrock+92}. The Pa7 line at 10049 is a lesser contributor whereas the HeI 10830, intrinsically about as strong as the stronger 9069\AA\ [SIII] line, is damped by the low quantum efficiency of CCDs at this wavelength and contributes negligibly in comparison to the other lines.
The fringing pattern produced by an extended source like the Orion Nebula varies across the field following the relative brightness distribution of the emission lines. This complicates its suppression through image processing.

To mitigate fringing, we created a fringe flat-field by filtering out both low and high spatial frequencies from our F850LP images. Then we combined the fringe flat with that obtained from the ACS fringe
optical model for the [S III] 9069{\AA} and 9532{\AA} lines of \cite{Walsh+2003}. This reduced the peak-to-peak amplitude of the fringe pattern from $\sim10\%$ to less than $2\%$, making our F850LP \flt\ images largely unaffected by fringing.

\subsection{ACS\ Drizzling}\label{section:datareduction:ACSdrizzling}
To remove cosmic rays and combine the images into a single mosaic aligned with a global astrometric reference frame we use the Multidrizzle software \citep{Fruchter+Hook2002, Fruchter+Sosey2009}. Multidrizzle takes as input the  \flt\ images and returns them as \drz~ {\it drizzled\/} images corrected for distortion, aligned on the common astrometric grid and cosmic-ray cleaned. Then it combines the \drz~ images into a single image, which usually represents the final \HST\  data product with optimal depth, cosmetic quality and improved sampling of the PSF.

For a survey area as large as the one imaged by our program, the alignment of the distortion corrected \drz\ images into a single astrometric frame is generally rather complex. The astrometric information originally stored in the original FITS\ image header, calculated on the basis of the nominal parameters of the instrument and telescope pointing, is not accurate enough. This because the absolute position of the \HST\ guide stars is typically known with an uncertainty of about $0.5-1\arcsec$, i.e. 10-20 ACS/WFC pixels. Guide star position errors affect not only the relative offset between the drizzled ACS images, but also induce a rotation error: as the observations are generally performed locking the Fine Guide Sensors (FGSs) on two different guide stars typically separated by about $25\arcmin$, there is an orientation error as large as $2\arcmin$ in position angle, of the order of one pixel at the edge of the ACS field. In general, therefore, the astrometric correction stored in the fits header as World Coordinate System (WCS) parameters must be corrected before drizzling.

In the case of our ACS mosaic, the large shifts between images and the relatively low number of point-like sources made the alignment procedure especially challenging.  Our final strategy has been based on the following steps:

\begin{enumerate}
\item First, a copy of the  F850LP and F775W \flt\ images was smoothed with a Gaussian filter and subtracted from the original ones to remove most of the non uniform background; SExtractor \citep{bertin1996} was then run to derive the position of the  sources;  the coordinates of unsaturated source in at least one of the two filters were matched and merged to reduce the possibility of contamination by clusters of cosmic rays. This resulted in an initial input catalog of sources;

\item We concentrated on the $F850LP$ images, as this filter yields the largest number of usable sources. The $X,Y$ positions of all  unsaturated point sources in the F850W \flt\ images were accurately measured using the IRAF 2-D Gaussian centering routine. 

\item  The measured  $X,Y$ positions were corrected for geometric distortion using the \cite{Anderson2006} distortion model,
known to be accurate to better than $\sim 0.1$ pixels, i.e. less than 5~mas, across the extent of the ACS/WFC  \flt\ images.

\item For each of the nine ACS strips, the distortion corrected $X,Y$ positions  of stars measured in adjacent visits were matched and placed on a common grid. Starting from the image at one extreme of the strip, the other images were registered using a linear plate model transformation of this type:
    \begin{eqnarray}
     x_2=\Delta x+ax_1+by_1\\
     y_2=\Delta y+cx_1+dy_1
    \end{eqnarray}
\noindent where the parameters $\Delta x, \Delta y, a, b, c, d$, giving the location of the tile on the strip, were determined through a minimization procedure; the typical RMS of the linear fit resulted about 1 milli-arcsecond.

\item A master catalog was built for each strip, listing all distortion corrected, registered and averaged $X,Y$ positions.

\item The sources on each master catalog with counterparts in the 2MASS catalog were identified and a least-square solution was found to derive their RA and DEC in the 2MASS\ system. 
\end{enumerate}


The shifts calculated for each strip were used to update the WCS information in the corresponding \flt\ header file. These new updated flat-fielded images were used as input to  MultiDrizzle. The same shifts derived for the F850LP filters were used for the other filters, since all exposures within a visit were taken without moving the telescope and  the positional offsets between different filters are generally smaller than a pixel.


The \drz images are corrected for cosmic rays. Cosmic ray removal was done in the driz\_{cr} step of Multidrizzle by comparing the value of a pixel in each input image to the ``min" or ``median" combination of the values in the other images.
This process is extremely sensitive to alignment errors and source variability. If the alignment is less than perfect, the very sharp PSF\ core of the (normally two in our case)\ detections may fall on different pixels and therefore be flagged as cosmic rays and removed in the following combination step. An early attempt of combining frames from multiple strips into a few large ``supertiles'' of $14,000\times13,000$ pixels resulted in images with less than optimal astrometric and photometric quality.  It was found that the best results in the image registration could  be obtained working on a strip-by-strip basis, i.e. with images taken with the same telescope orientation. To keep the size of the reduced images below 1GB, we had to split each final strip in two halves, left and right. The full set of drizzled images, produced with the standard north-up, east-left orientation, is thus composed of 18 mosaics per filter, each one corresponding to half of a strip.  The visits included in each strip are listed in Table~\ref{tab:strips}. The early set of $14,000\times13,000$ supertiles was used to produce the large color picture of the Orion nebula, as described in the Appendix. 

The celestial coordinates reported in our source catalog are in the absolute astrometric reference of 2MASS, with a precision of the order of a few tens of milli-arcsecond, as the the typical 2MASS precision for a single source is $\sim$100~mas \citep{skrutskie2006}. Their relative positions, however, preserve the centroid precision of ACS/WFC of a few milli-arcseconds (depending on the brightness of the sources, of the background and on the source morphology). Accurate astrometris analysis should therefore be performed only on sources belonging to the same strip.


\begin{table}
\begin{center}
\caption{Visits included in ACS drizzled strips\label{tab:strips}}
\begin{tabular}{ll}
\tableline
\tableline
{Strip} & {Visits}\\
\tableline
0R  &  01, 02, 03, 04, 05, 06, 07 \\
0L &  07, 08, 09, 0a, 0b, 0c, 0d \\
1R  &  11, 12, 13, 14, 15, 1e, 17 \\
1L  &  17, 18, 19, 1a, 1b, 1c, 1d \\
2R  &  27, 28, 29, 2a, 2b, 2c, 2d \\
2L  &  21, 22, 23, 24, 25, 26, 27 \\
3R  &  32, 33, 34, 35, 36, 27, 38 \\
3L  &  37, 38, 39, 3a, 3b, 3c, 3d \\
4R  &  41, 42, 43, 44, 45, 46, 37 \\
4L  &  45, 46, 47, 48, 49, 4a, 4b, 4c \\
5R  &  57, 58, 59, 5a, 5b, 5c, 5d \\
5L  &  52, 53, 54, 55, 56, 57, 58 \\
6L  &  67, 68, 69, 6a, 6b, 6c, 6d \\
6R  &  64, 65, 66, 67, 68, 69, 6a \\
7L  &  71, 72, 73, 74, 75, 76, 77 \\
7R  &  75, 76, 77, 78, 79, 7a, 7b \\
8L  &  84, 85, 86, 87, 88, 89, 8a \\
8R  &  85, 86, 87, 88, 89, 8a, 8b \\
\tableline
\end{tabular}
\end{center}
\end{table}

\subsection{WFPC2} \label{section:datareduction:WFPC2}
WFPC2 images were processed through the standard {\tt calwp2} calibration software implemented in the OTFR pipeline. The pipeline removes additive (bias, dark current) and multiplicative (flat field) effects, besides flagging static bad pixels and correcting for instrumental artifacts. As anticipated in Section~\ref{section:observations:CRremoval}, cosmic-ray corrections on the F336W images, taken as CR-SPLIT pairs, was performed using the same minimization routine used for the ACS data, whereas for the other filters we used a Laplacian filter  \citep{vandokkum01}.

We applied delta-flats for the WF4 chip of the  F336W images to correct for artifacts created by dust spots on the optics that had moved  after the standard flat-field frames had been taken. The delta-flat frames were created stacking all of the WF4 frames, excluding those containing bright stars or with bright areas of nebulosity. In total,  44 images were used for the first epoch and  50 for the second.  The frames were first divided by their mean value and then were median averaged with  $\pm2.5$ sigma clipping.  Photometry checks showed that the results obtained on the  images with and without the delta-flat corrections coincide within 2\%. We also removed bias jumps, mostly occurring in the second epoch (spring) data, following the guidelines reported on the WFPC2 website\footnote{http://www.stsci.edu/hst/wfpc2/wf4\_anomaly.html}.

\subsection{NICMOS} \label{section:datareduction:NICMOS}
NICMOS images were reduced using the 4.1.1 version of the CALNICA calibration pipeline, with the exception of custom dark reference files. The CALNICA procedure subtracts ADC bias level, accounts for signal accumulated at the time of the first read, corrects for detector nonlinearities, estimates signal rates and rejects cosmic rays fitting the sequence of non-destructive reads, performs dark and flat field correction, and removes static image artifacts.

The second part of the image processing has been performed using the CALNICB pipeline, which allows removing the background illumination pattern.  As an additional processing step, we corrected for vignetting by using the ratio of one of our vignetted exposures in the central region of the cluster with the $3\times 3$ NICMOS mosaic of the Trapezium obtained in 1998 \citep{luhman2000} to derive a first-order correction. A smooth function was fit to the vignetted edge in the ratio image and then expanded into a 2-D vignetting correction image (one per filter for F110W and F160W) that was applied to all affected exposures. A further ``delta-flat'' correction was applied by median averaging all images taken at the same FOM\ positions, rejecting those more affected by diffuse nebular emission or point sources.
After determining the offsets of individual images of each region by measuring the centroids of common stars,
all images belonging to the same dither group were distortion corrected and combined  using  WDRIZZLE  Version 3.4.1 into a $800\times800$ image, preserving the nominal scale of $0.2$~arcsec/pixel and with an alignment such that pixel $(129,129)$ of the central image of each group falls at pixel $(401,401)$ of the output image. A final cosmic rays and bad pixel rejection process was performed by using a custom rejection procedure  that compares minimum and median values of pixels commons to multiple frames. Absolute astrometry was obtained, for all fields but one, using both 2MASS (primary0 and ISPI (secondary) catalogs (See Section~\ref{section:photometry:NICMOS}).

\section{Photometry}
\label{section:photometry}

\subsection{ACS photometry}
\label{section:photometry:ACS}

\begin{figure}
\epsscale{1.0}
\plotone{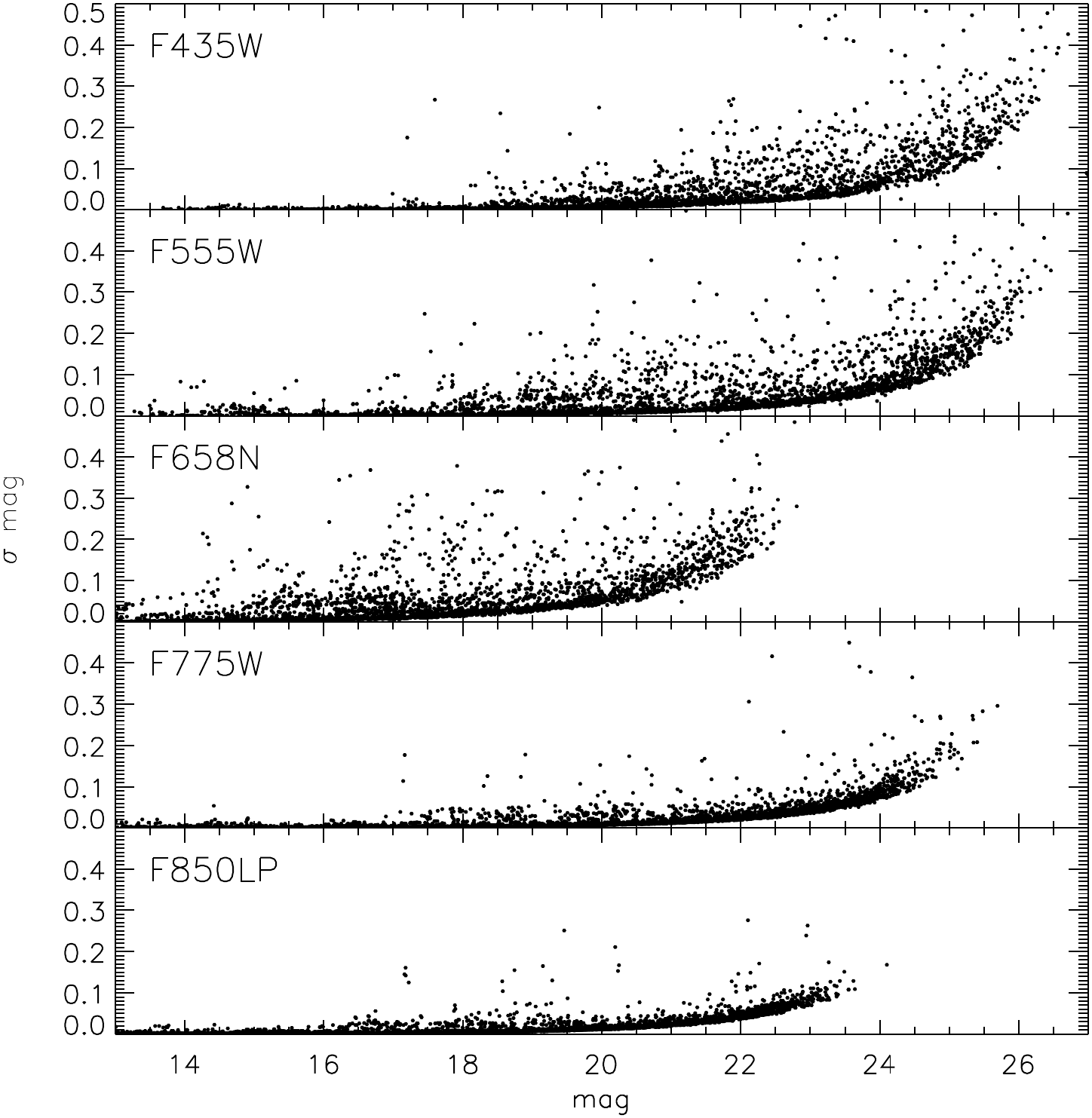}
\caption{Photometric errors as a function of magnitude for the 5 ACS filters.\label{Fig:ACSmagdmag}}
\end{figure}

\begin{figure*}
\epsscale{1}
\plotone{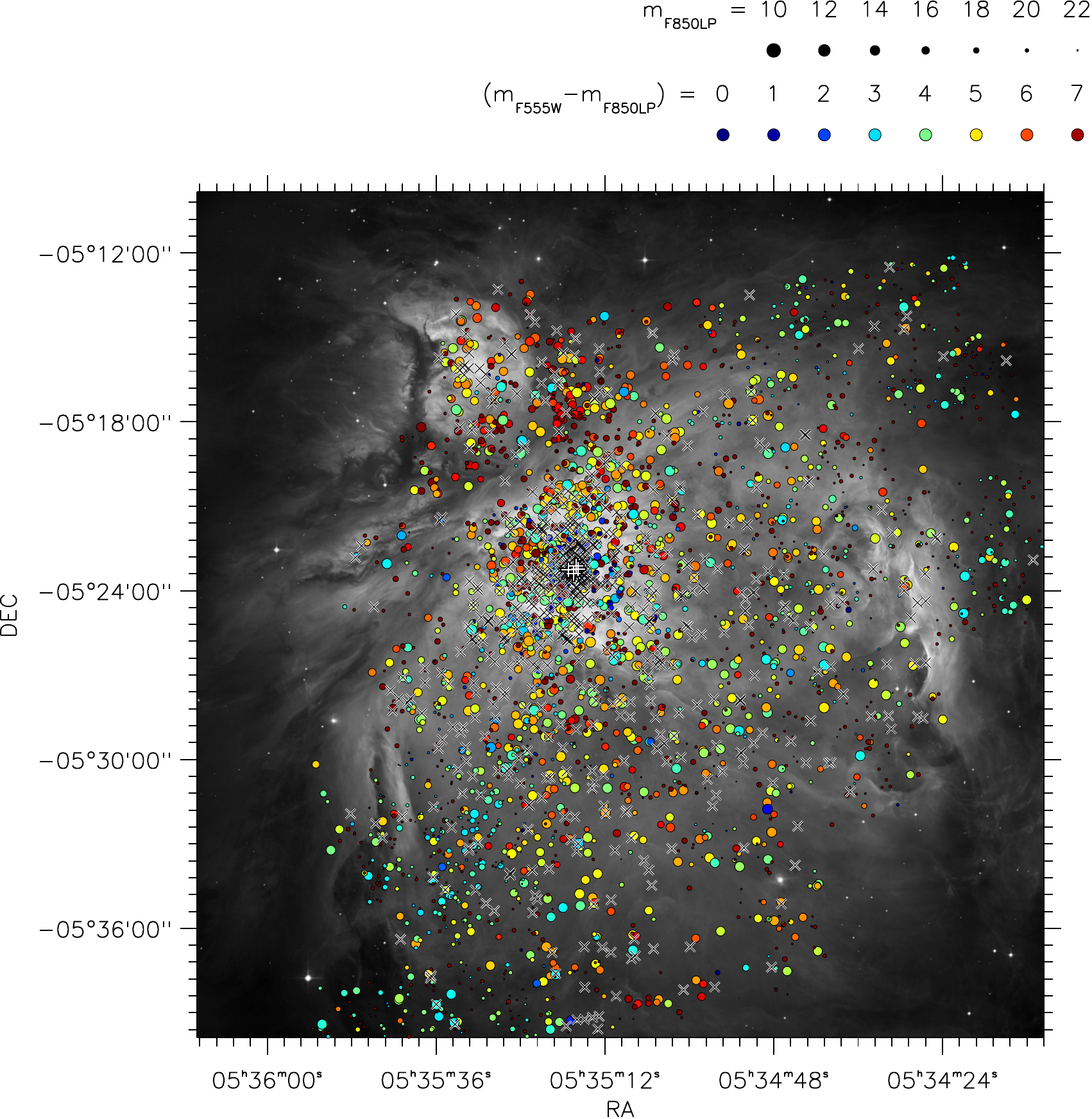}
\caption{Spatial distribution of the sources in the ACS photometry, overlaid on the image of the Orion Nebula produced from the ACS imaging (See Section \ref{section:results:colorimage} and the Appendix). The stars are color coded according to their color F555W-F850LP as shown in the upper label; sources detected only in the F850LP are plotted using the reddest color of the label (7). The size of the circles relates to the luminosity of each star. Crosses indicate either saturated sources ($+$) or objects detected with low signal-to noise ($\times$; $SNR\lesssim3$).\label{Fig:ACSradec}}
\end{figure*}

\begin{figure*}
\epsscale{1}
\plotone{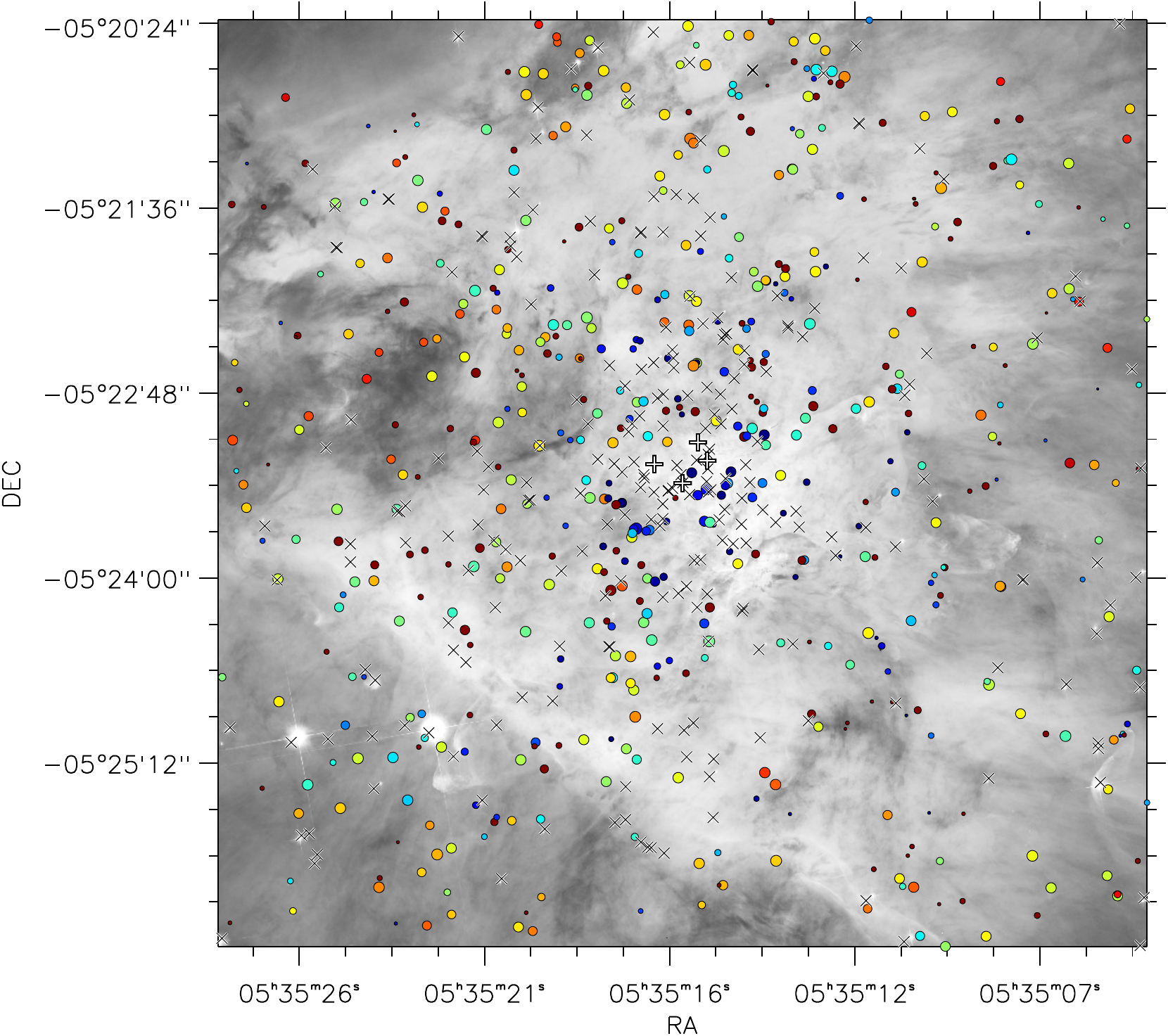}
\caption{Same as Figure \ref{Fig:ACSradec}, for the inner part of the region.\label{Fig:ACS_trapezium}}
\end{figure*}

In Section~\ref{section:datareduction:ACSdrizzling} we have described the construction of the ACS source catalog, containing a total of 8185 entries, including multiple detections. As anticipated in Section~\ref{section:datareduction:ACSFLT}, photometry extracted from our ACS drizzled images is prone to errors because of both source variability and the limited number of dither pointing used to produce the image. Having typically  only two exposures, often taken days apart, the cosmic-ray removal algorithm may flag for removal the PSF core of the brighter image.  To ensure photometric accuracy, our main photometric catalog has therefore been  extracted from the \flt\   images. This means that our source catalog will report, in most cases, multiple detections for each source, one for each visit.

As the \flt\ images are affected by cosmic rays, for each source and each filter we carefully tuned the aperture radius to avoid the presence of a cosmic ray. 
Aperture photometry was then obtained using a typical extraction radius of 10 pixels, occasionally reduced  to exclude a cosmic ray. Typical sky annuli were between 10 and 15 pixels, but when visual inspection showed strong spatial variability of the background, the sky annulus was adjusted accordingly to  get a more reliable estimate of the local background.

Having used a gain of 2 e-/ADU (see Section~\ref{section:observations:instruments-setup:ACS}), we could also extract the photometry of isolated saturated sources up to blooming distances of about 10 pixels. 
Our final catalog lists, for each source, the number of saturated pixels in the bleeding trails within the aperture and the maximum distance from the center.  Sources with larger bleeding trail usually require careful processing to optimize the extraction area and prevent the inclusion of nearby sources. While we do not provide their photometry, further analysis may allow to recover a few more bright sources.

Working with a range of extraction radii, we had to apply an aperture correction to our measures. We considered the possible errors introduced by directly applying the standard ACS aperture corrections of \cite{Sirianni+05}, as they were derived on drizzled, distortion-corrected images instead of our  \flt\ images. To address this point, we compared the stellar fluxes measured in our \flt\   images with those derived from the corresponding \drz\ images. In general, one expects \drz\ images to show broader PSFs cores than the \flt\ images, especially when only a few images are available. We found that for aperture radii larger than about 10 pixels the stellar fluxes extracted in the \drz\ and \flt\ images are  the same, allowing us to use the aperture corrections and  uncertainties tabulated by \cite{Sirianni+05}.
For aperture radii smaller than 10 pixels, the comparison showed some scatter. This is not surprising, considering that  our sample is not ideal to measure small systematic effects, due to the paucity of objects, the non-uniform background, and the possible contamination from circumstellar matter in young stellar objects. In order to measure the ratio between fluxes at radii $r<10$ pixels and $r=10$  pixels we selected ``mid-range'' sources, neither saturated or too faint in the \flt\ images, and averaged the results through a $\sigma$-clipping routine to reject outliers. The corrections derived are consistent (within $1 \sigma$) with the standard ACS photometric aperture corrections of \cite{Sirianni+05}. On these grounds, we assumed as a baseline the \cite{Sirianni+05} aperture corrections at all radii.

We also corrected our photometry for the ACS/WFC CCDs Charge Transfer Efficiency (CTE) using the correction formula of the Instrument Science Report ACS 2004-006 \citep{RiessMack04}. The CTE correction turned out to be very small, primarily because of the generally high background due to the strong nebular emission. Our final step was to apply the photometric zero-points. We adopted the standard VEGAMAG photometric system tabulated by \cite{Sirianni+05}.

In Figure~\ref{Fig:ACSmagdmag} we present the calibrated magnitudes versus their corresponding errors for the 5 ACS filters. Our $5\sigma$ sensitivity limit ($\sigma_{mag}\simeq 0.2$) are about $m[$F435W$]=26$, $m[$F555W$]=25.5$, $m[$F658N$]=22.5$, $m[$F775W$]=25.3$, $m[$F850LP$]=24.2$ in the Vega system.

In Figure~\ref{Fig:ACSradec} we show the spatial distribution of the ACS sources in the final catalog, color coded according to their color. In Figure~\ref{Fig:ACS_trapezium} we present the same, for the highly crowded central region of the ONC.

\subsection{WFPC2 Photometry} \label{section:photometry:WFPC2}
WFPC sources were identified by looking for counterparts to our ACS \citep{Ricci_atlas} and ISPI \citep{Robberto+2010} photometry. We preferred this approach, instead of considering all possible detections on the WFPC2 frames, in order to suppress the possible contamination from spurious effects in the aging CCD detectors. Both the ACS and ISPI catalogs have a significantly deeper detection limit, both due to the higher exposure times and because of the longer wavelengths -- therefore less affected by dust extinction -- covered by these two other surveys. 

Particular care has been used for close binaries, for which we checked the accuracy of the matching by comparing photometry estimate in different bands and relative stellar positions. 
\begin{figure}
\epsscale{1.0}
\plotone{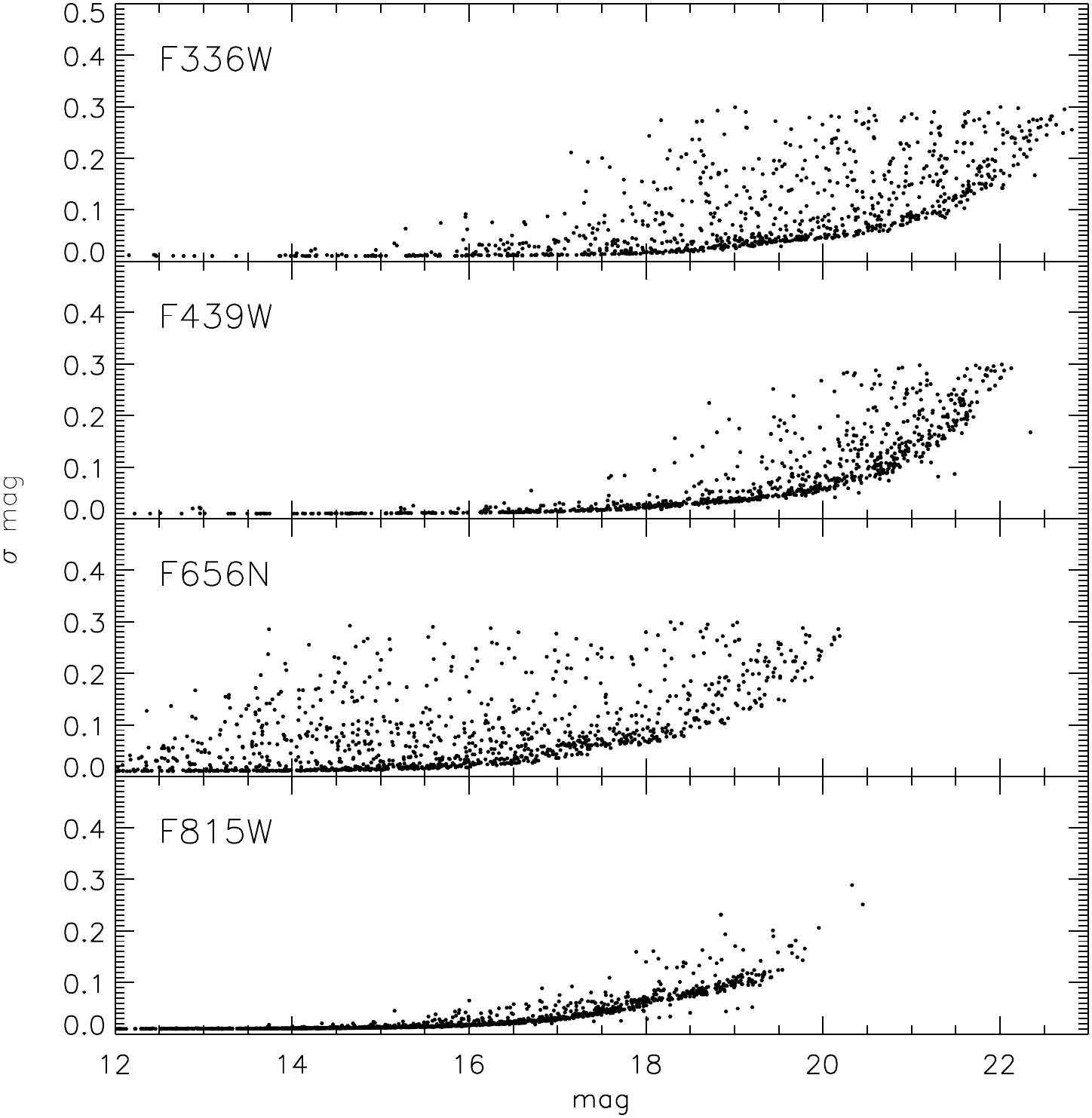}
\caption{Photometric errors as a function of magnitude for the 4 filters used with WFPC2.\label{Fig:WFPC2magdmag}}
\end{figure}
\begin{figure*}
\epsscale{1}
\plotone{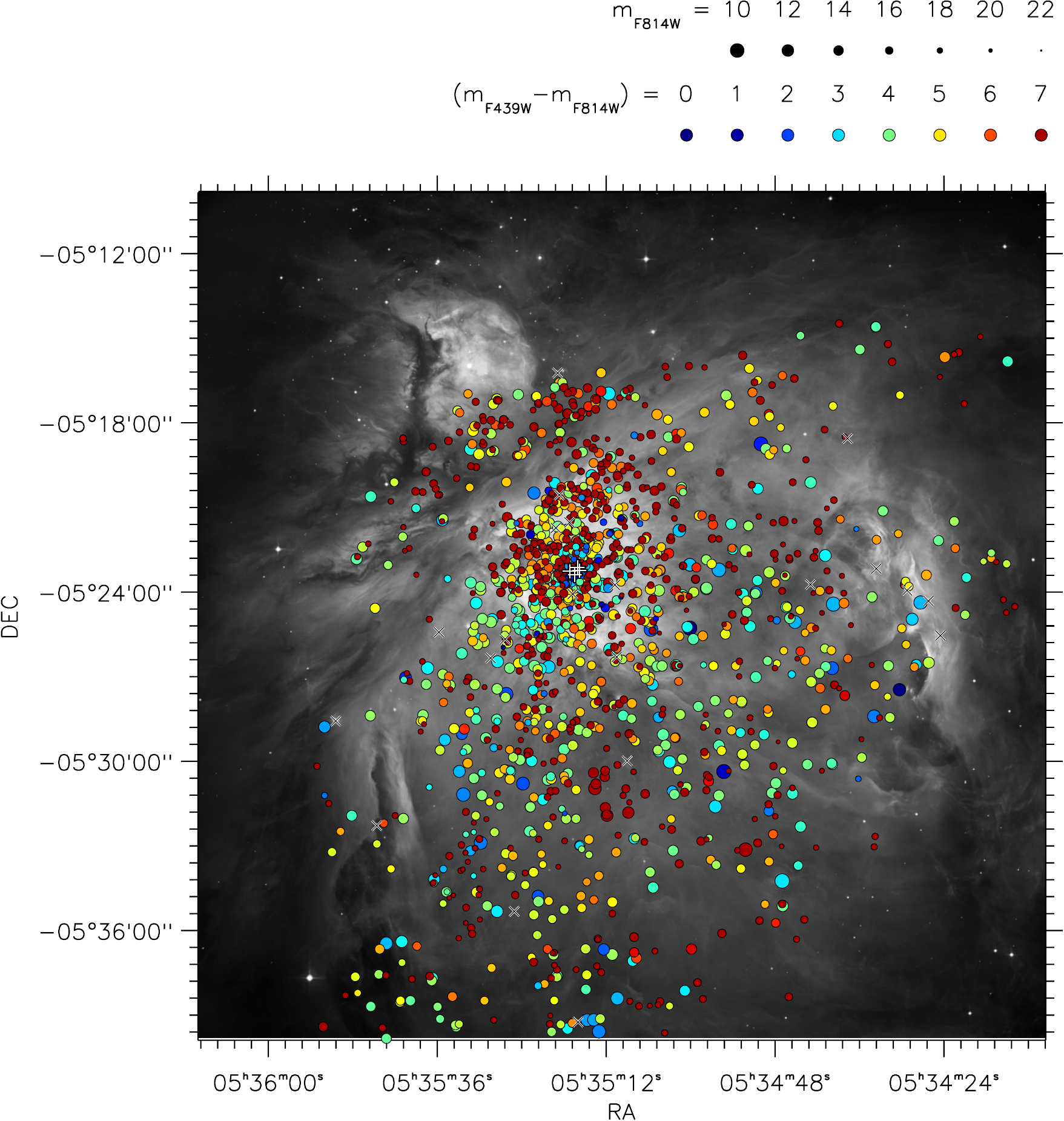}
\caption{Same as Figure \ref{Fig:ACSradec}, for the WFPC2 photometric catalog.\label{Fig:WFPC2radec}}
\end{figure*}

Having identified 1643 WFPC2 sources, we derived their photometry on the images corrected for cosmic rays.
We  extracted aperture photometry on both $0\farcs2$ and  $0\farcs5$ radius apertures, corresponding to 5 and 11 pixels on the WF1 and 2 and 5 pixels on WF2-4.  The sky annulus was always taken between $1\farcs0$ and $1\farcs5$, corresponding to 20 and 30 pixels with WF1 and 10 and 15 pixels with WF2-4.

We applied CTE\ correction to the measured counts following the recipe of \citet{Dolphin00} (also on http://purcell.as.arizona.edu/wfpc2\_calib/). From the same source we adopted the zero points to the Vega system, using the  values appropriate for a 14~e/adu gain (updated on Sep. 10, 2008).
For the F336W and F656N\ filters, not included in Dolphin's list, we used the zero points on the WFPC2 Data Handbook,
(http://www.stsci.edu/hst/wfpc2/analysis/wfpc2\_photflam.html)
applying the correction appropriate for 14~e/adu gain. Their values (for the representative PC chip) are provided in \ref{Tab:Exposures}. 

The aperture correction from $0\farcs1$ to $0\farcs5$ was taken from \citet{holtzman1995}, appropriate for the WF channels. For the F656N aperture correction, not listed by \citet{holtzman1995}, we assumed the F675W values.
Aperture correction between $0\farcs5$ and infinity was performed by  subtracting  0.1 magnitudes from the measured values,
as appropriate for the adopted zero points.
We added in quadrature the errors associated to the measured counts, zero point, CTE\ correction, aperture correction to the infinity and, in the case of the $0\farcs1$ apertures, to the aperture correction to the $0\farcs5$ radius. This last term, when present, usually dominates and for this reason the $0\farcs5$ photometry is  more accurate, except for the faintest sources detected in a few pixels. For the CTE\ correction we assumed an error 20\% of the amount of the correction itself.

In Figure~\ref{Fig:WFPC2magdmag} we present the calibrated magnitudes versus their corresponding errors for the 4 WFPC2 filters. Our $5\sigma$ sensitivity limits ($\sigma_{mag}\simeq 0.2$) are about $m[$F336W$]=22.2$, $m[$F439W$]=21.7$, $m[$F656N$]=19.5$, $m[$F816W$]=20.0$, in the Vega system.
In Figure~\ref{Fig:WFPC2radec} we show the spatial distribution of the WFPC2 sources in the final catalog, color coded according to their color.

\subsection{NICMOS Photometry} \label{section:photometry:NICMOS}

\begin{figure}
\epsscale{1}
\plotone{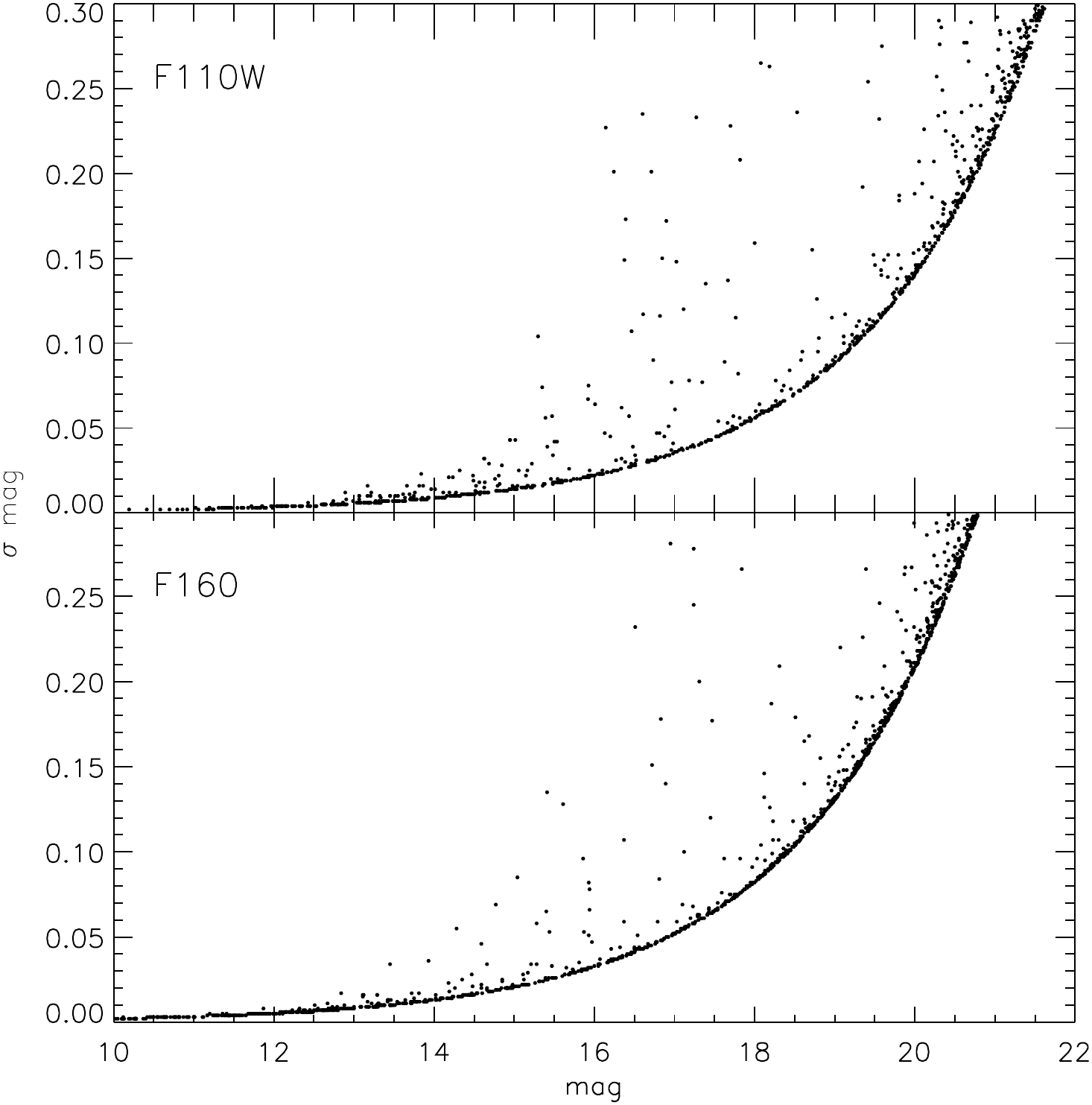}
\caption{Photometric errors as a function of magnitude for the 2 NICMOS filters.\label{Fig:NICMOSmagdmag}}
\end{figure}
\begin{figure*}
\epsscale{1}
\plotone{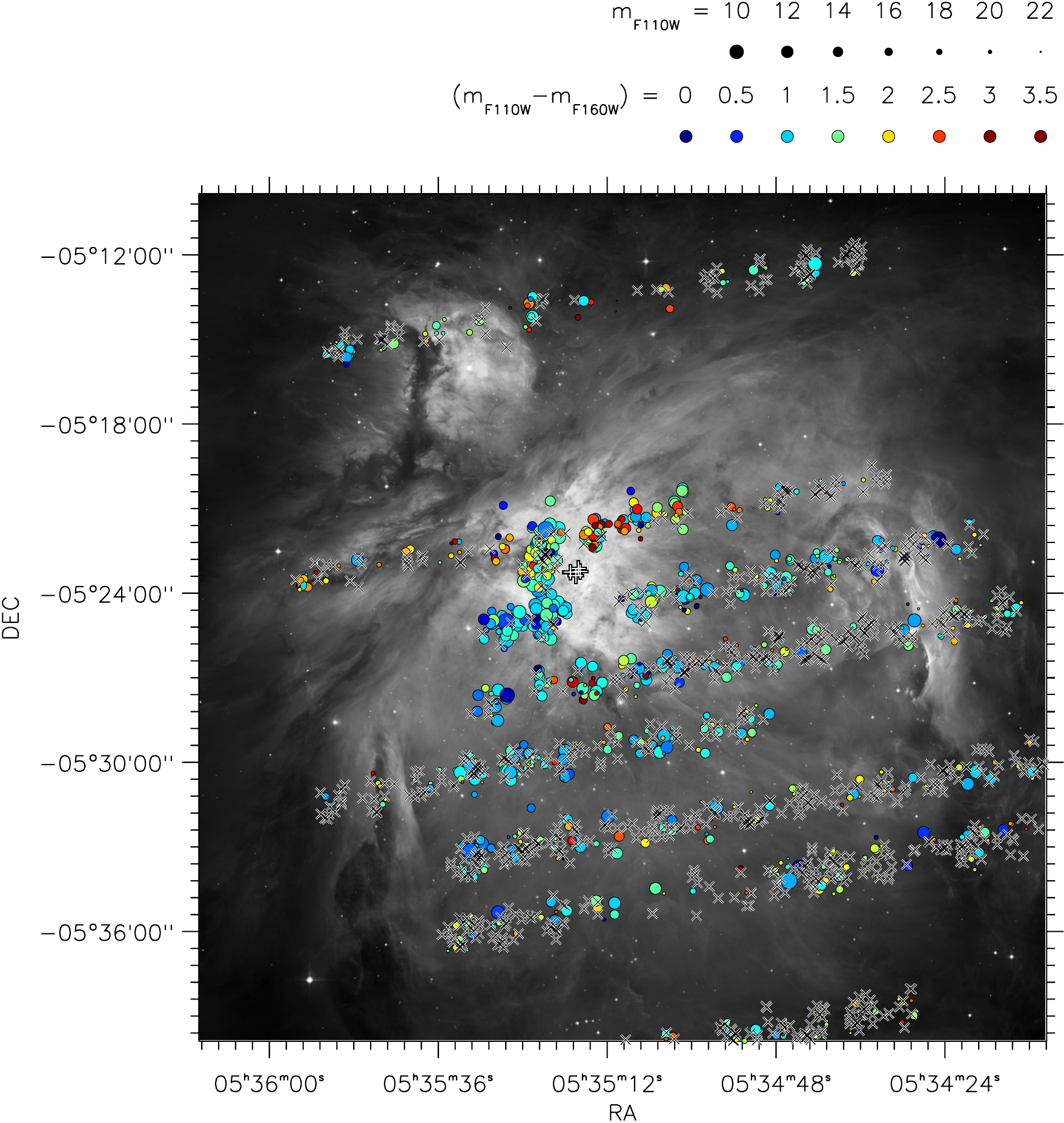}
\caption{Same as Figure \ref{Fig:ACSradec}, for the NICMOS photometric catalog.\label{Fig:NICMOSradec}}
\end{figure*}

Source identification and photometry in NICMOS images is made difficult by the strong and non-uniform background emission, especially at the center of the survey area, and by the undersampled PSF of the NICMOS Camera 3. To remove the large scale nebulosity, we applied a ring-median filter replacing each pixel with the median value taken in the surrounding annulus with inner and outer radius
of 7 and 11 pixels (corresponding to about $1\farcs$4--$2\farcs2$), respectively. The filtered image was subtracted from the corresponding original one to remove the large scale nebulosity and enhance the detection of point sources.

Each point source was visually inspected in the F160W image, deeper than the F110W image and with better PSF sampling. The compatibility of the radial profile of each source with the typical PSF was also inspected to exclude spurious detections due to nebulosity. We detect a total of 2116 objects and used the positions of the objects in the F160W images to extract the photometry in both bands.
Photometry was performed with an aperture radius of 2.5 pixels ($0\farcs5$) and a sky annulus between 10 and 15 pixels ($2\farcs0$--$3\farcs0$) for both filters. Bright isolated stars in different regions were used to determine aperture corrections with an estimated random error of 0.02 mag in both bands.
The zero points, adopted from the NICMOS Data Handbook \citep{Thatte+2009} 
are 22.50~mag (F110W)  and 21.66~mag (F160W) including aperture corrections.


Objects in common with 2MASS and with the ISPI near-IR survey of \citet{Robberto+2010} were used to calibrate the astrometry of each mosaic. For visit 70 no reference star could be found within the field of view. Coordinates for the three sources inthis field (entries 1389-1391, starting from 0) were obtained from the header information and are expected to be accurate to $\simeq 1\arcsec$, the average accuracy of the header coordinates derived from the other of calibrated frames. NICMOS astrometry is generally  in agreement with the ISPI astrometry to within $0\farcs1$ for 60\%  of the sources and within $0\farcs2$ for about 90\% of the sources.
Recently, \cite{Andersen+2011} converted the photometric data into the 2MASS system using the ground based data of \cite{Robberto+2010}. See their paper for a detailed discussion of how the completeness limit depends on the distance from the cluster center, and hence on the amount of nebulosity.

Figure~\ref{Fig:NICMOSmagdmag} shows the photometric error for the NICMOS sources against the corresponding magnitudes, for both filters. Our photometry is significantly deeper than the previous ISPI study, reaching a $5\sigma$ limit (or about 0.2 magnitudes of photometric error) of F110W$\simeq20.7$ mag, and F160W$\sim20$ mag.
In Figure~\ref{Fig:NICMOSradec}  we show the spatial distribution of the NICMOS sources in the final catalog, color coded according to their color.

\section{Data Products} \label{section:dataproducts}
In this section we summarize the data products we have publicly released, available for download as High Level Science Products in the Multi-mission Archive at the Space Telescope Science Institute ({\rm http://archive.stsci.edu/prepds/orion/}).

\subsection{ACS} \label{section:dataproducts:ACS}
The ACS dataset consists of three main products.  
\begin{enumerate}

\item
We provide the full set of 520 ACS \flt\ images, reprocessed according to version 5.1.1 of the CALACS pipeline, corrected for pixel area map and with reference celestial coordinates (CRVAL1 and CRVAL2 values in the fits headers) registered to match the absolute astrometry of the stars falling in the field. Note that these images like every other \flt\ image released by latest version of the OTFR pipeline, contain  SIP coefficients for the correction of the field distortion. However, only the ``direct''  $A_{nm}, B_{nm}$ coefficients are present, allowing to directly derive the Right Ascension and Declination corresponding to a certain  pixel (using e.g. the IDL {\sl xyad} procedure), but not viceversa. As the reverse coefficients,  $Ap_{nm}, Bp_{nm}$, are not present,  the transformation from celestial coordinates to distorted pixel coordinates may require a different (e.g. iterative) approach.

\item
We provide the full set of 90 ACS drizzled images, 9 strips per filter divided in a left (east) and right (west) part to maintain their image size below 1Gpix. Their naming convention is strip{\sl \#X\_filter}\_drz.fits, where {\sl \#} refers to the strip number (0 to 8), {\sl X} can be either L or R\ for the left or right part of each strip and {\sl filter} is the ACS\ filter name. For the correspondence between visits (\flt\ files) and strips (\drz\ files) see Table~\ref{tab:strips}.  Note that the ACS drizzling software does not use the SIP coefficients but more refined distortion tables internal to the tool.

\item
Finally, we provide the ACS source catalo with photometric data  for  3399 stars. Of these, 352 have been measured once, 2074 twice, 249 three times, 682 four times and 42 five times, for a total of 8185 entries. Table~\ref{Tab:ACS_catalog} lists the entries provided for each star.

\begin{table*}
\begin{tabularx}{400pt}{l|X}
\multicolumn{2}{c}{ACS source catalog\label{Tab:ACS_catalog}}\\
\hline
\hline
Column & Description \\
\hline
onc\_acs & Designation, or sequential entry number; identifies the observation of a source within a visit (orbit). \\
onc\_acs\_1 & First entry number. This field facilitates finding sources observed multiple times. It provides the entry number (onc\_acs) of the first appearance in the catalog. For example, entries onc\_acs=100, 101, and 102 are three observations of the same star (in different visits). They all have onc\_acs=100, as this is the first entry number attributed to this particular source. \\
x\_435 to x\_850 & Five columns for the x positions measured in the \flt\ images, ordered by wavelength (F435W, F555W, F658N, F775W, F850LP). For saturated filters, the average on the unsaturated ones is used. \\
y\_435 to y\_850 & Five columns for the y positions measured in the 5 \flt\ images. \\
m\_435 & F435W\ aperture photometry, in Vegamag, measured in the \flt\ image. If the star is undetected, this field takes the NULL value 99.9999; if saturated and not recovered, this value represents an upper limit. \\
dm\_435 & F435W\ aperture photometry error measured in the \flt\ image. If the star is undedetect, this value represent the  3$\sigma$ upper limit; if the star is saturated and not recovered, this field takes the NULL value 99.9999. \\
mf\_435 & F435W\ aperture photometry flag: 0=undetected; 1=detected; 2=saturated and recovered; 3=saturated and not recovered (upper limit); 4: undetected in the \flt\ image (flag=0) but measured in the \drz\ strip. \\
m\_555 & like the m\_435 field for the F555W\ filter. \\
dm\_555 & like the dm\_435 field for the F555W\ filter. \\ 
mf\_555 & like the mf\_435 field for the F555W\ filter.  \\
m\_658 & like the m\_435 field for the F658N\ filter.  \\
dm\_658 & like the dm\_435 field for the F658N\ filter. \\ 
mf\_658 & like the mf\_435 field for the F658N\ filter. \\
m\_775 & like the m\_435 field for the F775W\ filter. \\
dm\_775 & like the dm\_435 field for the F775W\ filter. \\ 
mf\_775 & like the mf\_435 field for the F775W\ filter. \\
m\_850 & like the m\_435 field for the F850LP filter.  \\
dm\_850 & like the dm\_435 field for the F850LP\ filter. \\ 
mf\_850 &  like the mf\_435 field for the F850LP\ filter. \\
rad\_435 to rad\_850 &  Five columns, one for each filter, for the extraction radius for aperture photometry in \flt\ image, in pixels. \\
csky\_435 to csky\_850 &  Five columns, one for each filter, for the sky counts in aperture photometry in \flt\ image. \\
ssky\_435 to ssky\_850 &  Five columns, one for each filter, for  the error of the sky counts in aperture photometry,in \flt\ image. \\
isky\_435 to isky\_850 & Five columns, one for each filter, for  the  inner radius of sky annulus in \flt\ image, in pixels. \\
osky\_435 to osky\_850 &  Five columns, one for each filter, for  the outer radius of sky annulus in \flt\ image, in pixels. \\
max\_435 to max\_850 & Five columns, one for each filter, for  the peak counts within extraction radius in \flt\ image. \\
spx\_435 to spx\_850 & Five columns, one for each filter, for  the number of saturated/bleeding pixels within the extraction aperture. \\
rspx\_435 to rspx\_850 & Five columns, one for each filter, for  the maximum distance of saturated pixels from the aperture center, usually along a bleeding trai, for each filter. If this value reaches 10 pixels, our maximum extraction radius, the star is considered saturated and is not recovered.   \\
type & source type, from visual inspection:\ 0) not measured; 1); detected in at least one filter and unresolved; 2)\ double (companion closer than $\approx 3$ pixels, one entry for both sources); 3) wide double (companion further than $\approx 3$ pixel, one entry for each source); 5)\ silhouette disk; 6) photoionized (proplyd or with other evidence of photoionization); 7) galaxy; 8) Herbig-Haro. \\
strip & \drz\ strip. \\
x\_strip & x coordinate on \drz\ strip. \\
y\_strip & y coordinate on \drz\ strip. \\
m\_435s to m\_850s & Five columns, one for each filter, for  the aperture photometry in \drz\ strip. Null value 0.000 is used for sources non measured. \\
dm\_435s to dm\_850s & Five columns, one for each filter, for  the aperture photometry error in \drz\ strip. Null value 0.000 is used for sources non measured. \\
rad\_435s to rad\_850s &  Five columns, one for each filter, for  the extraction radius for aperture photometry in \drz\ image, in pixels. \\
sky\_435s to sky\_850s &  Five columns, one for each filter, for  the sky counts  per pixel in \drz\ image. \\
dsky\_435s to dsky\_850s & Five columns, one for each filter, for  the  error on sky counts  per pixel in \drz\ image. \\
date & date of observation (UT). \\
time & time of observation (start of first exposure, F658N). \\
\hline
\multicolumn{2}{c}{Table 6 is published in its entirety in the electronic edition of the {\it Astrophysical Journal}.  A portion is shown here for guidance regarding its form and content.}
\end{tabularx}
\end{table*}

\end{enumerate}

\subsection{WFPC2} \label{section:dataproducts:WFPC2}
The WFPC2 dataset consists of two main products.  
\begin{enumerate}
\item
The full set of 416 WFPC2 images (the pairs of F336W images being coadded), processed according to version 15.4c of the OPUS pipeline, corrected for pixel area map and with reference celestial coordinates (CRVAL1 and CRVAL2 values in the fits headers) registered to match the absolute astrometry of the stars falling in the field. The \flt\ images released by the OTFR pipeline contain  SIP coefficients for the correction of the field distortion. However, only the ``direct''  $A_{nm}, B_{nm}$ coefficients are present, allowing to accurately derive the Right Ascension and Declination corresponding to a certain  pixel value (x,y) using e.g. the IDL {\sl xyad} procedure. The reverse coefficients,  $Ap_{nm}, Bp_{nm}$, instead are not present and the transformation from celestial coordinates to distorted pixel coordinates may therefore require a different, e.g. iterative, approach.

\item
The WFPC2 catalog provides photometric data, at least in one band, for 1643 sources in total. Among these, 1592 were previously detected in the ACS survey and 51 only by ISPI. For 1021 stars ($\sim$ 60 \%) $U$-band photometry is available. Finally, for 897 sources ($\sim$ 55 \% of the total), we have photometry in all $U$, $B$ and $I$ bands. For each of the 1643 stars listed in the WFPC2 catalog we provide the entries listed in Table~\ref{Tab:WFPC2_catalog}.

\begin{table*}
\begin{tabularx}{400pt}{l|X}
\multicolumn{2}{c}{WFPC2 source catalog\label{Tab:WFPC2_catalog}}\\
\hline
\hline
Column & Description \\
\hline
onc\_wfpc2 & Designation or entry number  in the WFPC2 catalog. \\
onc\_wfpc2\_1st & First entry number. Like the similar field in the ACS catalog, this field facilitates finding sources observed multiple times by provideing the entry number (onc\_wfpc2) of the first appearance in the catalog. \\ 
RAh/RAm/RAs & Three columns for the hours, minutes, and seconds of Right Ascension (J2000.0). \\
DE-/DEd/DEm/DEs & Four columns (the first one for the sign) for the Declination (J2000.0). \\
visit & Visit number \\
CCD & CCD detector of WFPC2; 1 is for the Planetary Camera, 2,3,4 are for the Wide Field Cameras. \\
x\_336 to x\_814 & Five columns, for the four filters F336W, F439W, F656N, and F814W, for the x position on the chip \\
y\_336 to y\_814 & Five column, one for each filter, for the corresponding y positions. \\
m\_336 & F336W  magnitude (Vegamag). \\
dm\_336 & Magnitude error in the F336W filter. \\
mf\_336 & F336W  aperture photometry flag: 0=not measured; 1=detected; 2=non detected; 3=saturated. \\
m\_439 & F439W (B-band) magnitude (Vegamag). \\
dm\_439 & Magnitude error in the F439W filter. \\
m\_656 &  F656N magnitude (Vegamag). \\
dm\_656 & Magnitude error in the F656N filter. \\
m\_814 &  F814W (magnitude (Vegamag). \\
dm\_814 & Magnitude error in the F814W filter. \\
rad\_336 to rad\_814 & Five column, one for each filter, for the aperture photometry extraction radius. \\
csky\_336 to csky\_814 &  Five column, one for each filter, for the sky counts for aperture photometry. \\
ssky\_336 to ssky\_814 & Five column, one for each filter, for the sky counts errors for aperture photometry. \\
max\_336 to max\_814 & Five column, one for each filter, for the source peak counts within extraction radius.  \\
date & Date of observation (UT). \\
time & Time of observation. \\
\end{tabularx}
\end{table*}

\end{enumerate}

\subsection{NICMOS} \label{section:dataproducts:NICMOS}
The NICMOS dataset also consists of two main products.  
\begin{enumerate}
\item The final drizzled images. 
\item The NICMOS source catalog, containing 2116 sources. The entries for each star are listed in Table~\ref{Tab:NICMOS_catalog}.
\end{enumerate}
\begin{table*}
\begin{tabularx}{400pt}{l|X}
\multicolumn{2}{c}{NICMOS source catalog\label{Tab:NICMOS_catalog}}\\
\hline
\hline
Column & Description \\
\hline
ENTRY & Source ID\  number in the NICMOS\ catalog. \\
visit & Visit ID. \\
x & x-position on the drizzled NICMOS tile. \\
y & y-position on the drizzled NICMOS tile. \\
RAh/RAm/RAs &  Three columns for the hours, minutes, and seconds of Right Ascension (J2000.0). \\
DE-/DEd/DEm/DEs & Four columns (the first one for the sign) for the Declination (J2000.0). \\
m\_110 & Magnitude in the F110W filter. 99.9999 is NULL value. \\
dm\_110 & Magnitude error in the F110W filter. If m\_110=NULL this is an upper limit. \\
mf\_110 & Magnitude flag in the F110W filter. 0=not measured, 1=measured; 2=undetected, with upper limit shown in dm\_110.\\ 
m\_160 & Magnitude in the F160W filter. 99.999 is NULL value. \\
dm\_160 & Magnitude error in the F160W filter. If m\_160=NULL this is an upper limit. \\
mf\_160 & Magnitude flag in the F160W filter. 0=not measured, 1=measured; 2=undetected, with upper limit shown in dm\_160. \\
\end{tabularx}
\end{table*}

\subsection{Ancillary ground based catalogs} \label{section:dataproducts:groundbased}
As anticipated in Sec.~\ref{sec:Introduction}, we have complemented our \HST\ survey with ground based observations at visible and near-IR wavelengths.  The main purpose of these observations was a) to measure sources that appeared saturated with the \HST, given the relatively long exposure times we had adopted, and b) to evaluate and minimize, as much as possible, the effect of source variability in the source colors, since the \HST\ cameras did not observe the same field at the same time. For this reason, the observations were carried out on the same nights, 1-2 January 2005, using WFI at the 2.2 telescope at La Silla for the optical survey and ISPI at the 4m telescope at Cerro Tololo for the near-IR survey. 

The optical observations, presented in \citet{dario2009}, have been taken in the $U$, $B$, $V$, $I$ broadband, 6200 Å TiO medium-band, and $H\alpha$ narrow-band filters with the WFI imager at the ESO/MPI 2.2 telescope at La Silla Observatory. The source catalog contains 2612 point-like sources in the $I$-band; 58\%, 43\%, and 17\% of them are also detected in $V$, $B$, and $U$-band, respectively. 1040 sources are identified in the $H\alpha$ band. 
The near-IR observations, presented in \cite{Robberto+2010}, have been obtained in the $J$, $H$, and $K_S$ filters. The catalog contains about 7800 sources, reaching 3$\sigma$ accuracy in the 2MASS system down to $J = 19.5$ mag, $H = 18.0$ mag, $K{_S} = 18.5$ mag, enough to detect planetary size objects ($M\simeq0.012$~M$_\odot$) 1 Myr old under $A_V\simeq$10 mag of extinction at the distance of the Orion Nebula, according to the \citet{Chabrier+00} models. 

The images associated to these ground based observations are also available as High-Level Science Product at the 
http://archive.stsci.edu/prepds/orion/. The tables with the photometry and derived physical quantities can be downloaded at no cost in machine readable format from the electronic version of the publications or from the CDS archive, respectively  http://iopscience.iop.org/0067-0049/183/2/261/ and http://cdsarc.u-strasbg.fr/viz-bin/Cat?J/ApJS/183/261 for \citet{dario2009}, and 
http://iopscience.iop.org/1538-3881/139/3/950/ and http://cdsarc.u-strasbg.fr/viz-bin/Cat?J/AJ/139/950 for \citet{Robberto+2010}.

\section{Selected results} \label{section:results}
In this section we briefly illustrate some of the results obtained so far from our Treasury Program. 

\subsection{ACS Color-Composite Image} \label{section:results:colorimage}
\begin{figure*}
\plotone{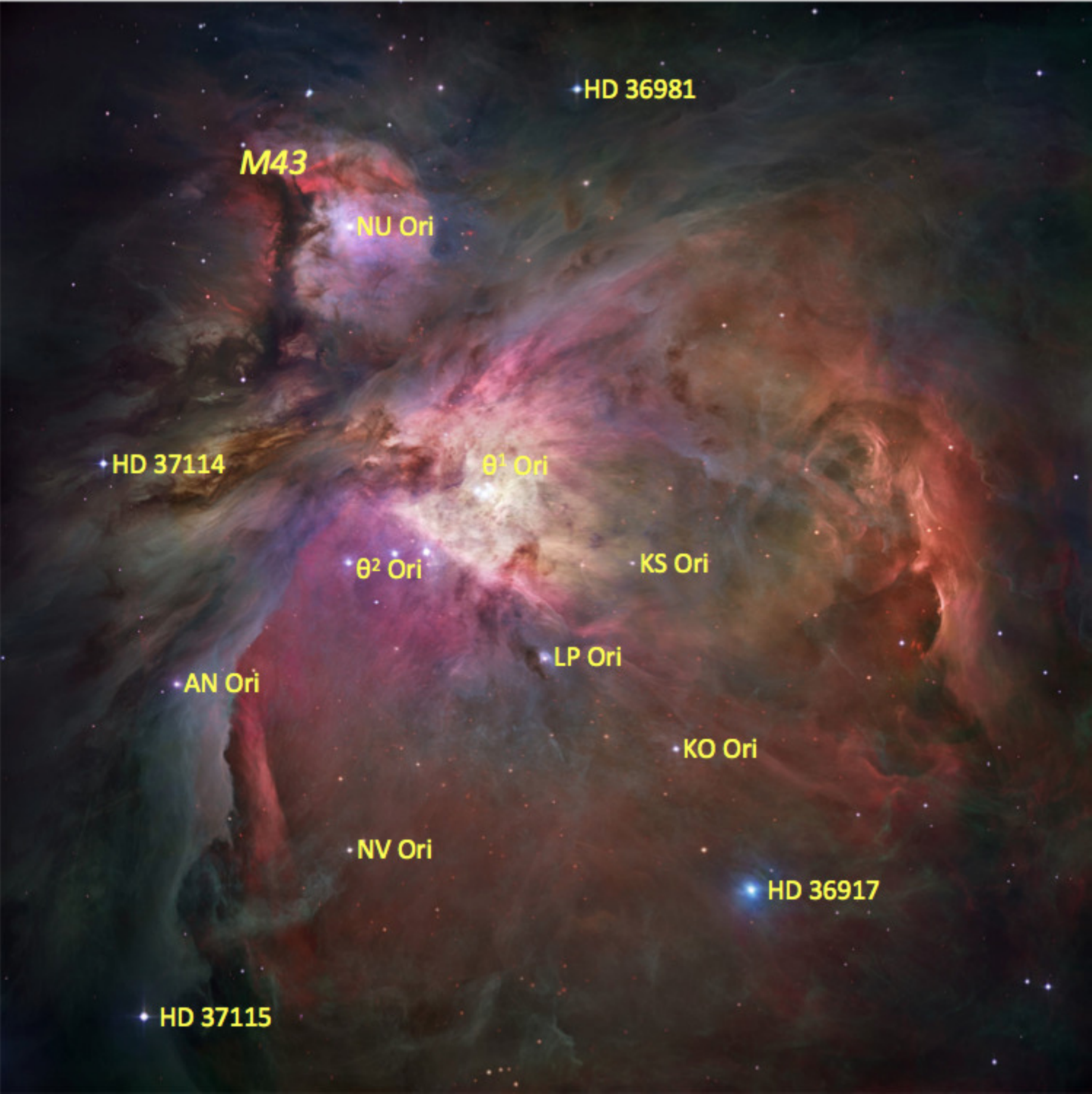}
\caption{ACS color composite image, together with the names of the main bright sources in the field of view\label{Fig:BigChart_ACSstars}}
\end{figure*}

\begin{figure*}
\plotone{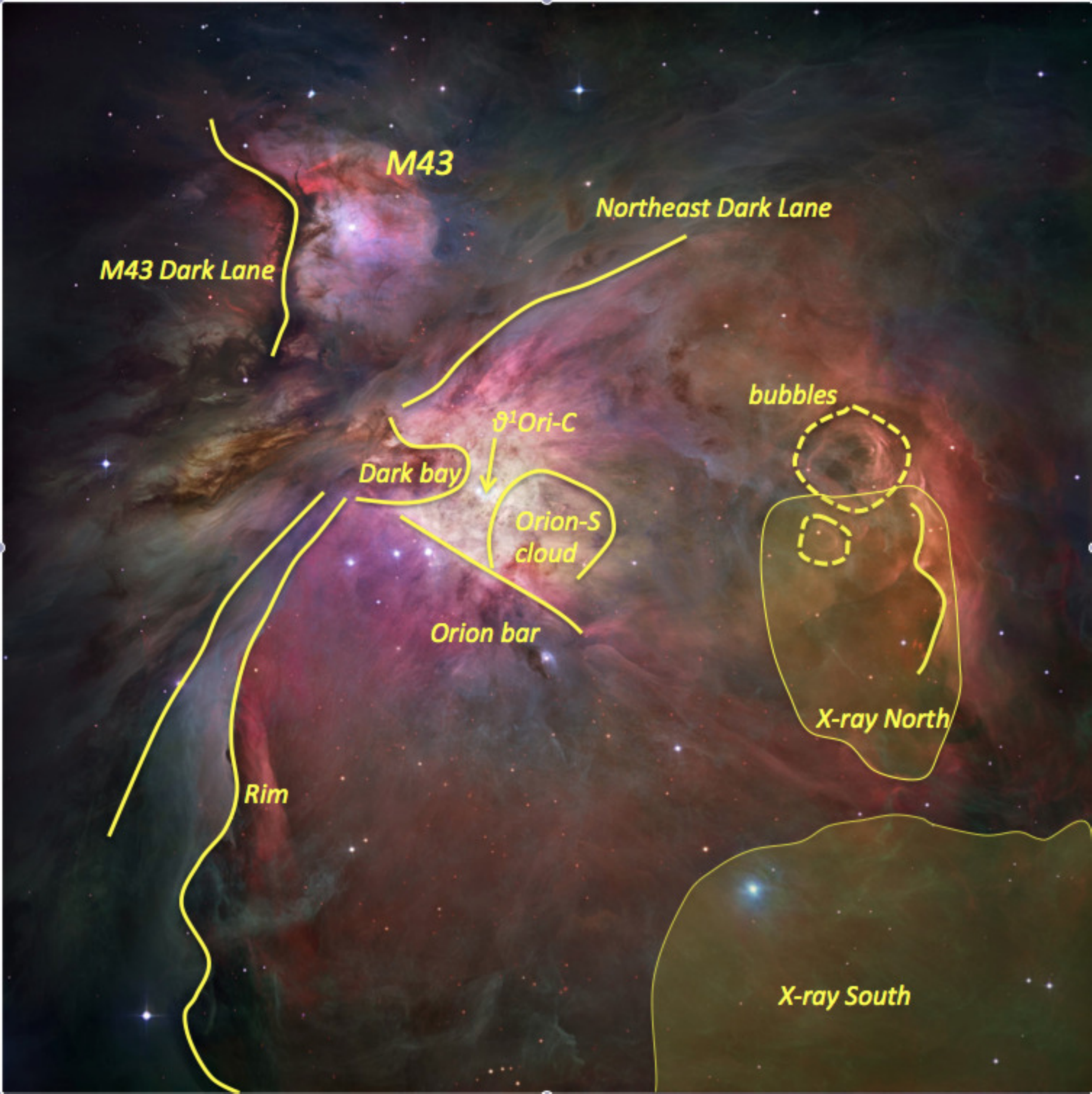}
\caption{Same as Figure \ref{Fig:BigChart_ACSstars}, but indicating the principal nebular features of the Orion Nebula according to \citet{ODellHarris2010}. See the text for details\label{Fig:BigChart_ACSdiffuse}}
\end{figure*}


One of the first products of our survey has been a multicolor image of the Orion Nebula. The image, available in both .pdf, .jpeg and .tiff format with resolution up to $16,000\times16,000$ pixels, can be downloaded from {\rm http://hubblesite.org/gallery/album/pr2006001a}. To facilitate comparison with other images and datasets, we have extracted the RGB planes of the $16,000\times16,000$ jpg image and converted them into FITS files adding basic astrometric information. They are available at the web site {\rm http://archive.stsci.edu/prepds/orion}. For accurate astrometric work, however, the drizzled strips at full resolution ($50\times50$~mas) presented in this paper and available at the same site should be used. In the Appendix we detail the complex image processing that has led to the production of the spectacular Hubble color image, shown in Figure~\ref{Fig:BigChart_ACSstars} together with the names of the main bright sources in the field of view.

M42 is well known for its complex morphology. The historic reference for the naming of the major features,  summarizing two centuries of early naked-eye observations, is \citet{Holden1882}. The  wealth of data now available allows to better understand this region, in particular the inner few arcminutes \citep[``Huygens region'', see e.g.][]{ODell2001a, ODell2001b, ODell2008}. The Orion Nebula is a fossil cavity carved by the expansion of an $HII$ region originally embedded within the Orion Molecular Cloud. The ionized blister has broken-out of the molecular cloud at the current epoch, providing to our vantage point a low-extinction view of the cluster of young, forming stars.  UV photons are still produced by the most massive stars, generating a  photoionized wind streaming from the molecular cloud interface and filling the cavity. \citet{ODellHarris2010} provide the most up-to-date model of the large structure of the nebula, based on the combined analysis of images and spectroscopy of emission-lines from multiple ions of various ionization potential energy. 
In Figure~\ref{Fig:BigChart_ACSdiffuse} we illustrate some of the most prominent features discussed by  \citet{ODellHarris2010}. In particular: a) the ``Bright Bar'', and escarpment of the ionization front nearly edge-on with respect to our vantage point; b) the Orion-South region, an active site of star formation identified as a dense cloud in front of the main ionization front  \citep{ODell+2009}; c) the M43 region around the B1V star NU~Ori, separated from $\theta^1$Ori-C by the optically thick Northeastern Dark Lane. Figure~\ref{Fig:BigChart_ACSdiffuse} also shows the approximate location of the two extended X-ray emitting regions discovered by \cite{Gudel+2008} and the outline of a couple of remarkable circular structures (``bubbles'') at the eastern side of the region.
An earlier 3-d reconstruction of the Orion Nebula has informed the production of the "fly-by" animation  included in the award-winning {\sl Hubble 3D} IMAX$^{\circledR}$ 3D documentary\footnote{http://www.imax.com/hubble/}.

\subsection{Color-Magnitude diagrams} \label{section:results:CMDs}
\begin{figure}
\epsscale{1}
\plotone{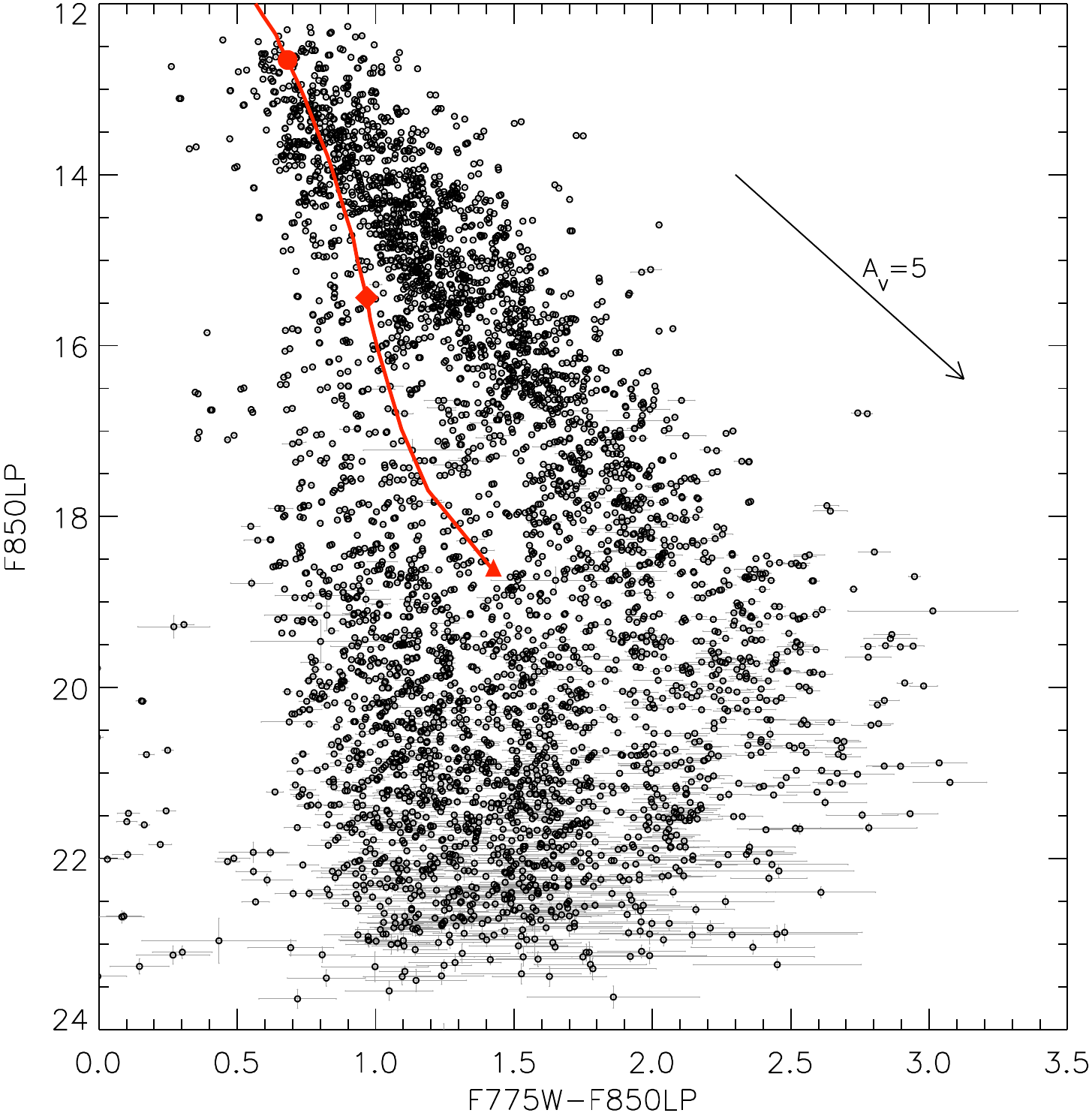}
\caption{$z$ versus $(i-z)$ color magnitude diagram from our ACS photometry. The red line represents a \citet{baraffe1998} 1~Myr isochrone for $A_V=0$. Along this line, the circle marks a mass of $0.5$~M$_\odot$, the diamond symbol the H-burning limit ($M=0.08$~M$_\odot$), and the triangle $M=0.02$~M$_\odot$. The arrow indicates a reddening vector of $A_V=5$~mag, assuming the extinction law of \citet{cardelli1989}.\label{Fig:CMD_ACS}}
\end{figure}

\begin{figure}
\epsscale{1}
\plotone{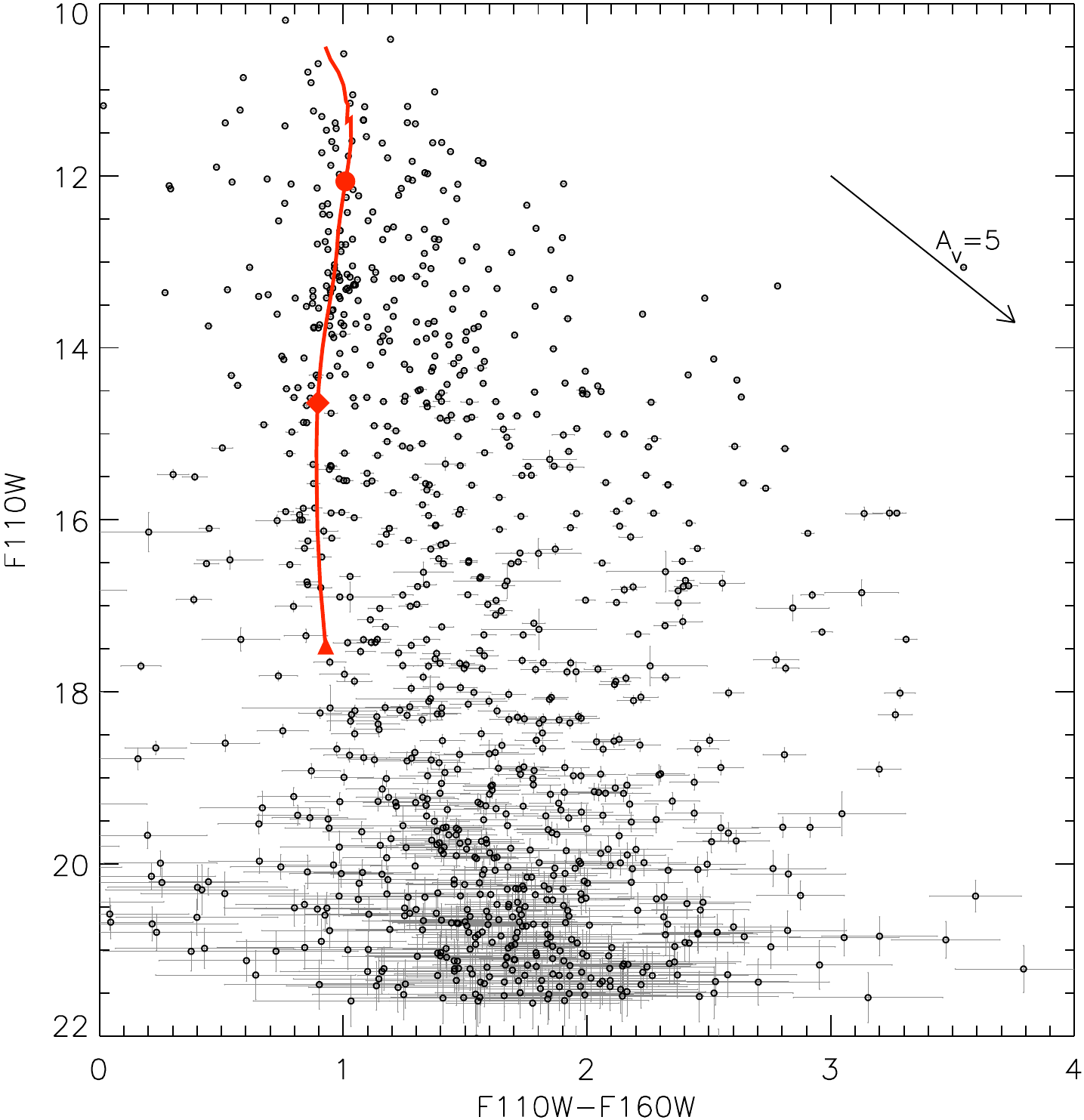}
\caption{Same as Figure \ref{Fig:CMD_ACS}, but for using the NICMOS photometry (F110W versus F110W-H160W).\label{Fig:CMD_NICMOS}}
\end{figure}

Color-color and color-magnitude diagrams allow to isolate stars with anomalous properties, like extreme blue colors, due either to light scattered off the disk or emitted by accretion, as well as foreground objects with very low reddening.
In Figure~\ref{Fig:CMD_ACS} we show the $z$ vs. $(I-z)$ color-magnitude diagram, which shows the clear separation between the young ONC sequence (the diagonal distribution of source located mostly above the isochrone) and the background population (the population of data points below the isochrone. The ONC stars are distributed  along the reddening direction, the density of points tracing the luminosity function, and therefore the initial mass function, of the cluster. The same diagram made for the NICMOS filter (Figure~\ref{Fig:CMD_NICMOS}  shows a much less distinct PMS locus, presubably because the IR filters penetrate better into the molecular cloud, yielding stellar members with a wider range in $A_V$. 
In both  Figure~\ref{Fig:CMD_ACS} and Figure~\ref{Fig:CMD_NICMOS} we also display a theoretical PMS isochrone. This is a 1~Myr isochrone from Baraffe et al (1998), converted in HST magnitudes through synthetic photometry assuming model spectra from \citet{2010arXiv1011.5405A}. The isochrone is merely shown for illustrative purpose, to highlight the depth of our photometric survery, reaching well down to planetary masses. It is well known that both evolutionary models and synthetic spectra are somewhat inaccurate in the VLMS and BD regime, and they should be calibrated empirically before deriving stellar properties from the observed quantities.
For the analysis of the NICMOS dataset and the derivation of the NIR IMF we refer to \citet{Andersen+2011}; see also Section \ref{section:results:other}.
\subsection{Circumstellar disks and proplyds} \label{section:results:proplyds}
\begin{figure*}
\plotone{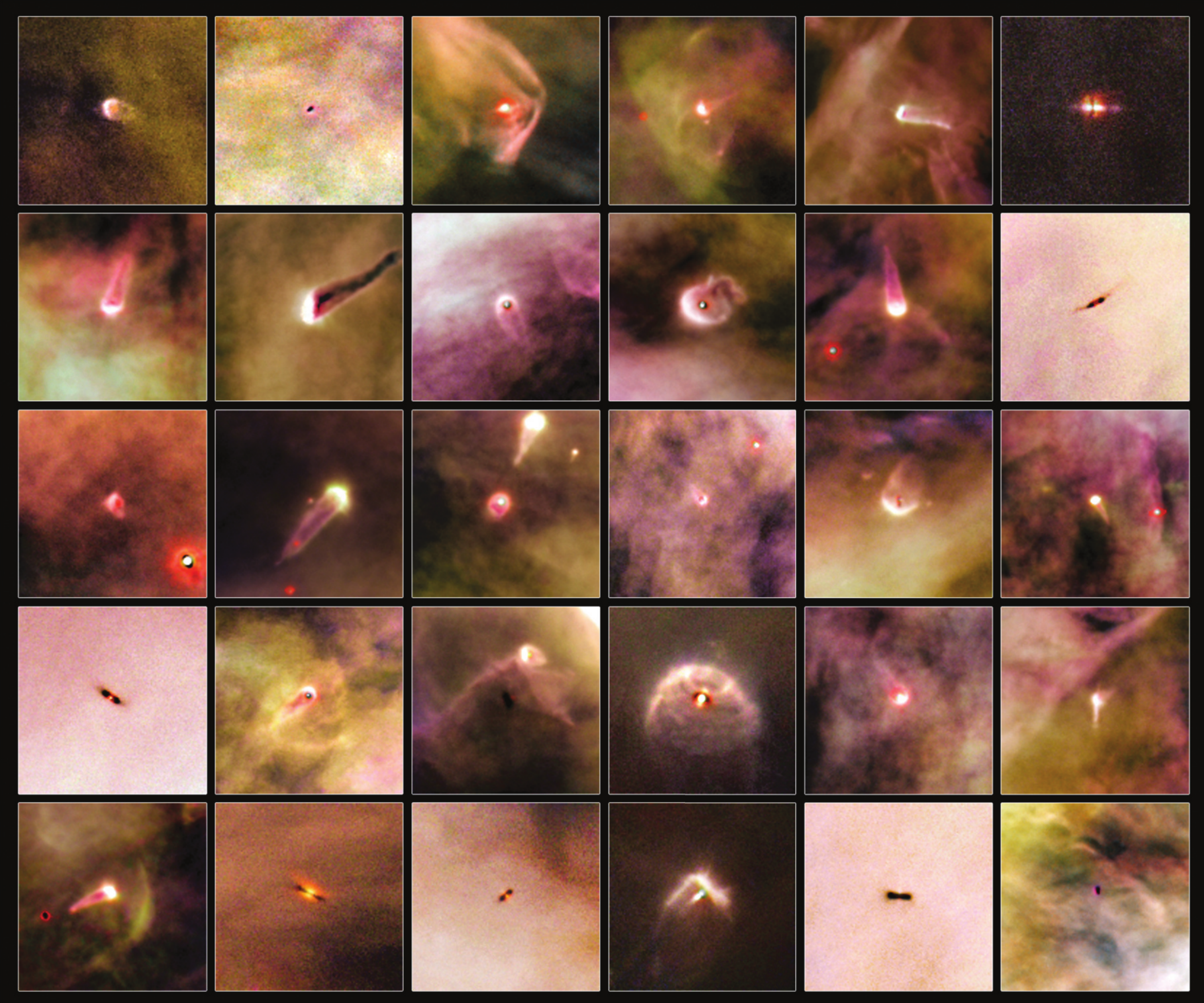}
\caption{A beautiful collection of protoplanetary disks in the ONC observed in our ACS imaging. In many cases they appear externally photoionized. Images are adapted from \citet{Ricci_atlas} and the ESA Press Release heic0917 ({\em Image credit: NASA, ESA and L. Ricci, ESO})\label{Fig:Ricci}}
\end{figure*}

The excellent angular resolution of our \HST\ survey allows us to resolve a large number of protoplanetary disks in Orion. \citet{Ricci_atlas} used the ACS images to obtain a census of all of such objects detected in our ACS imaging survey. The list counts 178 externally ionized protoplanetary disks \citep[\emph{proplyds}][]{odell-wen1994}, 28 disks seen only in absorption against the bright nebular background (\emph{silhouette disks}), 8 disks inferred by the presence of an opaque mid-plane between extended bipolar emission (\emph{bipolar nebulae} or \emph{reflection nebulae}), and five sources showing jet emission with no direct evidence of disk (either as a dark silhouette or through photo evaporated emission). As usual, many of these disks are associated with jets seen in H$\alpha$ and/or circumstellar  reflection emission seen in broadband filters. 
The work of \citet{Ricci_atlas} also resulted in an ESA Press release (heic0917)\footnote{available at  http://www.spacetelescope.org/news/heic0917/}. A color-composite ACS image of a sample of disks is presented in Figure \ref{Fig:Ricci}.

Using ACS data, two disks have been studied in detail. \citet{robberto+2008} have analyed the photo ionized proplyd 124-132. This remarkable systems appears as a photoevaporated disk surrounding a two pointlike sources separated by 0.15\arcsec, or about 60 AU at the distance of Orion. The authors show that the luminosity and colors of the sources are compatible with a pair objects of similar substellar mass ($\simeq0.04$~M$_\sun$) and about 1 Myr old. On the other hand, more massive stars under a large amount of extinction could also explain the observed fluxes; spectroscopic confirmation is needed to disentangle the two cases.  In any case, this system represent the first direct observation of a circumbinary disk undergoing photoevaporation. The second system investigated in detail is the well known dark silhouette disk 114-426 \citet{miotello+2012}. The high signal-to-noise ratio of the ACS images has allowed to clarify the complex morphology of the system and unveil evidence of photoevaporation. By comparing the opacity of the outer regions at different wavelengths, the authors have reconstructed the distribution of dust grains finding evidence for large (micron size) grains with a spatial gradient compatible with a photo-evaporative wind.

\subsection{Other studies} \label{section:results:other}
Our accurate multi-band photometry for thousand of ONC stars provides a unique dataset for studies of this young population. In the past few years preliminary photometry has been utilized for a number of investigations. In this section we highlight the main ones.
%
\citet{dario2009,dario2010} used the WFI ground-based optical photometry obtained with the MPG-ESO 2.2~m telescope in La~Silla as a follow-up to the \HST\ Treasury Program to study the stellar population of the ONC down to 0.1~M$_\odot$. In these works, the ACS photometry was used to estimate the photometric completeness of the ground based data, given the significantly deeper detection limit of the \HST\ survey. The second paper presents a new version of the Hertzsprung-Russel diagram of the cluster first derived by \citet{hillenbrand97}. The new estimates of stellar masses and ages indicate a peak age of $\sim$2-3~Myr and a flattening or turn-over of the IMF at about $0.2-0.3$~M$_\odot$. Recently, \citet{2012ApJ...748...14D} have used new WFI data to expand this study into the substellar regime.

A more detailed analysis of the age spread of the ONC, using the ACS data presented in this paper, has been performed by \citet{reggiani2011}. Using the bayesian analysis tool {\tt Chorizos} \citep{maiz-apellaniz2004}, they  estimate the extinction and accretion luminosity towards each source. From the isochronal ages, accounting for  all major sources of uncertainty through Monte Carlo modeling, they derive a mean cluster age of 2.2 Myr with a spread of  of few Myr, inconsistent with that of a coeval stellar population and in agreement with a star formation activity lasting between 1.5 and 3.5 Myr.

The NICMOS data were used by  \citet{Andersen+2011} to determine the ratio of low-mass stars to brown dwarfs  as a function of radial distance, out to about 1.5~pc. The comparison with the results previously obtained for the central 0.3~pc$\times$0.3~pc region suggests that the fraction of low-mass members of the cluster is mass segregated.

Following an early study of the ONC core by \citet{Robberto+2004}, who used WFPC2 data to derive accretion rates $\dot{M}$ in the core of the ONC for 40 stars,  \citet{manara+2012} used analyzed the new WFPC2 photometry to derive the mass accretion rate for $\sim$700 cluster members using both the $U$-band excess and $H\alpha$ emission. This is the largest sample of nearly coeval PMS sources to date for which $\dot{M}$ has been derived, and has allowed to  perform the most complete statistical study of the final phases of the stellar mass build-up.
The data show trends between the mass accretion rate  and the age and mass of the sources, with a mass accretion rate decaying more rapidly for lower stellar masses.

Two studies used the dataset presented in this paper to analyze the proper motion of selected sources. \citet{ODell2005} used the ACS and WFPC2 images as a second epoch, together with previous \HST\ observations of the ONC, to measure proper motions of three stars, JW~349, JW~355, and JW~451 that had been reported as high-velocity low-mass runaway stars. The combination of HST data does not show evidence of significant proper motion.
\citet{henney2007} also used our \HST\  images, together with high resolution spectra and radio maps, to study the nature of optical outflows in the vicinity of the Orion South region.

The combination of the HST data with those obtained by other great observatories like Chandra and Spitzer is still in an early phase. 
\citet{prisinzano2008} used the X-ray COUP survey \citep{getman2005} to select 41 Class 0-Ia candidate sources. The data from our \HST\ Treasury Program were used to trim the list by removing objects detected at visible wavelengths. The authors find that Class 0-Ia objects appear to be significantly less luminous in X-rays than the more evolved Class II stars with mass larger than 0.5 M$_\odot$.
\citet{aarnio2010} used the ACS and WFI fluxes, together with {\sc 2mass} and {\it Spitzer} IRAC+MIPS fluxes, to model the spectral energy distributions  of the 32 most powerful X-ray flaring sources from the {\it Chandra} Orion Ultradeep Project \citep[COUP][]{getman2005}. They were able to determine for each star the location of the inner edge of the circumstellar disk relative to the co-rotation distance from the star, based on each star's known rotation period from \cite{stassun1999}. In most cases the large magnetic loops are not anchored to the circumstellar disks, but appear as free-standing magnetic structures. 

The scientific potential of the dataset presented in this paper still has to be fully mined. Our team is working on the analysis of the sample of binary sources, the accurate reconstruction of individual SEDs and on the multicolor study of other prominent proplyds. Also the matching of the HST data with other datasets, both from Spitzer and Chandra and ground based telescopes relative e.g. to radial velocity, proper motions and source variability will provide unique information on the structure and evolution of the ONC and Pre-Main-Sequence evolution in general. By making the entire dataset available to the community we hope to stimulate research in these directions.

\section{Conclusion}
In this paper we have presented the observing strategy, data analsysis, data products and a summery of the scientific results obtained by the \HST\ Treasury Program on the Orion Nebula Cluster (HST GO-10246, PI M. Robberto). The survey, carried out simultaneously with all \HST\ imagers (ACS, WFPC2 and NICMOS), probed the Pre-Main-Sequence population of the Orion Nebula Cluster down to masses of the order of 1 Jupiter mass. The finally reduced images and the photometric databases, including the complementary ground based data obtained at La Silla and Cerro Tololo, are available to the community as high level data products on the Multimission Archive hosted by the Space Telescope Science Institute (http://archive.stsci.edu/preds/orion/). 

\acknowledgments
The authors thank Rusty Whitman and the TRANS development team at STScI for their support crafting the \HST\ visits pattern; Tony Roman was the Program Coordinator for STScI; Ilana Dashevsky, Alfred Schultz and the \HST\ TRANS development team for allowing NICMOS to use the FOM dithering; Ron Gilliland for pointing out the opportunity of using gain=2 settings with ACS. We also acknowledge the anonymous referee for useful comments.

Based on observations made with the NASA/ESA Hubble Space Telescope under the GO program \#10246, obtained at the Space Telescope Science Institute, which is operated by the Association of Universities for Research in Astronomy, Inc., under NASA contract NAS 5-26555. These observations are associated with program \#10246.

\appendix \label{section:appendix}

\section{Production of the \HST\ image} \label{section:appendix:HSTimage}
The set of ACS images presented in this paper has been used to obtain the most detailed images of the Orion Nebula to date. We describe here the process of producing a color composite image from the assembled Hubble data for distribution to the public, media and astronomers.

Working with over one billion pixels of high dynamic range ACS data presented practical challenges. An equivalent challenge was producing an image distinct from the gallery of existing photographs of this popular target, but not so unusual as to appear unrealistic. Compromises were necessary to achieve a result with an unwieldy volume of data and subjective choices, made for largely aesthetic reasons, were needed to arrive at the final product. The entire process was carried out in collaboration with the science team, keeping in mind a set of main goals:
1) Produce a visually striking image, with 2) Realistic overall appearance, tonality and color; 3) Render coolest stars red, hottest stars blue;
4) Represent the broad dynamic range in the data. 5) Render detailed structure present in the data; 6) Stay honest to the data.

\begin{figure*}
\epsscale{1}
\plotone{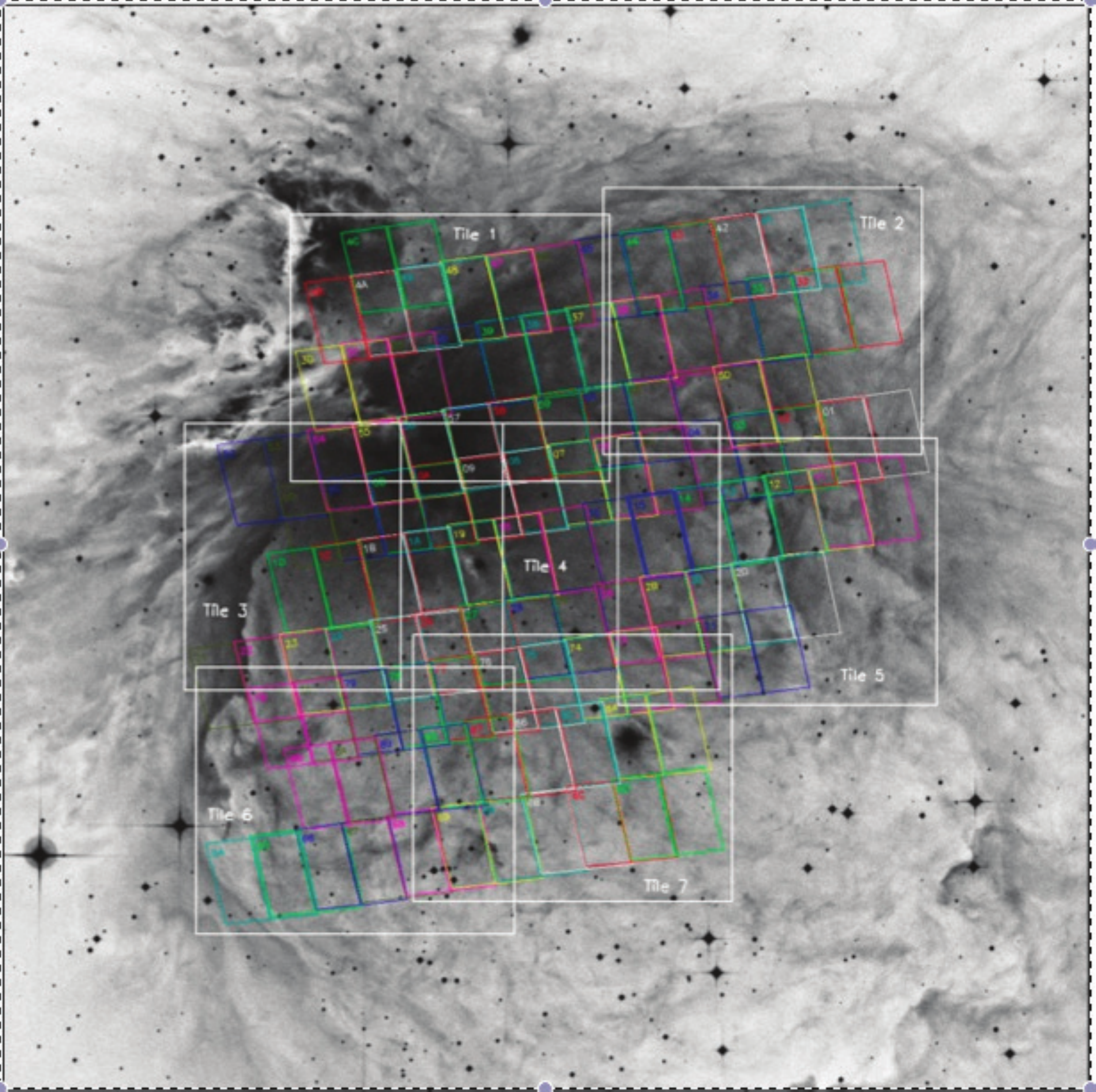}
\caption{Distribution of 104 ACS fields on the sky and outline of the 7 primary tiles used for the prodution of the \HST\ image,
superimposed on an AAT image of the Orion Nebula (David Malin Images). The primary tiles (white rectangles) represent the results of initial data reduction, seven equal-sized tiles of $13,000\times14,000$ pixels, or 174Mpx.\label{Fig:supertiles}}
\end{figure*}

The data used as input were the 7 overlapping tiles, each $13,000\times14,000$~px (174~Mpx) mentioned in Section~\ref{section:datareduction:ACSdrizzling} and shown in Figure~\ref{Fig:supertiles}. This amounts to 35 FITS files, 1.35~GB each. Fully assembled, accounting for overlap and some margin, they correspond to $32,567\times35,434$~px or 1.1~Gpx.

The remainder of this report explains these procedures in somewhat more detail. 

\subsection{Intensity scaling of each tile/filter}
\begin{figure*}
\epsscale{1}
\plotone{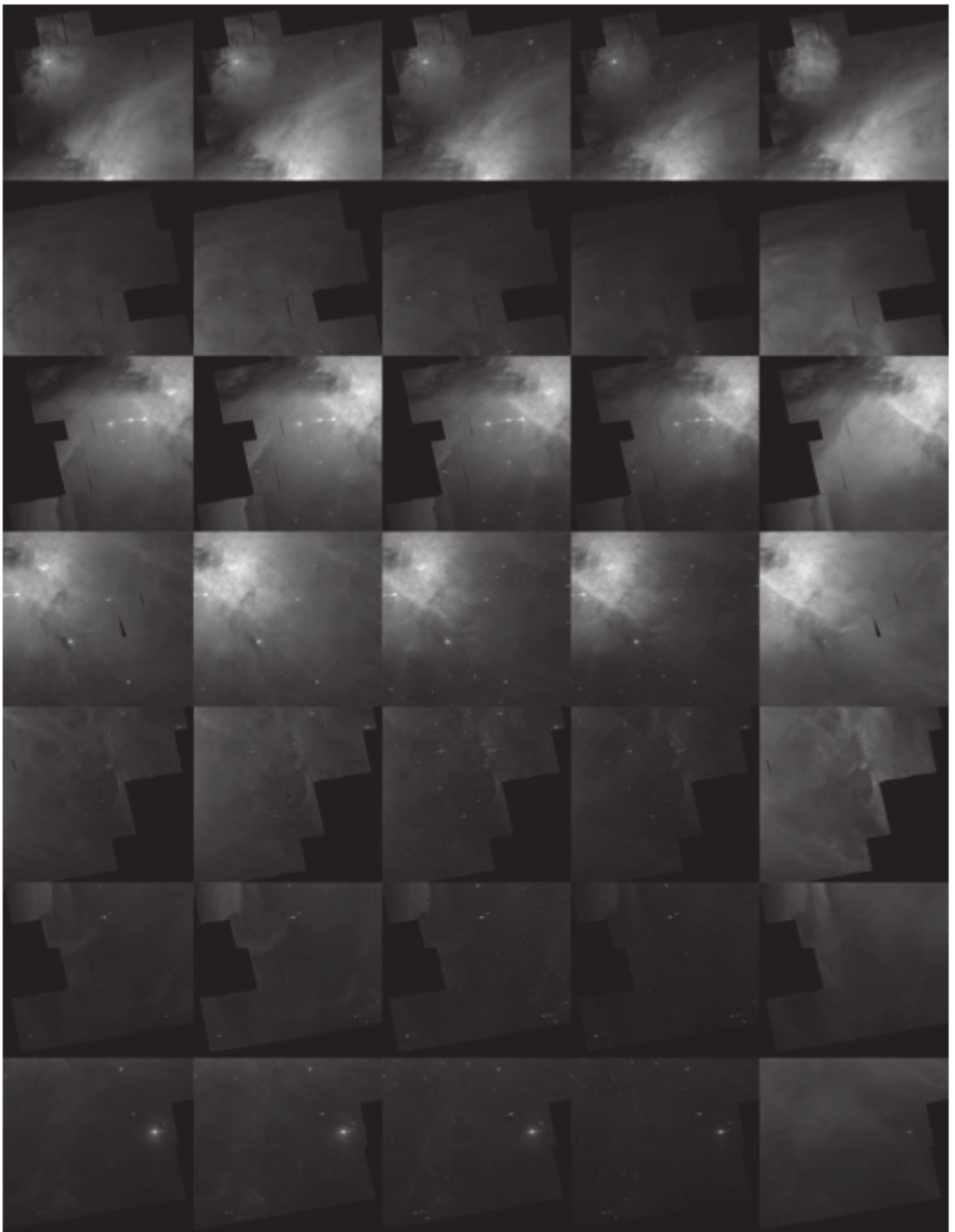}
\caption{Images resulting from initial intensity scaling to 16-bit, gray, flat Photoshop documents, each $13,000\times14,000$~px (174Mpx). Images in each row represent data from the five different filters, left to right: $B$, $V$, $i$, $z$, H$\alpha$. Images in each column represent the initial tiles from each filter.\label{Fig:35supertiles}}
\end{figure*}

The first step was to scale each of the 35 input files into an editable, 16 bits per channel (bpc) image. Since the ACS data were processed and calibrated consistently, the same intensity-scaling parameters could be used for each tile and filter. Parameters were adjusted to produce a wide, smooth range of tones, with detail apparent in the brightest (highlight) and darkest (shadow) regions across all of the images (Figure~\ref{Fig:35supertiles}).

\subsection{Retile of the full mosaic}
\begin{figure*}
\epsscale{1}
\plotone{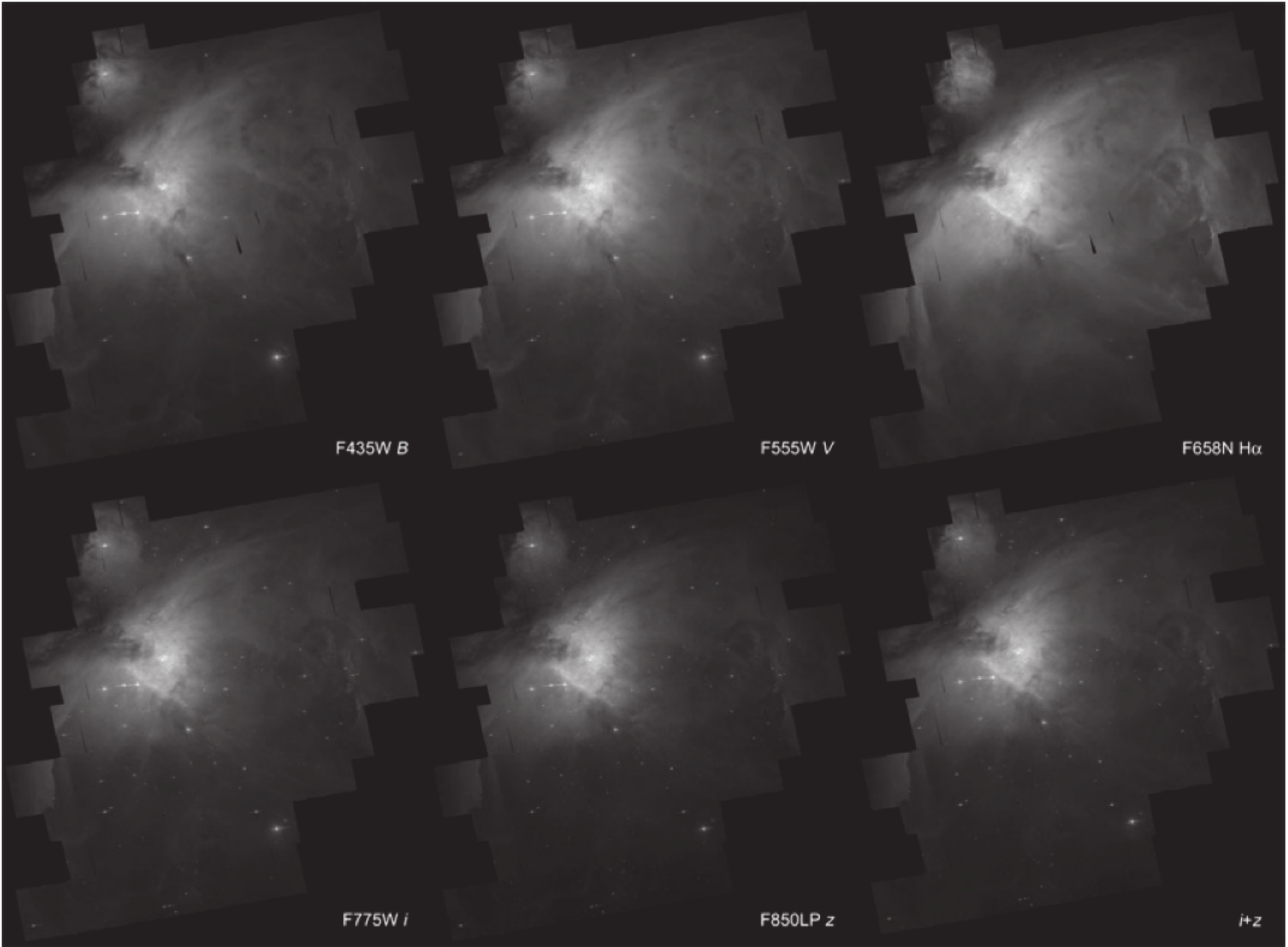}
\caption{Images retiled for each filter, 16-bit, gray, flat Photoshop documents, each $32,567\times35,434$~px (1.1~Gpx).\label{Fig:35superimages}}
\end{figure*}

The next step converted each set of the seven 16 bpc tiles, one set for each filer,i nto a single mosaic. 
The image size had to be reduced by a factor of two in each dimension because hardware and software limitations prevented working with the full-size, 16-bit images. While there was a strong desire to keep the image at the native pixel scale (0$\arcsec$.050/pixel), it soon became clear that this would have been highly impractical, given the state of the art in desktop processing at the time and the size of the full-scale image, $36,000\times36,000$~pixel or some 1.2~Gpx. Given the additional overhead of multiple image and adjustment layers, masks, etc., for image editing, anything larger than a $\simeq 16,000\times16000$~pixel proved prohibitive (Figure~\ref{Fig:35superimages}).

\begin{figure*}
\epsscale{1}
\plotone{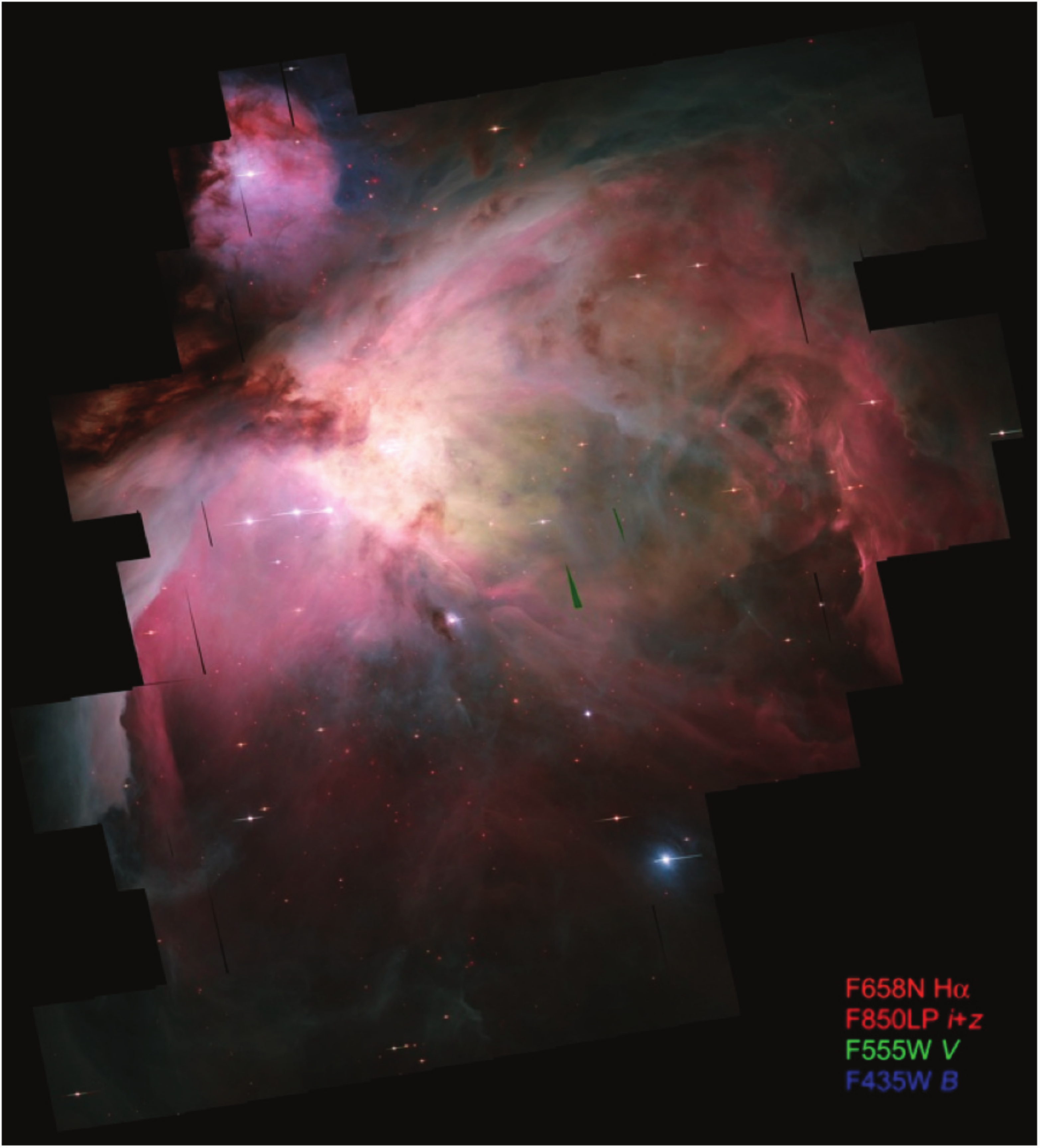}
\caption{Initial color composite image including only \HST\ ACS data, 16-bit RGB $32,567\times35,434$~px (1.1Gpx).\label{Fig:InitialColorComposite}}
\end{figure*}

The most straightforward way to reconstruct a color image assigns different images to the three additive primary color channels, red, green, and blue. However, it is possible to make a color composite with more than three constituent images using a layering paradigm to assign hue to individual gray scale images. After testing various combinations of filters and color assignments, the two reddest filters, F850LP and F775W were averaged into a single image. The final color adjustments were: F658N was rendered in red/orange, F850LP+ F775W in red, F555Win green, and F435W in blue (see Table \ref{Tab:assigned-colors-for-colorcomposite} and Figure~\ref{Fig:InitialColorComposite}).

\begin{table}
\begin{center}
\caption{Assigned colors to produce the color-composite image \label{Tab:assigned-colors-for-colorcomposite}}
\begin{tabular}{ccc}
\tableline
\tableline
{HST ACS Filter} & {ESO Filter} & {Assigned Color} \\
\tableline
F435W $B$ 			  & F435W 		&  blue  \\
F555W $V$ 			  & F555W 		& green  \\
F775W $i$+F850LP $z$     & [SII] 		          & red    \\
F658N $H\alpha$       & H$\alpha$ 	& orange\\
\tableline
\end{tabular}
\end{center}
\end{table}

\subsection{Combine with ground-based ESO data}
\begin{figure*}
\epsscale{1}
\plotone{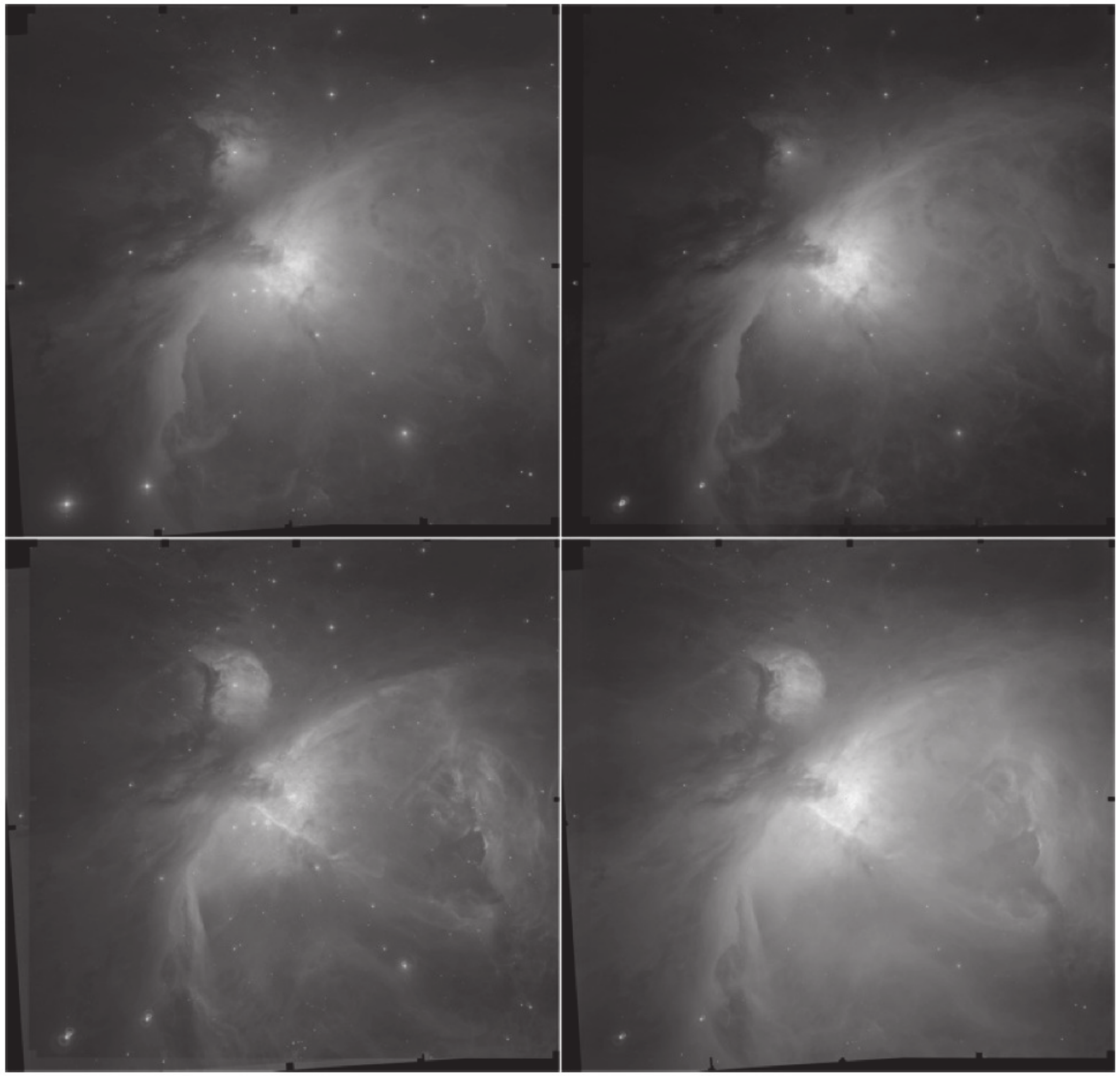}
\caption{Scaled MPG/ESO 2.2m WFI (ground-based) images, top-left: B; top-right: [O III], bottom left: [S II], bottom-right: H-alpha, 9064x8726px (75Mpx).
\label{Fig:ESO}}
\end{figure*}

The next step was the matching with ground based data. This was necessary because the ACS survey was not designed to fill out a rectangular pattern. The combination of stripes resulted in a mosaic with ragged edges and a few interior gaps. We felt that a clean, rectangular composition would be more appropriate for a public presentation of the image, as  the irregular edges would strongly distract viewers from the primary subject and finely detailed structure of the image.
The science team provided another set of data in five filters: together with the WFI  U, Band H$\alpha$ data presented by \citet{dario2009}, we used two unpublished images also taken at the MPG/ESO 2.2m telescope, one in sulfur ([S II]) and the other in the oxygen ([O III]) filter. These images matched closely enough the passbands of their closest ACS filters (Figure~\ref{Fig:ESO}).

\begin{figure*}
\epsscale{1}
\plotone{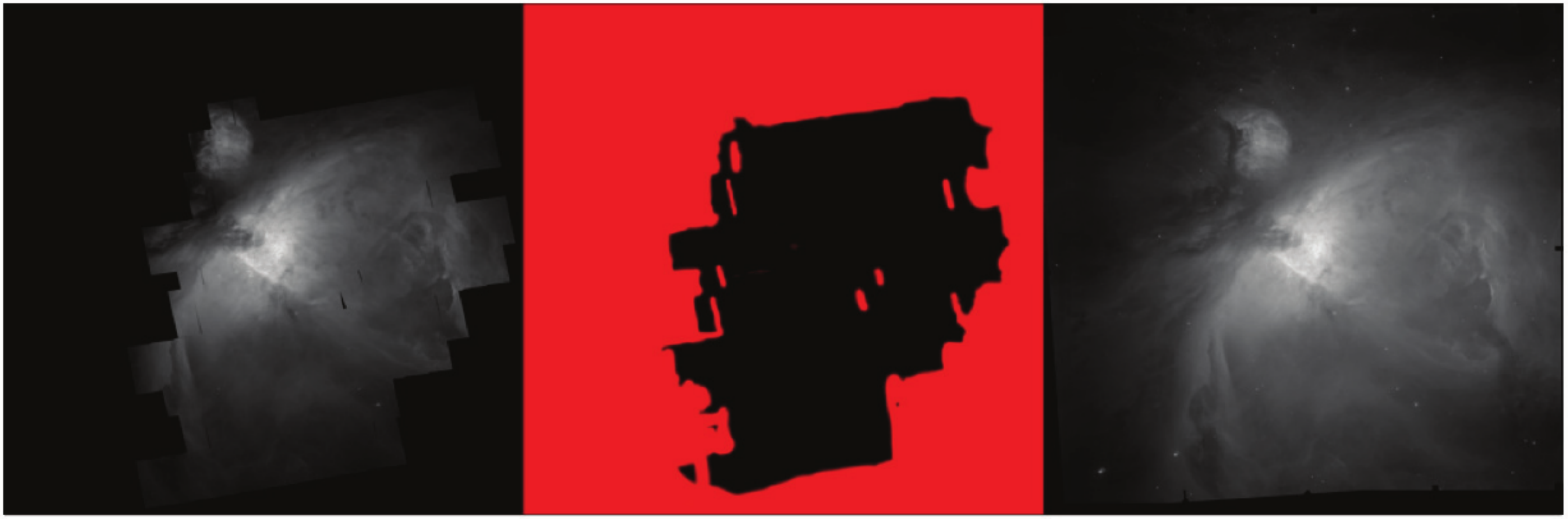}
\caption{Left: \HST\ ACS H-alpha mosaic, Center: mask to blend with ground data; Right: scaled ESO H-alpha image.
\label{Fig:Mask}}
\end{figure*}

The \HST\ data where then scaled from the full-dynamic-range FITS into 16-bit images. Slightly different scaling parameters were used for each image, though a logarithmic transformation was used for all. The goal was to balance shadow and highlight values with a consistent tonal range among the four images used.
The scaled ESO data were finally combined with the \HST\ data scaled at the ESO pixel size, 16-bits, separately for each filter combination. A mask was generated starting with the non-black pixels in each \HST\ image. This mask was modified to blend the two images smoothly (Figure~\ref{Fig:Mask}. Brightness adjustments were applied to match the images where they overlap. In addition to the large areas of blank data filled by the ESO images, areas were blended to fill in smaller gaps. After up-resampling the masked composite to the \HST\ pixel size, the \HST\ images were reincorporated to produce a flat, 16-bit version of each filter/color, cropped to the final orientation and size, $36,000\times36,000$~px (1.2Gpx).

\subsection{Composite, balance tonal range per filter, and apply color}
\begin{figure*}
\epsscale{1}
\plotone{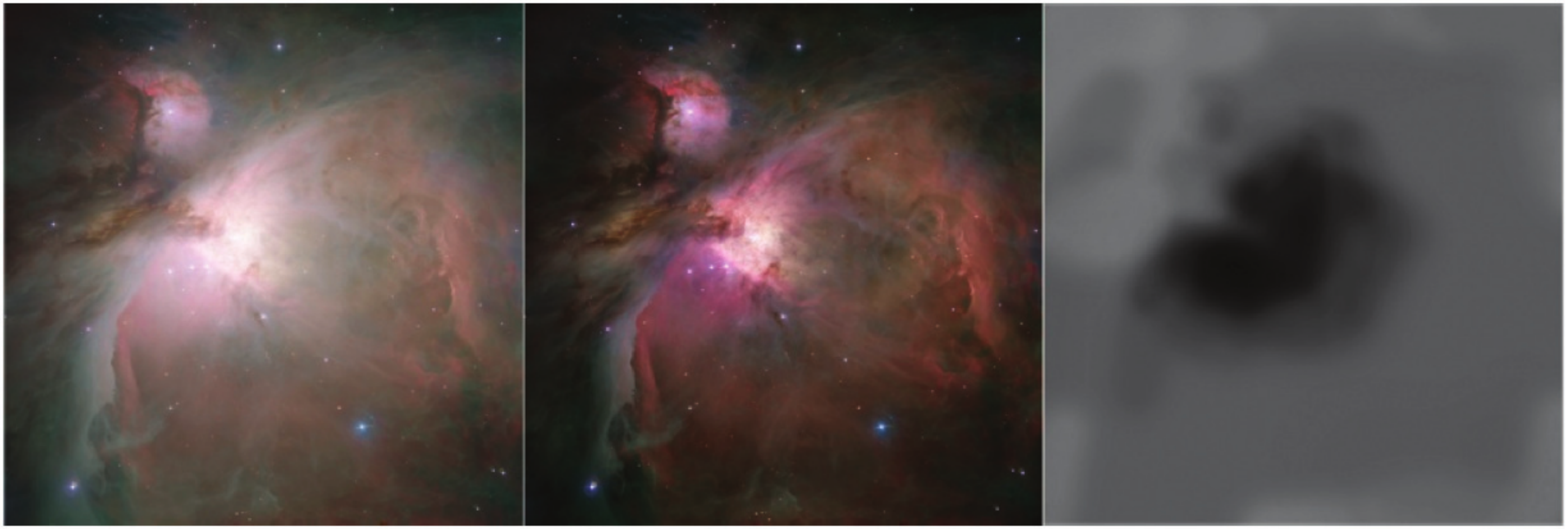}
\caption{Left: initial composite including both ACS and ESO data, Center: result after applying brightness, contrast, and color balance adjustments, Right: cumulative effect of adjustments and masks on a constant, gray image to illustrate smoothness of the adjustments applied based on the smoothed image brightness.
\label{Fig:adjustments}}
\end{figure*}

The next step was the creation of a color composite from the combined \HST\ and ESO images, based on the initial prototypes with the \HST\ data only.  Brightness, contrast, and color balance adjustments were applied at this stage, both to the individual filter components and to the combined image. Masks were used to apply adjustments to localized areas with the goal of preserving detail and maximizing local contrast across the image (Figure~\ref{Fig:adjustments}).

\subsection{Resize, adjust brightness, contrast, color}

At this point the image was downsized by 2$\times$ in each dimension, 18,000$\times$18,000 pixels (309~Mpx), before applying further adjustments to improve brightness, contrast and color. Again, masks were used to apply adjustments in selected regions to expand the overall tonal range. To further facilitate additional editing, the flat, 16-bit color composite was tiled into four pieces in a 2x2 array.

\subsection{Retouch cosmetic artifacts}

\begin{figure*}
\epsscale{1}
\plotone{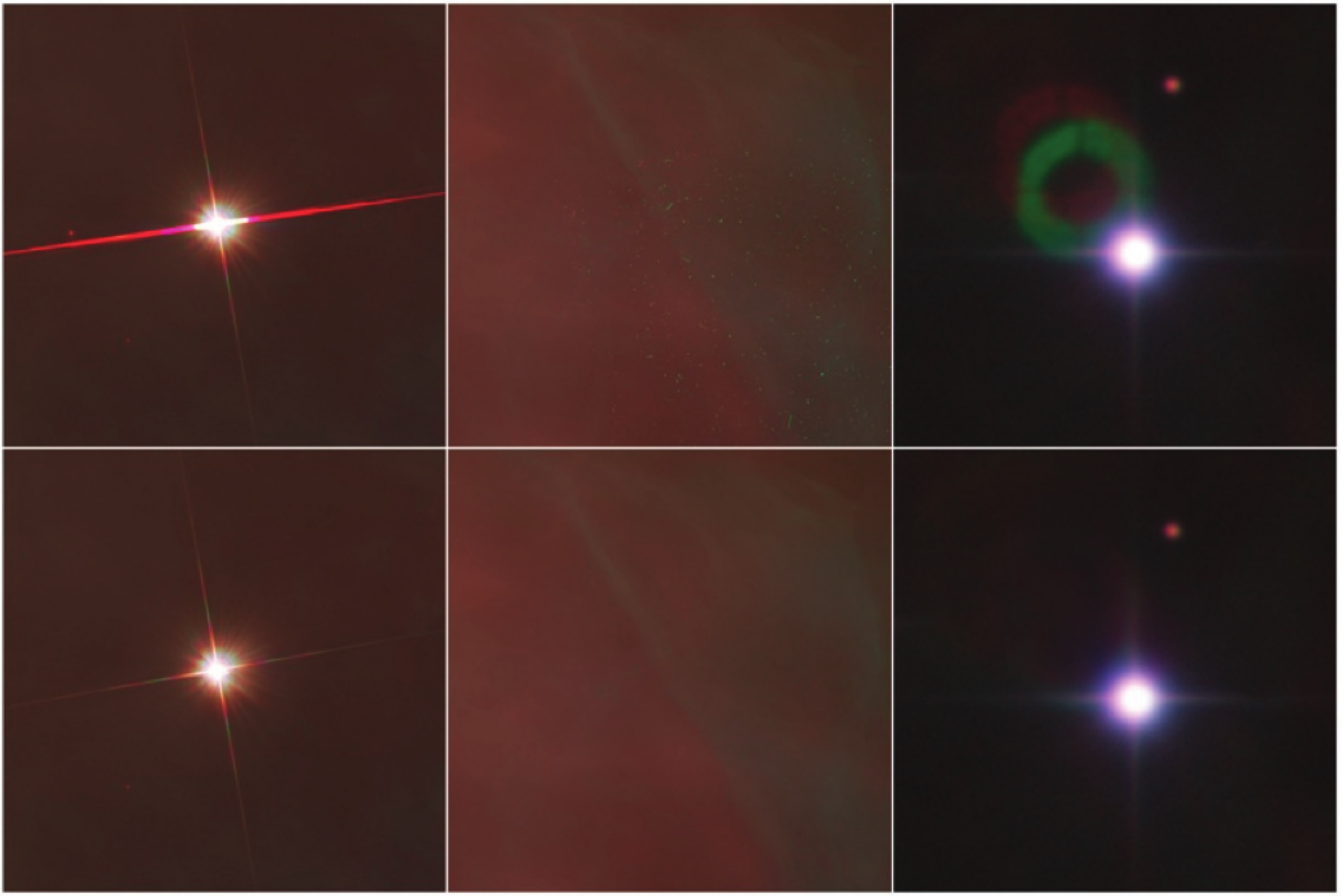}
\caption{Examples of cosmetic retouching on small sections of the adjusted color composite mosaic. Top: three examples of observational artifacts: CCD charge bleed, residual cosmic rays, and internal reflections. Bottom: retouched versions of the same image sections.
\label{Fig:cosmetic}}
\end{figure*}

Our last steps of the process required some retouching and restoration. It is known that 
telescopes and cameras introduce certain well-known artifacts into astronomical images such as:
\begin{itemize}
 \item CCD saturation bleeding
 \item Diffraction spikes
 \item Residual cosmic rays
 \item Internal reflections: primarily on brightest stars in the ground-based image, resulting from light reflecting, from filters and other optical elements within the instrument.
\end{itemize}
Most of these artifacts were removed from our Orion Nebula image using standard digital editing techniques, paying great attention to avoid changing the character of the underlying image (Figure~\ref{Fig:cosmetic}).

\subsection{Retile mosaic, final adjustments, sharpening and high-pass filter}

\begin{figure*}
\epsscale{1}
\plotone{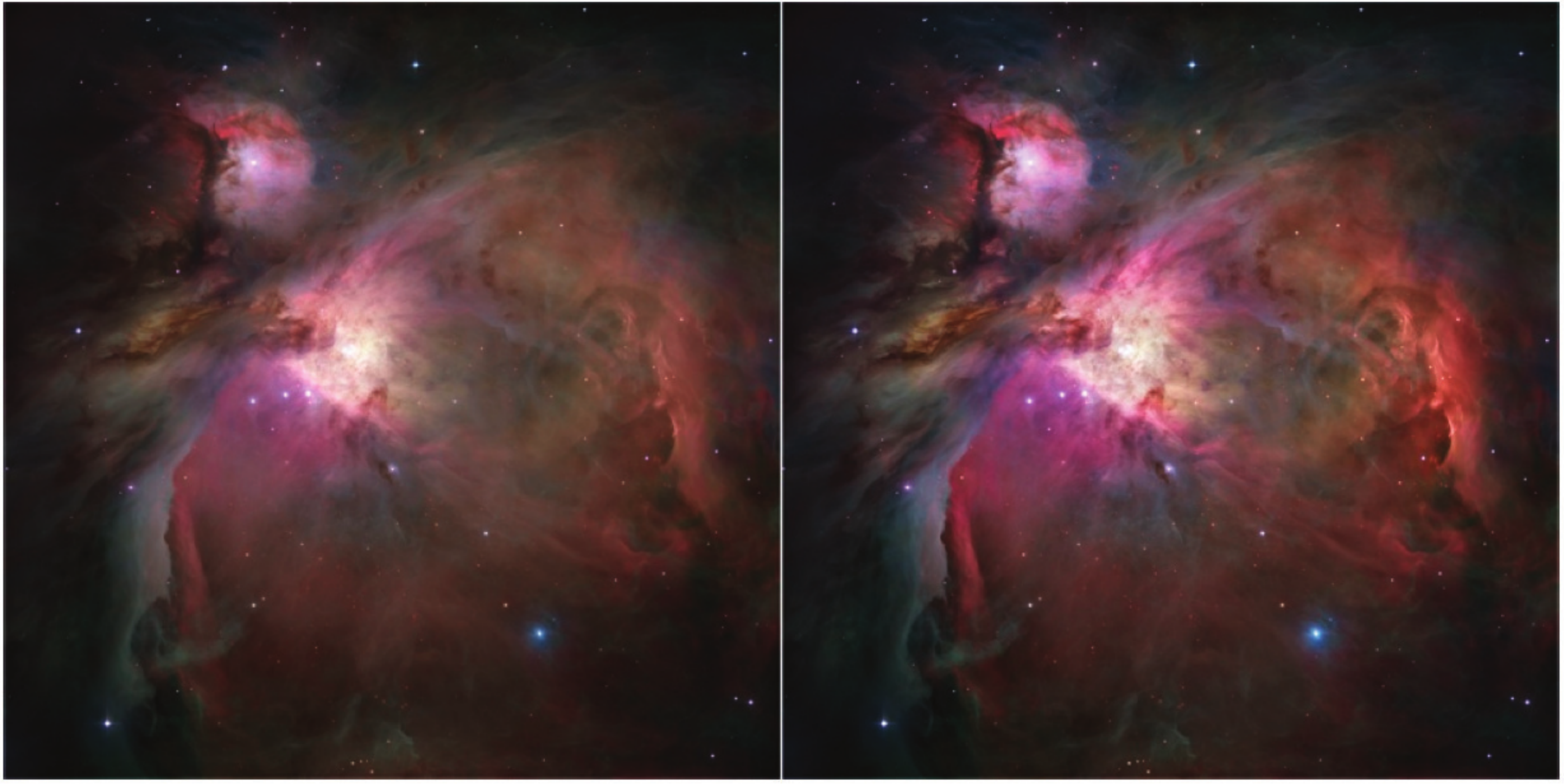}
\caption{Result of final adjustments. The intent was to enhance contrast throughout the image to make a crisper rendition but without sacrificing highlight or shadow detail, nor an overall sense of the very large range of brightness inherent in the data.
\label{Fig:final}}
\end{figure*}

After retouching, the four flattened tiles were reassembled into a full, flat mosaic. Some additional adjustments were applied, mostly to increase the overall contrast, enhancing detail in the shadows while taking care not to lose detail in the highlights.
A slight sharpening was applied to the final, flat image. Care was taken to do it in a minimal amount to avoid introducing visible sharpening artifacts. Finally, the overall contrast was slightly  enhanced by inserting a copy of the image to which a moderate high-pass filter had been applied at $60\%$ opacity (Figure~\ref{Fig:final}).

\end{document}